\documentclass[traditabstract,longauth]{aa} 
\usepackage{multirow}
\usepackage{xcolor}
\usepackage[breaklinks, colorlinks, citecolor=blue, linkcolor=blue]{hyperref}
\usepackage{caption}
\usepackage{subcaption}
\usepackage{threeparttable}
\usepackage{rotating}
\usepackage{rotfloat}
\usepackage{float}
\usepackage{moresize}
\usepackage{wrapfig}
\usepackage{multicol}
\usepackage{wrapfig}
\usepackage[normalem]{ulem} 
\usepackage{txfonts}
\usepackage[T1]{fontenc}
 \usepackage{graphicx}   
\usepackage{amsmath}     
 \usepackage{color}
\usepackage{amssymb}
\usepackage{epstopdf}
\usepackage{natbib}
\bibpunct{(}{)}{;}{a}{}{,}  
\usepackage{url}
\usepackage{booktabs}
\usepackage{supertabular}
\usepackage{longtable}

 \newcommand{\vsini}{$V \sin i_\star$}

\newcommand{\teff}{$T_{\rm eff}$}
\newcommand{\logg}{log\,$g_\star$}

\newcommand{\feh}{[Fe/H]}
\newcommand{\cah}{[Ca/H]}
\newcommand{\mgh}{[Mg/H]}
\newcommand{\nah}{[Na/H]}
\newcommand{\sih}{[Si/H]}

\newcommand{\kms}{km\,s$^{-1}$}
\newcommand{\ms}{m~s$^{-1}$}

\newcommand{\gc}{g~cm$^{-3}$}
\newcommand{\lgr}{$\log\,(R^\prime_{HK})$}

\newcommand{\Lsun}{$L_{\odot}$}                          
\newcommand{\Msun}{$M_{\odot}$}
\newcommand{\Rsun}{$R_{\odot}$}
 
\newcommand{\Fearth}{$F_{\oplus}$}
\newcommand{\mearth}{$M_{\oplus}$}
\newcommand{\rearth}{$R_{\oplus}$}

\newcommand{\mjup}{$M_\mathrm{J}$}

\newcommand{\Prot}{$P_{\rm rot}$}  
\newcommand{\mstar}{$M_\star$}
\newcommand{\lstar}{$L_\star$}
\newcommand{\rstar}{$R_\star$}
\newcommand{\rhostar}{$\rho_{\mathrm{*}}$}

\newcommand{\ticno}{229650439}

\newcommand{\targeta}{\mbox{TOI-1438}}

\newcommand{\targetb}{TOI-1438~b}
\newcommand{\targetc}{TOI-1438~c}

\newcommand{\bjdtdb}{\ensuremath{\rm {BJD_{TDB}}}}
 
 \newcommand{\logRHK}[1][]{$-4.925 \pm 0.013$} 
 
\newcommand{\smassariadne}[1][]{$0.876 \pm 0.038$}    
\newcommand{\smassariadnegrav}[1][]{$0.764\pm 0.127$}  
\newcommand{\smassparam}[1][]{$0.867 \pm 0.027$}    

\newcommand{\sradiusariadne}[1][]{$0.820 \pm 0.017$}    
\newcommand{\sradiusisochrones}[1][]{$0.x\pm 0.0x$}   
\newcommand{\sradiusparam}[1][]{$0.822\pm 0.018$}    
\newcommand{\sradiusspechmatch}[1][]{$0.81\pm 0.08$}    
\newcommand{\sradiusgaia}[1][]{$0.853^{+0.026}_{-0.016} $}   

\newcommand{\srhoariadne}[1][]{$2.2 \pm 0.2$}        
\newcommand{\srhospechmatch}[1][]{$x \pm x$}      
\newcommand{\srhoparam}[1][]{$2.2 \pm 0.2$}    
\newcommand{\srhotransitmodel}[1][]{$x \pm x$}     
\newcommand{\srhotic}[1][]{$x \pm x$}    
 
\newcommand{\steffariadne}[1][]{$5230\pm 60$}  
\newcommand{\sloggariadne}[1][]{$4.50\pm0.07$}  
\newcommand{\sfehariadne}[1][]{$0.03\pm0.06$}  
\newcommand{\Lumariadne}[1][]{$0.45 \pm 0.02$}  
\newcommand{\Avariadne}[1][]{$0.04 \pm 0.05$}  
 
\newcommand{\steffsme}[1][]{$5189 \pm 65$}   
\newcommand{\sloggsme}{$4.50\pm 0.05$}    
\newcommand{\scahsme}[1][]{$0.07\pm 0.06$}  
\newcommand{\sfehsme}[1][]{$0.04\pm 0.05$}    
\newcommand{\snasme}[1][]{$0.13\pm 0.06$}    
\newcommand{\smghsme}[1][]{$0.14 \pm 0.06$}    
\newcommand{\ssihsme}[1][]{$0.07 \pm 0.06$}    
\newcommand{\svsinisme}[1][]{$1.8 \pm 0.9$}  
\newcommand{\svmic}[1][]{$0.8$}  
\newcommand{\svmac}[1][]{$2.5$}  
\newcommand{\velsme}[1][$\mathrm{km\,s^{-1}}$]{$-29.21$}  

\newcommand{\steffspechmatch}[1][]{$5225 \pm 110$}   
\newcommand{\sloggspechmatch}[1][]{$4.53 \pm 0.12$}   
\newcommand{\sfehspechmatch}[1][]{$0.06\pm 0.09$}   

\newcommand{\sloggparam}[1][]{$4.52 \pm 0.03$}  

\newcommand{\steffgaia}[1][]{$5202^{+8}_{-4} $}       
\newcommand{\slogggaia}[1][]{$4.53 \pm 0.01$}  
\newcommand{\sfehgaia}[1][]{$-0.14 \pm 0.01$}  
\newcommand {\lstargaia}[1][]{$0.449\pm0.002$}  
\newcommand{\distancegaia}{$ 110.7\pm 0.1$}      
\newcommand{\parallaxgaia}{$9.0312\pm0.0111$}  
\newcommand{\velgaia}[1][$\mathrm{km\,s^{-1}}$]{$-29.40 \pm 0.34$}  
\newcommand{\pmra}{$-22.8246 \pm 0.0125$}
\newcommand{\pmdec}{$53.0721 \pm  0.0155$} 

\newcommand{\stefftfop}[1][]{$5259\pm 125$}   
\newcommand{\sfehtfop}[1][]{$0.10 \pm 0.06 $}   
\newcommand{\sloggtfop}[1][]{$4.57 \pm 0.09 $}  
\newcommand{\sradiustfop}[1][]{$0.816 \pm 0.046 $}   
\newcommand{\smasstfop}[1][]{$0.90 \pm 0.11$}  

\newcommand{\spectraltype}{K0V}  
\newcommand{\smassspectraltype}{0.88}  
\newcommand{\sradiusspectraltype}{0.81}  
\newcommand{\srhospectraltype}{2.3}  

\newcommand{\protvsini}{$22.8 \pm 11.4$}   
\newcommand{\prottess}{$23.0 \pm 1.4$}  

\newcommand{\agespechmatch}[1][]{$4.2\pm 1.5$}  
\newcommand{\ageparam}{$5.9\pm 4.2$}  
\newcommand{\ageariadne}[1][]{$2.6^{+4.2}_{-2.3}$}   
\newcommand{\ageallmodels}[1][]{$2 - 6$}

\newcommand{\Tzerob}[1][]   {$1683.6256 \pm 7\mathrm{e}{-4}$}
\newcommand{\Tzeroc}[1][]   {$1689.9136 \pm 1.4\mathrm{e}{-3}$}
\newcommand{\Tzerod}[1][]   {$3267^{+37}_{-54}$}
\newcommand{\Pb}[1][]   {$5.139670 \pm 3\mathrm{e}{-6}$}
\newcommand{\Pc}[1][]   {$9.428089 \pm 1\mathrm{e}{-6}$} 
\newcommand{\lnPd}[1][]   {$7.9^{+0.2}_{-0.3}$} 
\newcommand{\Pd}[1][]   {$2767^{+573}_{-756}$} 
\newcommand{\Pdyear}[1][]   {$7.6^{+1.6}_{-2.4}$} 
\newcommand{\eb}[1][ ]   {$0$} 
\newcommand{\ebfree}[1][ ]   {$0.039^{+0.019}_{-0.039}$} 
\newcommand{\wb}[1][]   {$90$}
\newcommand{\cosib}[1][]   {$0.0660^{+0.0019}_{-0.0021}$}
\newcommand{\bb}[1][ ]   {$0.956\pm0.004$} 
\newcommand{\arb}[1][ ]   {$14.5\pm0.4$} 
\newcommand{\rrb}[1][ ]   {$0.03410\pm0.0013$}
\newcommand{\kb}[1][] {$3.8\pm0.7$} 
\newcommand{\ecfree}[1][ ]   {$0.16^{+0.08}_{-0.09}$} 
\newcommand{\wc}[1][]   {$199^{+20}_{-22}$}
\newcommand{\ed}[1][ ]   {$0.25^{+0.08}_{-0.11}$} 
\newcommand{\wwd}[1][] {$18^{+7}_{-10}$}
\newcommand{\cosic}[1][]   {$0.0417^{+0.014}_{-0.018}$}
\newcommand{\bc}[1][ ]   {$0.902^{+0.009}_{-0.008}$} 
\newcommand{\arc}[1][ ]   {$21.6^{+0.7}_{-0.6}$} 
\newcommand{\rrc}[1][ ]   {$0.0308\pm0.0008$}
\newcommand{\kc}[1][] {$3.5\pm0.7$} 
\newcommand{\kd}[1][] {$35^{+3}_{-5}$} 
\newcommand{\mpb}[1][]  {$9.4 \pm 1.8$} 
\newcommand{\mpc}[1][]  {$10.6 \pm 2.1$} 
\newcommand{\mpdearth}[1][]  {$673 \pm 80$} 
\newcommand{\mpdjup}[1][]  {$2.1 \pm 0.3$} 
\newcommand{\rpb}[1][]   {$3.04 \pm 0.19$}
\newcommand{\rpc}[1][]   {$2.75 \pm 0.14$} 
\newcommand{\Tperib}[1][] {$x$}  

\newcommand{\ecosd}[1][] {$0.26^{+0.16}_{-0.19}$}  
\newcommand{\esind}[1][] {$0.25\pm0.08$}  
\newcommand{\ecosc}[1][] {$-0.39^{+0.09}_{-0.11}$}  
\newcommand{\esinc}[1][] {$-0.13^{+0.11}_{-0.17}$}  

\newcommand{\eblim}[1][] {$0.058$}  
\newcommand{\eclim}[1][] {$0.24$}  

\newcommand{\prvb}[1][]  {$x$} 
\newcommand{\ib}[1][]   {$86.21^{+0.12}_{-0.11}$} 
\newcommand{\ic}[1][]   {$87.61^{+0.10}_{-0.08}$} 
\newcommand{\ab}[1][]   {$0.0553\pm0.0015$} 
\newcommand{\ac}[1][]   {$0.083\pm0.003$} 
\newcommand{\ad}[1][]   {$3.6\pm0.8$} 
\newcommand{\depthbSC}[1][]   {$x$} 
\newcommand{\RMbSC}[1][]   {$x$} 
\newcommand{\insolationb}[1][]   {$145\pm10$} 
\newcommand{\insolationc}[1][]   {$65\pm4$} 
\newcommand{\tsmb}[1][ ]   {$67\pm18$}
\newcommand{\tsmc}[1][ ]   {$36\pm9$}
\newcommand{\denstrb}[1][]   {$x$}  
\newcommand{\densspb}[1][]   {$x$} 
\newcommand{\Teqb}[1][]   {$971\pm11$} 
\newcommand{\Teqc}[1][]   {$794\pm9$}
\newcommand{\ttotb}[1][]   {$1.04^{+0.05}_{-0.03}$} 
\newcommand{\tfulb}[1][]   {$0.65\pm0.05$}
\newcommand{\ttotc}[1][]   {$1.67\pm0.04$} 
\newcommand{\tfulc}[1][]   {$1.21\pm0.05$} 
\newcommand{\tegb}[1][]  {$x$} 
\newcommand{\deltamagb}[1][]  {$x$} 
\newcommand{\denpb}[1][]   {$1.8\pm0.5$} 
\newcommand{\denpc}[1][]   {$2.9\pm0.7$} 
\newcommand{\grapb}[1][]   {$997\pm232$}
\newcommand{\grapc}[1][]   {$1425\pm297$}
\newcommand{\grapparsb}[1][]   {$x$}  
\newcommand{\jspb}[1][] {$24\pm5$} 
\newcommand{\jspc}[1][] {$37\pm8$} 
\newcommand{\qoneSC}[1][]   {$x$} 
\newcommand{\qtwoSC}[1][]   {$x$}  
\newcommand{\qsum}[1][]   {$0.63\pm0.09$}  

\newcommand{\HARPSN}[1][]   {$-29476^{+10}_{-6}$} 
\newcommand{\jHARPSN}[1][]  {$4.4\pm0.4$} 
\newcommand{\HIRES}[1][]   {$-15^{+9}_{-7}$} 
\newcommand{\jHIRES}[1][]   {$5.0^{+0.8}_{-1.1}$}

\begin{document} 
 
 \titlerunning{TOI-1438: A  rare system with two sub-Neptunes and a tentative long-period Jupiter-like planet}
   \title{TOI-1438:  A  rare system with two short-period sub-Neptunes and a tentative long-period Jupiter-like planet  orbiting a  \spectraltype~star}
\author{Carina~M.~Persson\inst{\ref{OSO}} 
\and
Emil~Knudstrup\inst{\ref{Chalmers},\ref{sac}}  
\and
Ilaria~Carleo\inst{\ref{IAC},\ref{ULL}}  
\and 
Lorena~Acu\~na-Aguirre\inst{\ref{Max-Planck Heidelberg}}  
\and 
Grzegorz~Nowak\inst{\ref{Torun Polen},\ref{IAC},\ref{ULL}} 
\and
Alexandra~Muresan\inst{\ref{Chalmers}}  
\and
Dawid~Jankowski\inst{\ref{Torun Polen}} 
\and
Krzysztof~Go\'zdziewski\inst{\ref{Torun Polen}} 
\and
Rafael~A.~Garc\'ia\inst{\ref{CEA Saclay}}   
\and
Savita~Mathur\inst{\ref{IAC}, \ref{ULL}}  
\and
Dinil~B.~Palakkatharappil\inst{\ref{CEA Saclay}}  
\and
Lina~Borg\inst{\ref{INSA},\ref{CEA Saclay}}  
\and
Alexander~J.~Mustill\inst{\ref{Lund}}  
\and
Rafael~Barrena\inst{\ref{IAC}, \ref{ULL}}  
\and
Malcolm~Fridlund\inst{\ref{OSO}, \ref{Leiden}}   
\and
Davide~Gandolfi\inst{\ref{Torino}}  
\and
Artie~P.~Hatzes\inst{\ref{Tautenburg}}  
\and
Judith~Korth\inst{\ref{Geneve}}  
\and
Rafael~Luque\inst{\ref{Chicago}, \ref{SaganFellow}}   
\and
Eduardo~L.~Mart\'in\inst{\ref{IAC}, \ref{ULL}}    
\and
Thomas~Masseron\inst{\ref{IAC}, \ref{ULL}}  
\and
Giuseppe~Morello\inst{\ref{Andalucia}, \ref{Palermo}}  
\and
Felipe~Murgas\inst{\ref{IAC}, \ref{ULL}}  
\and
Jaume~Orell-Miquel\inst{\ref{IAC}, \ref{ULL}, \ref{Austin}}  
\and
Enric~Palle\inst{\ref{IAC}, \ref{ULL}}  
\and
Simon H. Albrecht\inst{\ref{Aarhus}}  
\and
Allyson~Bieryla\inst{\ref{Harvard}}  
\and
William~D.~Cochran\inst{\ref{McDonald}}  
\and 
Ian~J.~M.~ Crossfield\inst{\ref{Kansas}}  
\and
Hans~J.~Deeg\inst{\ref{IAC}, \ref{ULL}}  
\and 
Elise~Furlan\inst{\ref{Caltech}}  
\and
Eike~W.~Guenther\inst{\ref{Tautenburg}}  
\and 
Steve~B.~Howell\inst{\ref{NASA Ames}}  
\and
Howard~Isaacson\inst{\ref{Berkeley}}  
\and
Kristine~W.~F.~Lam\inst{\ref{Berlin}}  
\and
John~Livingston\inst{\ref{Tokyo}, \ref{NRAO}, \ref{Sokendai}}  
\and
Rachel~A.~Matson\inst{\ref{Naval}} 
\and
Elisabeth~C.~Matthews\inst{\ref{Max-Planck Heidelberg}}  
\and
Seth~Redfield\inst{\ref{Wesleyan}}  
\and
Joshua~E.~Schlieder\inst{\ref{Goddard}}   
\and
Sara~Seager\inst{\ref{KavliMIT}, \ref{AstrophysicsMIT}, \ref{MITAeronautics}}  
\and
Alexis~M.~S.~Smith\inst{\ref{Berlin}} 
\and
Keivan~G.~Stassun\inst{\ref{Vanderbilt}}   
\and
Joseph~D.~Twicken\inst{\ref{SETI}, \ref{Ames}}  
\and
Vincent~Van~Eylen\inst{\ref{Mullard}}  
\and 
Cristilyn~N.~Watkins\inst{\ref{Harvard}}  
\and
Lauren~M.~Weiss\inst{\ref{Notre Dame}}  
 }     
 
 \offprints{carina.persson@chalmers.se}
    \institute{Chalmers University of Technology, Department of Space, Earth and Environment, Onsala Space Observatory, SE-439 92 Onsala, Sweden.\label{OSO}   
       \email{\url{carina.persson@chalmers.se}}
   \and Chalmers University of Technology, Department of Space, Earth and Environment, 412 93, Gothenburg, Sweden.\label{Chalmers} 
   \and Stellar Astrophysics Centre, Department of Physics and Astronomy, Aarhus University, Ny Munkegade 120, DK-8000 Aarhus C, Denmark\label{sac}  
   \and   Instituto de Astrof\' isica de Canarias (IAC), C. Via Lactea S/N, E-38205 La Laguna, Tenerife, Spain.\label{IAC} 
   \and Universidad de La Laguna (ULL), Departamento de Astrof\'isica, E-38206 La Laguna, Tenerife, Spain.\label{ULL}  
 \and Max-Planck Institut für Astronomie, Konigstuhl 17, 69117, Heidelberg, Germany.\label{Max-Planck Heidelberg}  
 \and Institute of Astronomy, Faculty of Physics, Astronomy and Informatics, Nicolaus Copernicus University, Grudziadzka 5, 87-100 Toru\'n, Poland.\label{Torun Polen}  
 \and Universit\'e Paris-Saclay, Universit\'e Paris Cit\'e, CEA, CNRS, AIM, 91191, Gif-sur-Yvette, France. \label{CEA Saclay}  
 \and INSA Lyon - Institut National des Sciences Appliqu\'ees, France \label{INSA}  
 \and Lund Observatory, Division of Astrophysics, Department of Physics, Lund University, Box 118, 221 00, Lund, Sweden. \label{Lund}  
  \and Leiden Observatory, University of Leiden, PO Box 9513, 2300 RA, Leiden, The Netherlands. \label{Leiden}   
 \and Dipartimento di Fisica, Università degli Studi di Torino, via Pietro Giuria 1, I-10125, Torino, Italy. \label{Torino}  
\and Th\"uringer Landessternwarte Tautenburg, Sternwarte 5, 07778, Tautenburg, Germany. \label{Tautenburg}  
\and Observatoire astronomique de l'Universit\'e de Gen\'eve, Chemin Pegasi 51, 1290 Versoix, Switzerland.\label{Geneve}  
\and Department of Astronomy \& Astrophysics, University of Chicago, Chicago, IL 60637, USA.\label{Chicago} 
\and NHFP Sagan Fellow.\label{SaganFellow}  
\and Instituto de Astrofísica de Andalucía (IAA-CSIC), Glorieta de la Astronomía s/n, 18008 Granada, Spain  \label{Andalucia}   
\and INAF - Osservatorio Astronomico di Palermo, Piazza del Parlamento, 1, 90134 Palermo, Italy. \label{Palermo}  
 \and Department of Astronomy, University of Texas at Austin, 2515 Speedway, Austin, TX 78712, USA. \label{Austin}  
 \and Department of Physics and Astronomy, Aarhus University, Ny Munkegade 120, 8000 Aarhus C, Denmark. \label{Aarhus}    
 \and Center for Astrophysics \textbar \ Harvard \& Smithsonian, 60 Garden Street, Cambridge, MA 02138, USA.
 \label{Harvard}  
 \and McDonald Observatory, The University of Texas, Austin Texas USA.\label{McDonald}  
\and{Department of Physics and Astronomy, University of Kansas, Lawrence, KS, USA}. \label{Kansas}  
\and NASA Exoplanet Science Institute, Caltech/IPAC, Mail Code 100-22, 1200 E. California Blvd., Pasadena, CA 91125, USA. \label{Caltech}
\and NASA Ames Research Center, Moffett Field, CA 94035, USA. \label{NASA Ames}  
\and  501 Campbell Hall, University of California at Berkeley, Berkeley, CA 94720, USA. \label{Berkeley}  
\and  Institute of Space Research, German Aerospace Center (DLR), Rutherfordstr. 2, 12489 Berlin, Germany. \label{Berlin} %
\and Astrobiology Center, 2-21-1 Osawa, Mitaka, Tokyo 181-8588, Japan.\label{Tokyo}  
\and 
National Astronomical Observatory of Japan, 2-21-1 Osawa, Mitaka, Tokyo 181-8588, Japan.\label{NRAO}  
\and 
Department of Astronomy, The Graduate University for Advanced Studies (SOKENDAI), 2-21-1 Osawa, Mitaka, Tokyo, Japan.\label{Sokendai}  
\and  U.S. Naval Observatory, Washington, D.C. 20392, USA.\label{Naval}  
\and Astronomy Department and Van Vleck Observatory, Wesleyan University, Middletown, CT 06459, USA.\label{Wesleyan}    
\and NASA Goddard Space Flight Center, 8800 Greenbelt Rd., Greenbelt, MD 20771, USA.\label{Goddard}  
\and Department of Physics and Kavli Institute for Astrophysics and Space Research, Massachusetts Institute of Technology, Cambridge, MA 02139, USA. \label{KavliMIT}  
\and Department of Earth, Atmospheric and Planetary Sciences, Massachusetts Institute of Technology, Cambridge, MA 02139, USA.\label{AstrophysicsMIT}   
\and Department of Aeronautics and Astronautics, MIT, 77 Massachusetts Avenue, Cambridge, MA 02139, USA.\label{MITAeronautics}   
\and Department of Physics \& Astronomy, Vanderbilt University, Nashville, TN 37235, USA.  \label{Vanderbilt}   
\and SETI Institute, Mountain View, CA 94043 USA.\label{SETI}      
\and NASA Ames Research Center, Moffett Field, CA 94035 USA.\label{Ames}   
\and  Mullard Space Science Laboratory, University College London, Dorking, Surrey, RH5 6NT.\label{Mullard}    
\and Department of Physics and Astronomy, University of Notre Dame, Notre Dame, IN 46556, USA. \label{Notre Dame}  
}   

   \date{Received  28 April 2025; accepted 31 July 2025} 
 
  \abstract
  {We present the detection and characterisation of the \targeta~multi-planet system   discovered   by the  Transiting Exoplanet Survey Satellite (TESS).  
To confirm the planetary nature of the  candidates and determine their masses,  we collected a series  of follow-up observations including    high-spectral resolution  observations with HARPS-N and HIRES over a period of five~years. 
  Our combined modelling  shows that the \spectraltype~star hosts   two transiting sub-Neptunes with  $R_\mathrm{b} = $ \rpb~\rearth, $R_\mathrm{c} = $ \rpc~\rearth, $M_\mathrm{b} = $  \mpb~\mearth, and $M_\mathrm{c} = $\mpc~\mearth. The orbital periods of planets~b and c are 5.1 and 9.4~days, respectively,   corresponding to instellations of  \insolationb~\Fearth~and \insolationc~\Fearth. 
  The bulk densities are  \denpb~\gc~and \denpc~\gc, respectively, suggesting a volatile-rich   interior composition. By combining the planet and stellar parameters, we were able to compute a set of  planet interior structure models. Planet b presents a high-metallicity envelope that can accommodate up to 2.5~\% in H/He in mass, while planet c cannot have more than 0.2~\% as H/He in mass. For any composition of the core considered (Fe-rock or ice-rock), both planets would require a volatile-rich envelope. 
  In addition to the two planets, the radial velocity (RV) data  clearly reveal a third   signal,   likely coming from a non-transiting planet, with an   orbital period of \Pdyear~years  and an RV semi-amplitude of \kd~\ms. Our best-fit model finds a minimum mass of \mpdjup~\mjup~and an eccentricity of \ed.  
However, several RV activity indicators also show  strong signals at similar periods, suggesting  this signal might (partly) originate from   stellar   activity.    
More data over a longer period of time are needed to conclusively determine the nature of this   signal.
If it is confirmed as a triple-planet system, \targeta~would be one of the few detected systems to date characterised by an architecture with two small, short-period planets and one massive, long-period planet, where the inner and outer systems are separated by an orbital period ratio of the order of a few hundred. 
  }

   \maketitle

\begin{table}[!t]
\caption{Basic parameters for \targeta.}
\begin{center}
 \resizebox{0.9\columnwidth}{!}{
\begin{tabular}{lll} 
\hline\hline
     \noalign{\smallskip}
Parameter    & Value   \\
\noalign{\smallskip}
\hline
\noalign{\smallskip}
\multicolumn{2}{l}{\emph{Main identifiers}} \\
\noalign{\smallskip}
TIC  & \ticno \\
2MASS &  J18434182+7456152 \\
WISE &  J184341.73+745615.9 \\
TYC & 4442-00562-1 \\
UCAC4 & 825-019611 \\
Gaia DR2   &    2268101727131230464     \\
APASS & 60356991 \\
\noalign{\smallskip}
\hline
\noalign{\smallskip}
\multicolumn{2}{l}{\emph{Equatorial coordinates  (epoch 2015.5)}} \\
\noalign{\smallskip} 
R.A. $(J2000.0)$  & $18\fh43\fm 41\fs71$     \\
Dec. $(J2000.0)$  & +74$\fdg 56\farcm16\farcs2$    \\
\noalign{\smallskip}
\hline
\noalign{\smallskip}
\multicolumn{2}{l}{\emph{Magnitudes}} \\
Johnson $ B$ &   $11.972\pm0.1560$     \\
Johnson $V$  & $10.957\pm0.014$      \\
TESS & $10.3237\pm 0.006$ \\
 $G \tablefootmark{a}$  &   $10.8555\pm0.0001$ \\
 $G_{RP}\tablefootmark{a}$      &   $10.2670\pm0.0003$ \\
$G_{BP} \tablefootmark{a}$      &   $11.2812\pm0.0008$ \\
$J$  &   $9.584\pm 0.018$     \\
$H$    &   $9.191\pm0.017$      \\
$K$   &     $9.09\pm0.02$   \\
WISE $W1$   &     $9.054\pm0.023$   \\
WISE $W2$   &     $9.116\pm0.020$   \\
\noalign{\smallskip}
\hline
\noalign{\smallskip}  
Parallax$\tablefootmark{a}$  (mas) &\parallaxgaia\,   \\  
Distance$\tablefootmark{a}$  (pc) &\distancegaia\,   \\  
Radial velocity$\tablefootmark{a}$ (\kms) & \velgaia   \\   
$\mu_{RA}\tablefootmark{a}$ (mas~yr$^{-1}$) & \pmra   \\   
$\mu_{Dec}\tablefootmark{a}$ (mas~yr$^{-1}$) & \pmdec   \\  
Ruwe$\tablefootmark{a, b}$ & 0.853\\
\noalign{\smallskip}
\hline
\noalign{\smallskip}
\multicolumn{2}{l}{\emph{This work}} \\
\mstar~(\Msun) & \smassariadne \\
\rstar~(\Rsun)  & \sradiusariadne \\
\rhostar~(\gc)  & \srhoariadne \\
\lstar~(\Lsun)  & \Lumariadne \\ 
\teff~(K)  & \steffariadne \\
$P_{\rm rot}/\sin i_\star$~(d) & \protvsini \\
\logg~(dex)   &  \sloggariadne  \\
\feh~(dex)   & \sfehariadne \\
\cah~(dex)   & \scahsme\\
\mgh~(dex)   & \smghsme \\
\nah~(dex)  & \snasme \\
\sih~(dex)    & \ssihsme \\
\vsini~(\kms) &\svsinisme\\
\lgr\,$\tablefootmark{c}$  & \logRHK \\
Spectral type & \spectraltype \\
    \noalign{\smallskip} \noalign{\smallskip}
\hline 
\end{tabular}
}
\tablefoot{
\tablefoottext{a}{Gaia DR3.}
\tablefoottext{b}{Renormalised unit weight error (Gaia).}
\tablefoottext{c}{The average \ion{Ca}{II} chromospheric activity index.}
}
\end{center}
\label{Table: Star basic parameters}
\end{table}

\section{Introduction} \label{Section: introduction}
 One of the major discoveries made with the  \textit{Kepler} space telescope 
 \citep{2010Sci...327..977B, 2014PASP..126..398H} 
is the broad diversity of exoplanets. 
 For instance, super-Earths (\mbox{$R \approx 1.1 -  1.8~R_{\oplus}$}) and 
 sub-Neptunes  (\mbox{$R \approx 1.8 -  3.5~R_{\oplus}$})
 have been shown to represent a large fraction of all exoplanets 
  in the solar neighbourhood
 \citep{2010Sci...327..977B, 2012ApJS..201...15H, 2018AJ....156..264F}, 
 although they are not represented in the Solar System. 
 While fascinating, this diversity  
 presents challenges for  theories of planet formation and evolution. 
 
To investigate  exoplanets and their system architectures, accurate and precise 
estimates of exoplanet masses and radii play a fundamental role. 
Among the most powerful tools is    transit photometry performed from space 
combined with high-spectral resolution radial velocity (RV) measurements 
to obtain both radii and masses and hence bulk densities. 
\textit{Kepler} and the 
Transiting Exoplanet Survey Satellite  \citep[TESS; ][]{2015JATIS...1a4003R}, 
together with  high-resolution spectrographs 
 like HARPS \citep{Mayor2003HARPS} 
and  its sibling HARPS-N \citep[][]{Cosentino2012},  
have enabled the detection and characterisation of  the majority of 
the almost 6000~confirmed exoplanets.

Unfortunately, only about one-fifth of all detected exoplanets benefit from  
radius measurements taken from transit photometry and masses 
from either  RV or transit timing variations (TTVs). 
Many of these planets are larger than Neptune and found in single-planet systems, 
while many of the small planets  have high uncertainties in derived masses and radii. 
There are only  162~well-characterised small exoplanets in 108 systems 
with minimum two planets with 
radii $< 4$~\rearth~and uncertainties $< 21$~\% and 7~\% in mass and radius, respectively, 
based on the NASA Exoplanet Archive\footnote{\url{https://exoplanetarchive.ipac.caltech.edu}} 
as of  12~April~2025.  Of these, about
half are sub-Neptunes and the other half consists of super-Earths. 
Many more well-characterised small planets are needed
to constrain the planet formation, system architecture, and interior models.

Most multi-planetary systems discovered   
to date have a compact architecture with many intra-system 
similarities. Planets in systems without  outer long-period giants tend 
to be small with similar sizes and masses  
and  regularly spaced in nearly circular and coplanar orbits 
\citep[e.g.][]{2017ApJ...849L..33M, 2018AJ....155...48W, 2023AAS...24131603W, 2024A&A...692A.122M}. 
The architecture of   systems  with  both  inner small  short orbital period 
planets  and long orbital period  giants is   so far not well known due 
to observational biases and challenges
with very few resulting characterised systems. However, several studies 
have suggested a correlation
between inner small planets and outer giants  
\citep{Uehara2016, Bryan2019, Bonomo2023, 2025arXiv250106342V}. 

In this paper, we present the discovery and characterisation of the \targeta~multi-planet system   
discovered by   TESS. 
This system has two sub-Neptunes with short orbital periods orbiting a \spectraltype~star and one
 tentative long-period massive planet   as   suggested by a  trend in    
 RV measurements obtained with the HIRES instrument at the Keck Observatory    \citep{2025arXiv250106342V}.  
 Our  KESPRINT\footnote{KESPRINT is an international 
consortium devoted to the characterisation and research of exoplanets discovered with space-based missions, \url{https://kesprint.science}.}  
collaboration conducted   85 
follow-up  RV observations with HARPS-N   of this system  over a period of 
five years  to characterise the planets 
and the host star supported by the HIRES RVs, the TESS photometry, 
ground-based photometry with KeplerCam, and speckle imaging with 
Gemini \citep{Scott2021,Howell_Furlan_2022}. 
The basic parameters of the host star  are listed in Table~\ref{Table: Star basic parameters}.

We present the observations  in Sect.~\ref{Section: Observations}  and the data analysis in  
Sect.~\ref{Section: Data analysis}. 
In  Sect.~\ref{Section: Discussion}, we discuss the 
planet parameters,  the interior structure   models of planets~b and c, and   
the architecture of the system. We 
end the paper with conclusions in   Sect.~\ref{Section: Conclusions}.

\section{Observations} \label{Section: Observations}
 
    \subsection{TESS photometry} \label{Section: TESS photometry}
\targeta~was observed with TESS in 38 sectors from 2019 to 2024 
(sectors $14-20$ in 2019, sectors 21 and $23-26$ in 2020, 
sectors $40-41$ and 47 in 2021, sectors $48-51$, $53-58,$ and 
60 in 2022, sector 73 in 2023, sectors $74-76$, $78-81$, and $83-86$ in 2024) 
with a cadence of 120~seconds. The data from Sector 74 were omitted from 
the analysis in Sect.~\ref{Section: transit and RV modelling} as the light curve 
was quite noisy  (the root mean square, RMS) was four times larger compared to the other sectors).   
We downloaded the light curves  processed by the   TESS Science Processing 
Operations Center \citep[SPOC; ][]{jenkins2016} at NASA Ames Research Center  
from the Mikulski Archive for Space Telescopes 
(MAST\footnote{\url{https://mast.stsci.edu/portal/Mashup/Clients/Mast/Portal.html}}).  
   This pipeline 
   identifies and corrects   instrumental signatures in the  flux to produce  
    Pre-search Data Conditioning Simple Aperture Photometry (PDCSAP) data. 
   The result is a cleaner dataset with fewer systematics  
   \citep{Smith2012, Stumpe2012, Stumpe2014}. 
The SPOC conducted a transit search of the combined light curve from sectors $14-16$
on 26~October~2019 with an adaptive, noise-compensating matched filter 
\citep{2002ApJ...575..493J, 2010SPIE.7740E..0DJ, 2020ksci.rept....9J}, 
producing a threshold crossing event 
(TCE) with a 5.14~d period. The TESS Science Office (TSO) subsequently issued an 
alert for \mbox{TOI-1438.01} (with a planet radius of $\approx 2.5$~\rearth~on) 14~November~2019
\citep{2021ApJS..254...39G}. Likewise, SPOC conducted a transit search of the combined
light curve from sectors $14-19$ on 24~January~2020, producing a TCE with 9.43~d 
period; the TSO subsequently issued an alert for \mbox{TOI-1438.02} (with a radius of 
approximately $\approx 2.3$~\rearth~on) 21~February~2020 \citep{2021ApJS..254...39G}. 
Diagnostic tests
were conducted to determine the planetary nature of the signals reported in the 
data validation reports \citep[DVI;][]{2018PASP..130f4502T,Li:DVmodelFit2019}, 
available for download from MAST as well from the {\tt exofop-TESS} 
website\footnote{\url{https://exofop.ipac.caltech.edu/tess/target.php?id=229650439}}. 
 Additional tests  were performed for the combined sectors reported in subsequent DVRs 
 showing no concerns about   contaminating sources in the aperture of the SPOC pipeline. 
 In total, we identified   
 150  transits of \mbox{TOI-1438.01} (hereafter, \targetb) and  79 transits of 
 \mbox{TOI-1438.02} (hereafter, \targetc)  in the analysed TESS sectors (excluding sector~74).

\subsection{Ground-based photometry follow-up with KeplerCam}  
We observed a full transit window of  \targetc continuously for 203~min 
with a cadence of 120~s in Sloan $i'$ band on 
\mbox{UTC~15~May~2021} from KeplerCam on the 1.2-m telescope at 
the Fred Lawrence Whipple Observatory. 
The $4096\times4096$ Fairchild CCD 486 detector has an image scale 
of $0\farcs672$ per $2\times2$ binned 
pixel, resulting in a $23\farcm1\times23\farcm1$ field of view and 
the differential photometric data were 
extracted using {\tt AstroImageJ} \citep{Collins:2017}. 

The \targetb~SPOC pipeline transit depth of \mbox{$770-790$}~ppm is 
generally too shallow to reliably detect with ground-based observations, 
so we therefore checked for possible nearby eclipsing binaries (NEBs) 
that could be contaminating the 
TESS photometric aperture and causing the TESS detection. 
To account for possible contamination from the wings of neighbouring 
star PSFs, we searched for NEBs 
out to $2\farcm5$ from \targeta. 
If fully blended in the SPOC aperture, a neighbouring star that is fainter than the target star by 8.8 
magnitudes in the TESS-band could produce the SPOC-reported flux 
deficit at mid-transit (assuming a 100~\% eclipse). 
To account for possible TESS magnitude uncertainties and possible delta-magnitude differences 
between TESS-band and Sloan $i'$, we included an extra 0.5~magnitudes 
fainter (down to TESS-band magnitude of 18.0). 
We calculated the RMS of each of the 34 nearby stellar light curves 
(binned in 10~min bins) that met 
our search criteria. We find that the values are smaller by at least a 
factor of 5 compared to the required 
NEB depth for each respective star. 
We then visually inspected each neighbouring star's light curve to 
ensure that there are no obvious eclipse-like signals. 
Our analysis ruled out an NEB blend as the cause of the SPOC pipeline 
planet~b detection in the TESS data. 
The NEB light curve data are available on the {\tt EXOFOP-TESS} 
website.
 
\subsection{Gemini speckle imaging} \label{Gemini speckle imaging from ground}  
Close stellar companions (bound or in the line of sight) have the potential to confound 
exoplanet discoveries in a number of ways. For instance,  
the detected transit signal might be a false positive due to an eclipsing 
binary on the visual companion. 
Even real planet discoveries will yield incorrect stellar and exoplanet parameters 
if a close companion exists and  is unaccounted for \citep{ciardi2015, Furlan_Howell_2017}. 
Additionally, the presence of a close companion star leads to the non-detection 
of small planets residing in the same exoplanetary system. 
Given that nearly one-half of solar-like stars are in binary or multiple star systems 
 \citep{Matson2018},  high-resolution imaging provides crucial information toward 
 our understanding of exoplanetary formation, dynamics, and evolution \citep{Howell2021}.

  \begin{figure}[!ht]
 \centering
        \resizebox{\hsize}{!}
        {\includegraphics[]{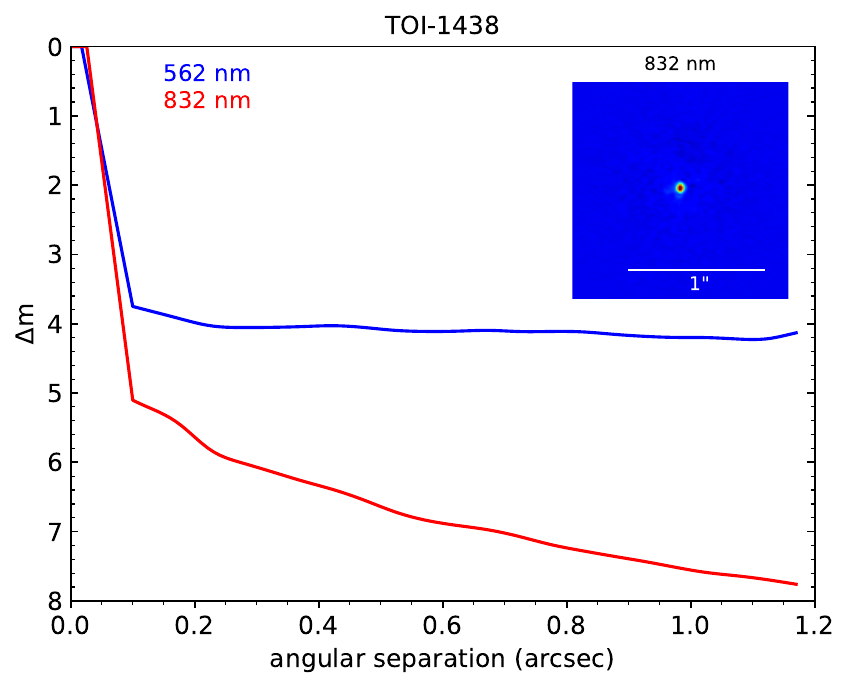}} 
   \caption{Final results of the Gemini North speckle imaging of \targeta. The blue and red curves show the $5\,\sigma$ contrast curves in 562~nm and 832~nm filters, respectively,  as a function of the angular separation out to 1.2\arcsec. The inset shows the reconstructed 832~nm image with a 1\arcsec~scale bar. \targeta~was found to have no close companions from the diffraction limit out to 1.2\arcsec~and   within the magnitude contrast levels achieved.}
      \label{Figure: speckle with Gemini}
 \end{figure}

\begin{table*}
\centering
 \caption{Comparison of spectroscopic  parameters of \targeta~derived by different methods.}   
 \label{Table: stellar spectroscopic parameters}
\begin{tabular}{llccccccc }
 \hline
     \noalign{\smallskip}
Method  & $T_\mathrm{eff}$  & $\log(g)$ & [Fe/H]   & [Ca/H]  & [Mg/H]  & [Na/H]   & [Si/H]  & \vsini    \\  
& (K)  &(dex) &(dex)&(dex)&(dex)&(dex)&(dex) &(\kms)  \\
    \noalign{\smallskip}
     \hline
\noalign{\smallskip} 
 {\tt {SME}}$\tablefootmark{a}$   & \steffsme & \sloggsme   &  \sfehsme  &\scahsme  &\smghsme &\snasme& \ssihsme  &\svsinisme  \\
 {\tt {SpecMatch-Emp}}   & \steffspechmatch & \sloggspechmatch   &  \sfehspechmatch  &\ldots&\ldots&\ldots &\ldots &\ldots  \\
{\tt {ARIADNE}}$\tablefootmark{b}$  &  \steffariadne& \sloggariadne       &     \sfehariadne  &\ldots&\ldots&\ldots &\ldots   &\ldots  \\
   Gaia DR3 &\steffgaia &\slogggaia &\sfehgaia &\ldots&\ldots&\ldots &\ldots  &\ldots   \\
\noalign{\smallskip} 
\hline 
\end{tabular}
\tablefoot{
\tablefoottext{a}{Adopted as priors for the stellar mass and radius modelling in Sect.~\ref{Subsection: Stellar modelling}.} 
\tablefoottext{b}{Posteriors from  Bayesian model averaging  with ARIADNE.}
}
\end{table*}

  \begin{table}
 \centering
 \caption{Comparison of stellar mass and radius of \targeta~derived by different methods compared to values of a typical \spectraltype~star.}
  \label{Table: stellar mass and radius}
 \resizebox{\columnwidth}{!}{ 
\begin{tabular}{lccc}
 \hline\hline
     \noalign{\smallskip}
Method    & $M_\star$  &     $R_\star$   & $\rho_\star$     \\
  & ($M_{\odot}$)  & ($R_{\odot}$)  & (g~cm$^{-3}$)  \\
    \noalign{\smallskip}
     \hline
\noalign{\smallskip} 
{\tt {ARIADNE}}$\tablefootmark{a}$    & \smassariadne& \sradiusariadne & \srhoariadne     \\
Torres & $0.900 \pm 0.060$ & $0.872 \pm 0.061$ & $1.9 \pm 0.4$ \\
{\tt {PARAM 1.3}}$\tablefootmark{b}$    & \smassparam    & \sradiusparam & \srhoparam \\
\spectraltype$\tablefootmark{c}$    & \smassspectraltype& \sradiusspectraltype   & \srhospectraltype  \\ 
      \noalign{\smallskip} \noalign{\smallskip}
\hline 
\end{tabular}
}
\tablefoot{
\tablefoottext{a}{{\tt {ARIADNE}} \citep[][]{2022MNRAS.513.2719V} computes the stellar parameters from a combination of  the SED fitting and MIST isochrones \citep{2016ApJ...823..102C}. 
Adopted as the final parameters for the modelling in Sect.~\ref{Section: transit and RV modelling}.} 
\tablefoottext{b}{\citet{daSilva2006}.}         
\tablefoottext{c}{{\url{https://www.pas.rochester.edu/~emamajek/EEM_dwarf_UBVIJHK_colors_Teff.txt}}.} 
}
\end{table}

\targeta~was observed on 2022~May~13~UT using the ‘Alopeke speckle instrument on the Gemini North 8-m telescope \citep{Scott2021,Howell_Furlan_2022}.  
‘Alopeke provides simultaneous speckle imaging in two bands (562 and 832~nm) with output data products including a reconstructed image with robust contrast limits on companion detections. 
Twelve sets of $1000 \times 0.06$~s images were obtained and processed in our standard reduction pipeline 
\citep{howell2011}. 
Figure \ref{Figure: speckle with Gemini} shows our final contrast curves and the 832~nm reconstructed speckle image. 
We find that \targeta~is a single star with no companion brighter than $5-8$ magnitudes below that of the target star from the 8-m telescope diffraction limit (20~mas) out to 1.2\arcsec. 
At the distance of \targeta, these angular limits correspond to spatial limits of 2.2  to 132~au.
 
\subsection{Radial velocity follow-up with HARPS-N} \label{Subsection: Radial velocity follow-up with HARPS-N}  
Within the KESPRINT consortium program, we observed \targeta~with the high resolution spectrograph HARPS-N \citep{2014SPIE.9147E..8CC} 
mounted on the 3.58-m Telescopio Nazionale Galileo (TNG) of Roque 
de los Muchachos Observatory in La Palma\footnote{Observing programs: CAT19A\_162 (15 spectra), CAT21A\_119 (8), CAT22A\_111 (23) - PI: Nowak; ITP19\_1 (2) - PI: Pall{\'e}; A41TAC\_49 (5) - PI: Gandolfi; CAT20B\_80 (1) - PI: Casasayas; CAT23A\_52 (16), CAT23B\_74 (10), CAT24B\_20 (5) - PI: Carleo).} 
covering wavelengths between 378~nm and 691~nm at a spectral 
resolution of $R\approx  115,000$. 
We followed this target between 17~March 2020 and 15~March 2025 
collecting 85~RVs  with exposure 
times of $1500-3300$~s and an 
average signal-to-noise ratio (S/N) of 51 at 550\,nm.  
We reduced the data through the Yabi web application (\citealt{yabi})\footnote{\url{http://ia2-harps.oats.inaf.it:8000/}} 
running the offline version of the HARPS-N data reduction 
software ({\tt DRS}) and choosing a K5 mask 
template and a cross-correlation function (CCF) width of 
30~\kms~with a step of 0.25~\kms. 
The HARPS-N absolute RVs measured with the {\tt DRS} have uncertainties in 
a range $0.9-5.6$~\ms~with a median value of 
1.6~\ms~and an RMS of 22.2~\ms~about the mean value. We 
measured  the differential line width (dLW), H$\alpha$,  Na\,D1, and Na\,D2 indexes with 
the {\tt Serval} \citep{Zechmeister2018} code and 
the template matching technique. The \mbox{HARPS-N} relative RVs measured with {\tt Serval} have 
uncertainties in a range $0.9-3.7$~\ms~with a median value of 
1.4~\ms~and an RMS of 22.3~\ms~about the mean value. 
 The RVs are listed in Tables~\ref{Table: HARPSN RVs drs} -- \ref{Table: HARPSN RVs serval}  together with  activity 
indicators from {\tt DRS} and {\tt Serval}.
 
 \subsection{Radial velocity follow-up with HIRES} 
 \label{Subsection: Radial velocity follow-up with HIRES}  
We collected 20~spectra with the High Resolution Echelle Spectrometer \citep[HIRES;][]{Vogt1994} at the Keck Observatory over two years. 
With a resolving power of $\approx 60,000$ and wavelength range from $374-970$~nm, 
HIRES provided precision RVs with a median uncertainty of 1.7~\ms~from spectra with an S/N of 150 for \targeta. 
The RMS of the HIRES RVs is 18.9~\ms~about the mean value. 
Using the iodine technique \citep{Butler1996} and the standard California Planet Search setup and RV extraction technique  \citep{Howard2010}, the RVs were combined with other RV datasets with an additional offset parameter in the RV model. A survey-wide assessment of HIRES RVs of TESS targets, including \targeta, can be found in \citet{Polanski2024} and the results in \citet{2025arXiv250106342V}. 
The HIRES RVs are listed in Table~\ref{Table: HIRES RVs}.
 
\section{Data analysis} \label{Section: Data analysis}

\subsection{Stellar properties } \label{Subsection: Stellar modelling} 
We used the empirical code     {\tt SpecMatch-Emp}   \citep{2017ApJ...836...77Y}   that compares the observed 
optical spectrum with a dense library of well-characterised 
stars to analyse the co-added high-resolution HARPS-N spectrum. 
We then proceeded with more detailed modelling using 
Spectroscopy Made Easy\footnote{\url{http://www.stsci.edu/~valenti/sme.html}} 
 \citep[{\tt SME};][]{vp96, pv2017}, a software that fits observations to computed synthetic spectra. We chose the MARCS stellar atmosphere grid \citep{Gustafsson08}, retrieved  atomic and molecular line data from
VALD\footnote{\url{http://vald.astro.uu.se}}  \citep{Ryabchikova2015}, and used the output from  {\tt SpecMatch-Emp} as initial values in the modelling. We followed the procedure in \citet{2018A&A...618A..33P} modelling one parameter at a time. The micro- and macro-turbulent velocities were kept fixed to \svmic~\kms~\citep{Bruntt2010b} and \svmac~\kms~\citep{Doyle2014}, respectively.
The final results  are   
 tabulated in Table~\ref{Table: stellar spectroscopic parameters} and are in good agreement with  {\tt SpecMatch-Emp} matching a typical \spectraltype~star.  The effective temperature and surface gravity are also consistent with Gaia~DR3 which, however, report a   lower iron abundance.

\begin{figure}
    \centering
    \includegraphics[width=\linewidth]{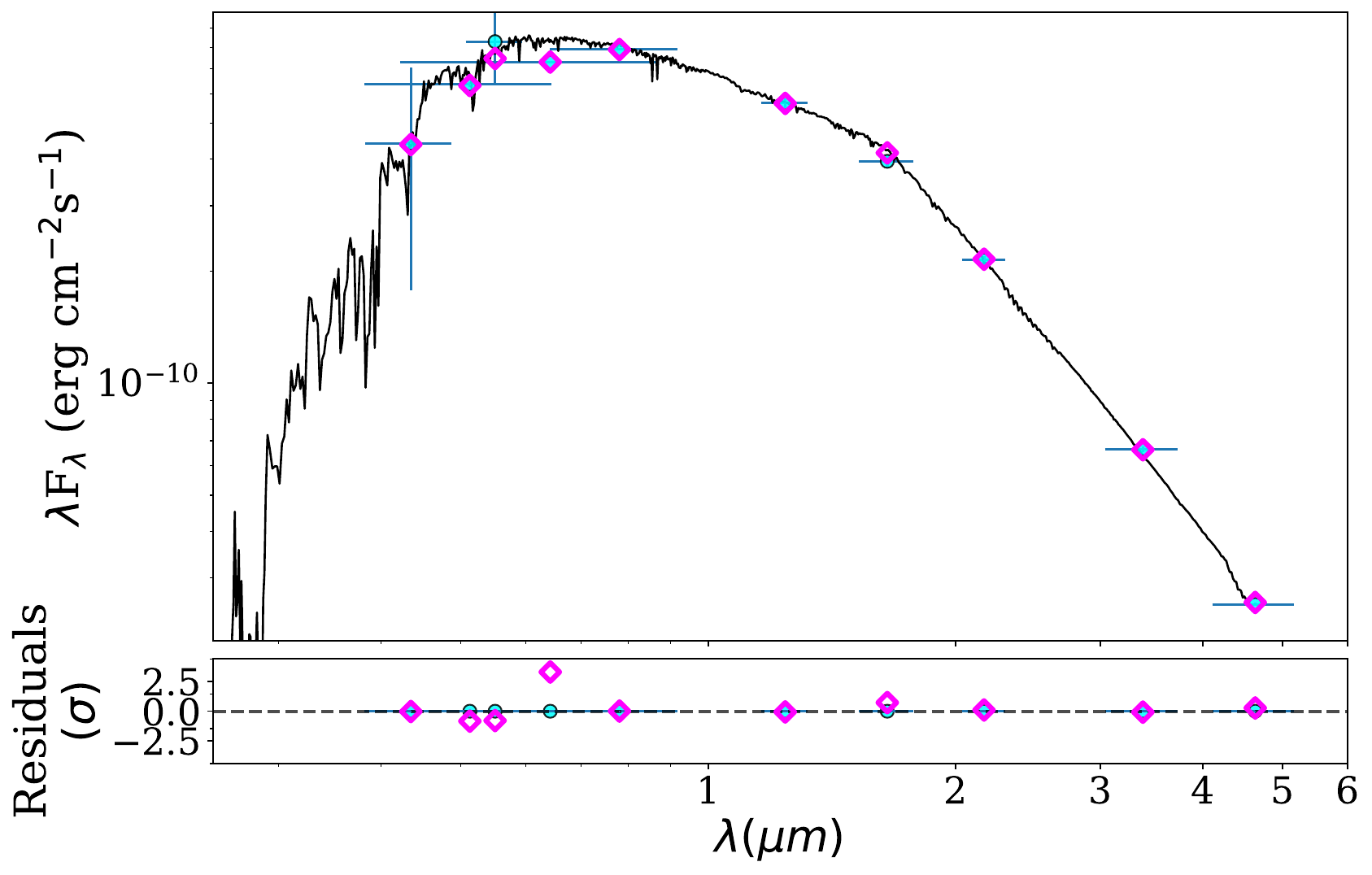}
  \caption{SED of \targeta~and the  model with highest probability \citep{Castelli2004}
   are plotted together with the   synthetic photometry  (magenta diamonds) and the observed 
photometry   (blue points). The vertical error bars outlines the   $1\,\sigma$ uncertainties and the horizontal bars   the 
effective width of the passbands. The residuals of the fit (normalised to the errors of the photometry) are shown 
in the lower panel.}
\label{Figure: SED}
\end{figure}

  In order to derive an empirical measurement of the stellar radius, we used   the {\tt {ARIADNE}}\footnote{\url{https://github.com/jvines/astroARIADNE}} \citep[][]{2022MNRAS.513.2719V} software to perform an analysis of the broadband spectral energy distribution (SED) of the star 
together with the {\it Gaia\/} DR3 parallax and the {\tt {SME}} results   as priors. 
The SED was fitted based on  
four atmospheric model grids,  
{\tt {Phoenix~v2}} \citep{2013A&A...553A...6H}, {\tt {BtSettl}} \citep{2012RSPTA.370.2765A}, 
\citet{Castelli2004}, and \citet{1993yCat.6039....0K},  
to the observed  broadband photometry  bandpasses  $G$, $G_{\rm BP}$, and $G_{\rm RP}$ (Gaia DR3),  
the Johnson $B$ and $V$   magnitudes (APASS), 
the infrared    $JHK_S$ magnitudes (2MASS) and W1-W2 magnitudes ({\it WISE}) listed in Table~\ref{Table: Star basic parameters}. An upper limit on  the visual extinction,   $A_V$, was obtained from the 
 dust maps of \citet{1998ApJ...500..525S}.    
A weighted average of the    stellar radius was computed with Bayesian model averaging  based on the  relative probabilities of each model. 
The SED model is shown in Fig.~\ref{Figure: SED}. 
The  luminosity becomes \Lumariadne~\Lsun~and  the extinction is consistent with zero \mbox{($A_\mathrm{V} =$ \Avariadne)}.   
The stellar mass is also modelled with {\tt {ARIADNE}}  based on the MIST \citep{2016ApJ...823..102C} isochrones and 
the output from the SED model. The results are listed in Table~\ref{Table: stellar mass and radius}. 

To check the derived stellar parameters, we  used the online applet {\tt {PARAM1.3}}\footnote{\url{http://stev.oapd.inaf.it/cgi-bin/param_1.3}} 
\citep{daSilva2006} with the {\tt SME} spectroscopic parameters as priors, the Gaia DR3 parallax and the 
apparent visual magnitude, along with   the {\tt PARSEC}  
\citep{2012MNRAS.427..127B} isochrones. Finally, we also compared our results with the  \citet{Torres:2010} 
calibration equations computed with the {\tt ARIADNE} posterior spectroscopic parameters.

The results of the different models are listed in Table~\ref{Table: stellar mass and radius}   with the corresponding stellar densities, 
which are used to constrain our transit model in Sect.~\ref{Section: transit and RV modelling} and comparison 
values of a typical \spectraltype~star.  
The values  are  in excellent agreement and we chose    the {\tt ARIADNE} results as our final adopted values  
also listed in Table~\ref{Table: Star basic parameters}. Some key stellar parameters   required to obtain the absolute planet radii and masses from  the transit and 
RV modelling, as well as the instellation and equilibrium temperature, are also listed in Table~\ref{Table: Orbital and planetary parameters}. 
Our derived stellar parameters are also in excellent agreement   with the   values listed on    the {\tt EXOFOP-TESS} website.

The age of a main sequence star such as \targeta~is often very difficult to   estimate accurately.  We obtained    ages of \mbox{\agespechmatch~Gyr} with
 {\tt SpecMatch-Emp}, \ageparam~Gyr with {\tt {PARAM1.3}}, and  \ageariadne~Gyr with {\tt {ARIADNE}}     
 based on the  MIST \citep{2016ApJ...823..102C} isochrones.  
  Based on the derived projected rotational period, \mbox{$P_{\rm rot}/\sin i_\star$},   in Sect.~\ref{Subsection: Stellar rotation period} 
 and different gyrochronology (rotation-age) relations   \citep{Barnes:2007, 2008ApJ...687.1264M, 2019AJ....158..173A, 2023ApJ...952..131M}, 
 we obtained ages from 1.8 to 5.4~Gyrs. 
Finally, we also estimate the age using the magnetochronology relations from \citet{2008ApJ...687.1264M}. 
This relation, combined with our mean value of 
\mbox{\lgr \,=\, \logRHK}~(Sect.~\ref{Subsubsection: Frequency analysis of   RVs and activity indicators}) results 
 in an age range from 5.0 to 5.5~Gyr. Thus, the different estimates points to an age between $\sim2$ and 6~Gyr.

\subsection{Stellar rotation period} \label{Subsection: Stellar rotation period}
We computed a first estimate of the projected rotation period, 
$P_{\rm rot}/\sin i_\star$ from the spectroscopic \vsini~together with \rstar\ and 
obtained  $P_{\rm rot}/\sin i_\star = 22.8 \pm 11.4$~d. Given the large error
 bars and the unknown
value of $i_\star$, we proceeded with an analysis of the TESS photometric data.

For the analysis of stellar rotation in the TESS light curves, especially for long rotation periods, it is crucial to preserve low-frequency stellar signals. Unfortunately, the PDCSAP flux includes filtering and CBV corrections that removes systematics, which can also eliminate long stellar rotation variability. Thus, we used SPOC \citep{Twicken10, 2020ksci.rept....6M} light curves 
as well as Quick Look Pipeline \citep[QLP,][]{2020RNAAS...4..204H,2020RNAAS...4..206H, 2022RNAAS...6..236K} SAP\_FLUX  data  
to investigate the signature of the stellar rotation.  This implies that particular caution is required, as the long-period variability observed in TOI-1438 may be significantly affected by TESS systematics, as seen in Sector 74. Exoplanet analyses rely on PDCSAP light curves, which are optimised for high-frequency signals; this approach allows the inclusion of a larger amount of data, as low-frequency variability is not critical for such studies.

Starting with 38~TESS sectors, 
we built light curves with a 30-min cadence, where the gaps between 
sectors (longer than 1~yr) have been 
removed, assigning to the first cadence of the new campaigns the 
time of the last cadence of the previous 
campaign plus 30~min. This is justified because we expect active regions
 to lose coherence during gaps of one to two years \citep[for intensive tests on TESS-like observations see,][]{2024tsc3.confE..32G}. Moreover, to 
 enable a reliable analysis of rotation periods up to 30 days, we required the use of segments of continuous TESS data spanning at least 90 days. 
 Hence, we  independently analysed chunks of continuous sectors 
(for example sectors $14-21$, $23-26$, $47-51$, and so on) and     
removed  the sectors  dominated by TESS systematics, such as sectors 57 and 74.

For all of the datasets, we tried different stitching methods  \citep{2024tkas.confE..93P} 
by using the mean of the sectors, the mean of half sectors, fitting 
different polynomials at the end and beginning of the continuous 
sectors, finding the middle point of the gap, and looking for the 
offset as done in \citet{2011MNRAS.414L...6G}. We also  combined the 
different methods using Bayesian techniques as done in 
\citet{2014MNRAS.445.2698H},  using the pyTADACS pipeline 
\citep[under development,][]{2024tsc3.confE..32G}. For all these 
light curves, we   interpolated the gaps using a multi-scale 
discrete cosine transform following inpainting techniques 
\citep{2014A&A...568A..10G, 2015A&A...574A..18P}.

We applied different methods on   the resulting light curves  to search for the stellar rotation period: 
periodogram, time-frequency analysis with wavelets \citep[e.g.][]{1998BAMS...79...61T,2010A&A...511A..46M}, the 
auto-correlation function \citep[e.g.][]{2013MNRAS.432.1203M}, 
and the composite spectrum that is a 
combination of the wavelet power spectrum and the ACF \citep{2017A&A...605A.111C}. 
These different methods in our rotation pipeline enabled us to find the most reliable rotation periods
 \citep{2015MNRAS.450.3211A}. This procedure was successfully  
 applied to the \emph{Kepler} data 
 \citep{2019ApJS..244...21S,2021ApJS..255...17S} providing a 
 catalog of more than 55,000 rotation periods. 

However, the analysis of the various TESS light curves of \targeta~did not yield a reliable rotation period. Modulations with periods of approximately 14 and 27 days were detected most of the time that correspond 
to the sector and half-sector periodicities suggesting that we were not able to properly detrend the light curves. 
Only the analysis of the first continuous segment (sectors $14 -21$) yielded a possible rotation period of \prottess~d  (obtained from the wavelet analysis),  
which does not seem to be due to any instrumental effect.  We want to emphasise that only small gaps were interpolated in this segment.
Given that the TESS observations span more than 5~yr, 
it is possible that this could be due to changes in the magnetic activity cycle of the star.

   \begin{figure}[!ht]    
\includegraphics[scale=0.33]{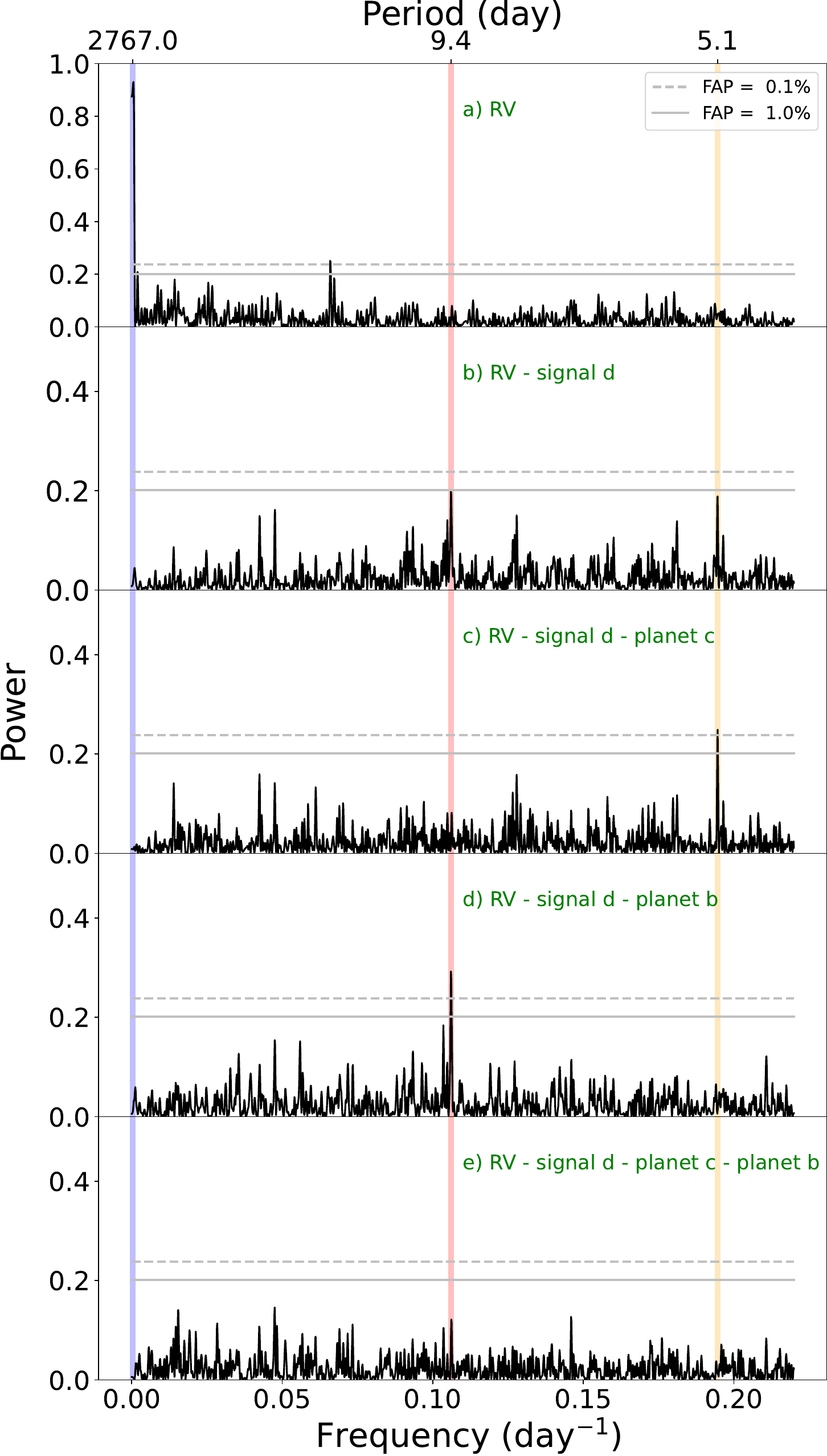}
   \caption{GLS periodogram of the \mbox{HARPS-N} and HIRES RV data. The GLS of the raw RVs are shown in the top panel and the below panels have  the best-fitting models subtracted  for a given planet or signal~d, as described in Sect.~\ref{Section: transit and RV modelling}. From right to left, the   orange,  red, and blue vertical lines  denotes  the  orbital periods of planets~b and c, and signal~d, respectively.
   }
      \label{Figure: carina periodogram RVs}
 \end{figure}

   \begin{figure}[!ht]
   \centering
  \includegraphics[scale=0.33]{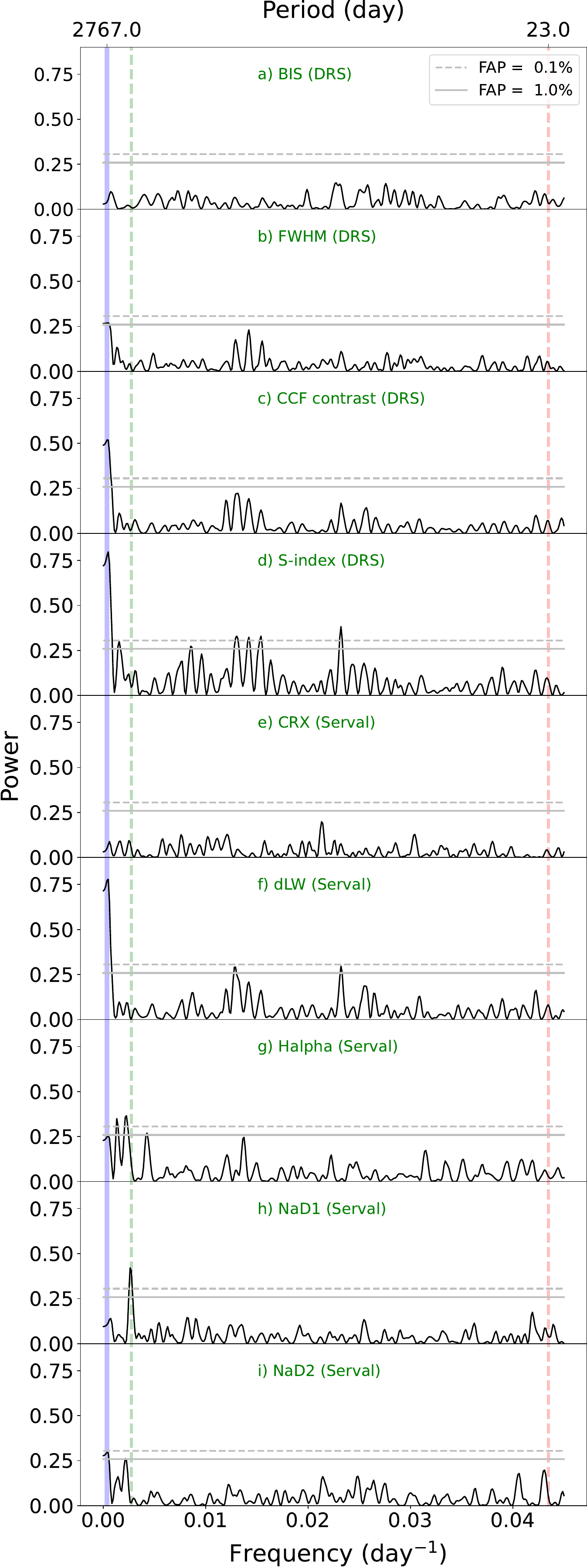}
   \caption{GLS periodogram of  \mbox{HARPS-N}  data with activity indicators from DRS and  Serval as indicated in the legends. The preliminary stellar rotation period of 23~days and Earth's orbital period of 365~days are marked with red and green vertical dashed   lines, respectively,   and signal~d with a blue vertical thick line.}
      \label{Figure: carina periodogram activity indicators}
 \end{figure}

Thanks to the ground-based follow-up of this star providing the 
activity index $R^\prime_{HK}$ from the \ion{Ca}{II} H \& K  
lines (see Sect.~\ref{Subsubsection: Frequency analysis of   RVs 
and activity indicators}), we checked the magnetic activity level 
of the star during the TESS observations. We found that the TESS 
sectors where the overall variance of the light curve is larger \citep[which could be related to a larger activity, e.g.][]{2016A&A...596A..31S} 
correspond to a high level of surface magnetic activity retrieved 
from $R^\prime_{HK}$. This suggests that the low-frequency 
modulation we see in the TESS light curves could be of magnetic 
origin related to activity. A longer, continuous dataset is required to obtain a reliable $P_{\mathrm{rot}}$  for this star.

The  preliminary value of \prottess~d is, however,  in agreement with the rotation period computed from \rstar~and \vsini.
Assuming that the  value derived from TESS photometry is robust, 
we estimated the stellar inclination from the rotation period in
combination with the spectroscopically derived  \vsini~and $R_\star$  following \citet{2020AJ....159...81M}. 
Using \mbox{\vsini$=1.8\pm0.9$~\kms}, 
\mbox{$R_\star=0.820\pm0.017$~R$_\odot$}, and 
\mbox{\Prot \, =  \prottess~d}, 
the stellar inclination becomes   \mbox{$i_\star=64^\circ$$^{+26}_{-11}$}. This is  consistent with the picture of a well-aligned system.

\subsection{Frequency analysis of the RVs and activity indicators} \label{Subsubsection: Frequency analysis of   RVs and activity indicators} 
We performed a frequency analysis of the HARPS-N time series data to search for  signals from the planets and the 
host star in the RV measurements and activity indicators. 
Figure~\ref{Figure: carina periodogram RVs} shows the generalised Lomb-Scargle \citep[GLS;][]{Zech09} 
periodograms of the \mbox{HARPS-N} and HIRES RVs. The false alarm probabilities (FAPs) were computed using 
the bootstrap technique  \citep{1997A&A...320..831K}. A peak is considered to be significant if its \mbox{FAP $< 0.1$~\%}. 
In the top panel of Fig.~\ref{Figure: carina periodogram RVs}, we show the periodogram of the raw RVs, 
which clearly displays a long-period signal marked with a blue vertical line which we hereafter refer to as signal~d. The origin may be a
wide orbiting, massive companion, either (sub-)stellar or planetary. 

We subtracted our best fitted  model of the   long-period signal of 2767~d (see Sect.~\ref{Section: transit and RV modelling} and Table~\ref{Table: Orbital and planetary parameters})  
and plot the resulting periodogram in the second top panel, revealing two peaks at the orbital periods of planets~b 
and c  (5.1 and 9.4~days, respectively) found in the transit data, marked by   vertical red and orange lines. 
Further below we show the periodograms after also subtracting the best-fitted model for planet~c  and b in the third and fourth panels, respectively,  
where the   signals of planets~b and c  are significantly detected in respective plot. Residuals after subtraction of both planets
and signal~d are plotted in the bottom panel. 

Figure~\ref{Figure: carina periodogram activity indicators} shows the GLS periodograms  
 of several activity indicators. 
No signals are detected at the orbital periods of planet~b or c, hence we focus the figure on the  short frequencies 
since  a long-period signal of the order of thousands of days (similar to signal~d)    is   seen 
in several of the activity indicators. We mark the orbital period of signal~d with a vertical blue line for comparison, as well as 
the preliminary stellar rotation period of 23~days  and Earth's orbital period of 365~days with red and green vertical dashed   lines, respectively.  The long-period signal is most prominent in the 
 S-index,  dLW, CCF contrast, and   Na\,D1.    

A counterpart to the RV signal in the activity indicators  would typically suggest that this 
phenomenon is of stellar origin, possibly stemming from long-period magnetic  cycles. 
The long-period peak in Na\,D1, however,  coincides with Earth's orbital period (0.00274~days$^{-1}$)  which suggests contamination by telluric features in this activity indicator.  
A correlation plot of RV against  dLW is shown in Fig.~\ref{Fig: correlation plot RV vs dLW}  which may suggest that dLW is $\sim 90^\circ$ out of phase,   hence possibly anti-correlated with RVs.

Subtracting the long-period signal~d of 2767~d from the \mbox{S-index}, dLW, and CCF contrast   does not remove the signal or reveal the stellar rotation period. This may indicate that there are several long-period signals. However, the exact periodicity of the signals is difficult to determine, as the baseline is only  1824~d and  thus the signals are not resolved at low frequencies. 

We   searched for chromatic changes in the RVs to rule out a planetary origin of signal~d. If  signal~d is due to an orbiting planet, it should not be chromatic 
but consistent across different wavelength ranges. This is supported by the non-correlation of the CRX, defined as the RV gradient as a function of wavelength, with RVs shown in    Fig.~\ref{Figure: Fig: correlation plot RV vs CRX}. In addition, we
 extracted the RVs from the HARPS-N spectra  
from different parts of the spectra; “blue” ($387.5 - 455.7$~nm), “yellow” ($453.9 - 549.3$~nm) , and “red” ($548.2 - 691.1$~nm) wavelengths. 
Signal~d was seen in the RVs extracted from all three wavelength ranges 
and was consistent with that observed from using the whole wavelength range supporting a planetary origin.

The \targeta~system may thus have an outer planet that induces the  variation  seen in the RV data as well as a 
 long-term stellar magnetic cycle as detected in the activity indicators, but with negligible effects on the RVs. 
 This scenario is supported by the large semi-amplitude $K$ and the fact that \targeta~is not a very active star in addition to our analysis of the RV jitter in Sect.~\ref{Subsection: signal d}.

     \begin{figure}[!ht]
\includegraphics[scale=0.3]{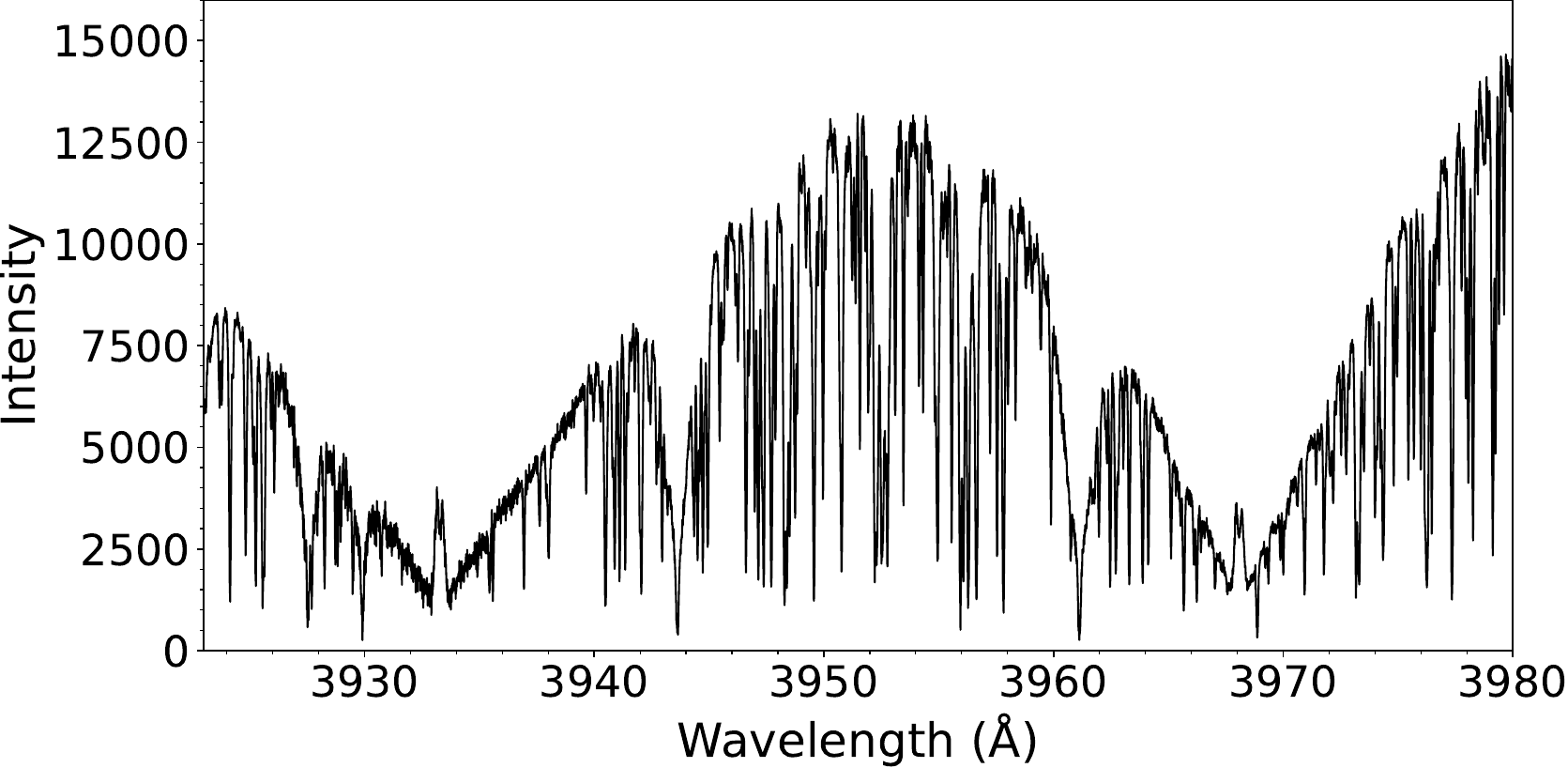} 
   \caption{Emission in the line cores of the \ion{Ca}{II} H \& K absorption lines in the co-added HARPS spectrum of \targeta. }
      \label{Figure: CaHK emission}
 \end{figure}

Compared with other solar-like stars from the Mount Wilson survey \citep{1978ApJ...226..379W, Baliunas1995}, a cycle period of about 2000 days would put the star on the active branch. 
When searching for signs of activity in the co-added HARPS-N spectrum, 
we identified only   weak  emission in the line cores of the \ion{Ca}{II} H \& K absorption
lines shown in Fig.~\ref{Figure: CaHK emission}.
Furthermore, the average \ion{Ca}{II} chromospheric activity  index \lgr~is  \mbox{$-4.925 \pm 0.013$}, which  
is comparable to the Sun at minimum activity \citep{2008ApJ...687.1264M,BoroSaikia2018}. 

However, since we cannot   rule out stellar activity as a possible origin of signal~d with certainty, 
the planetary nature of signal~d cannot be 
confirmed and remains a tentative discovery.  
More observations over a   longer period of time is required to   
resolve the issue and disentangle the signals.

We also searched the TESS light curves for transits originating from this potential planet without success. 
The expected time of mid-transit   seems to fall at a time where TESS was not observing 
this system. Furthermore, assuming  coplanarity with planets~b and c, the impact parameter of planet~d would be 
$>30$, clearly suggesting that   transits cannot be detected.

\subsection{Joint transit and radial velocity modelling} \label{Section: transit and RV modelling}
We modelled the TESS light curves using the {\tt{batman}} package 
\citep{batman} to extract information on the orbital period ($P$), 
the mid-transit time ($T_0$), the planet-to-star radius ratio 
($R_{\rm p}/R_\star$), the scaled orbital separation ($a/R_\star$), 
and the orbital inclination ($i$) of both transiting planets. 
Despite the wealth of TESS photometry for this system, the transits 
are quite shallow and  have a high impact 
parameter ($b=\cos i\times   a/R_\star$), which entails that some parameters 
are difficult to constrain. 
 Following \citet{seager03}, we therefore  constrained  $a/R_\star$ by a 
 Gaussian prior on the stellar density   derived from our spectral analysis.

\begin{figure*}
    \centering
    \includegraphics[width=0.89\textwidth]{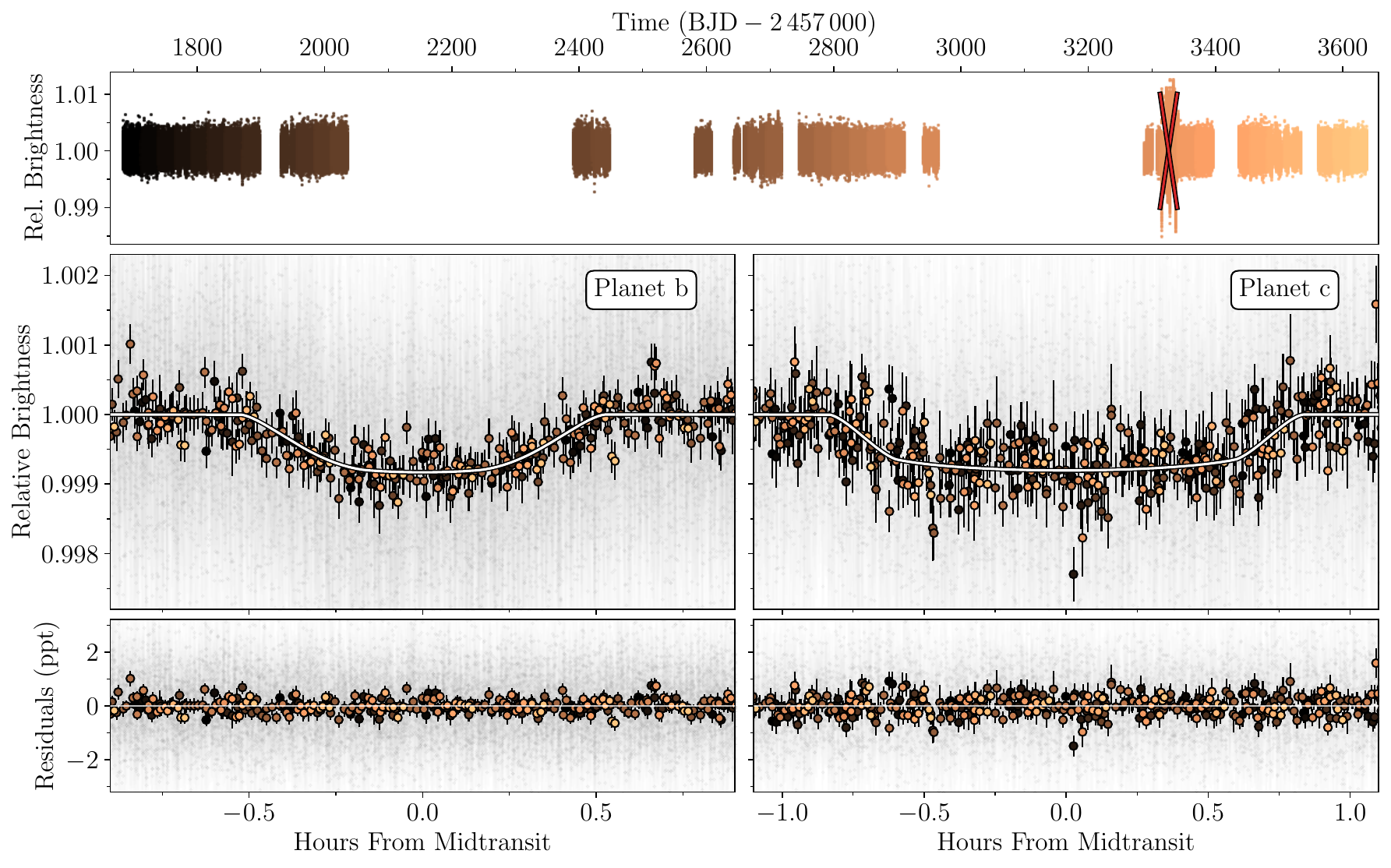}
    \caption{TESS photometry of \targeta. Top: The de-trended TESS time series colour-coded according to sector; from black in Sector 14 to light copper in Sector 81. The data from Sector 74 is under the red cross and was excluded from the analysis. In total, there are   150 and 79 transits of planets~b and c, respectively (excluding Sector~74). Middle: The phasefolded light curves for planets~b (left) and c (right). The unbinned data are shown as small, grey markers, and the larger markers are $\sim$10~min binned data colour-coded as in the top panel. The best-fitting models are shown as the white lines. Bottom: Residuals from subtracting the best-fitting models from the data.}
    \label{fig:lcs}
\end{figure*}

To account for photometric noise (instrumental and stellar) that might 
skew the estimation of the transit parameters, 
we detrended the light curves before fitting using Gaussian Process 
(GP) regression utilising the {\tt{celerite}} library \citep{celerite}.
We used a Matérn-3/2 kernel for our GP which is characterised by two 
hyperparameters; an amplitude 
($A$) and a timescale ($\tau$). This was done in an iterative manner, 
where we subsequently used the best-fitting transit parameters to 
filter out the transits when detrending. 
The detrended light curves for all TESS sectors are shown in Fig.~\ref{fig:lcs}. 
During fitting, we only included data in an interval with pre-ingress and 
post-egress around each mid-transit time.

We modelled the RVs as a sum of Keplerian orbits \citep[e.g.][]{Murray2010} to obtain information 
on the RV semi-amplitude ($K$) and, hence, the masses of the planets, as well as the orbital eccentricity ($e$) 
and argument of periastron ($\omega$). 
We extracted the HARPS-N RVs  from the \texttt{DRS} pipeline and 
\texttt{Serval}.  The results from using either set of RVs were fully consistent with a slightly lower RMS for 
the \texttt{DRS} RVs, and we thus chose to use these in the joint transit and RV analysis. 
 
\begin{figure*}
    \centering
    \includegraphics[width=0.89\textwidth]{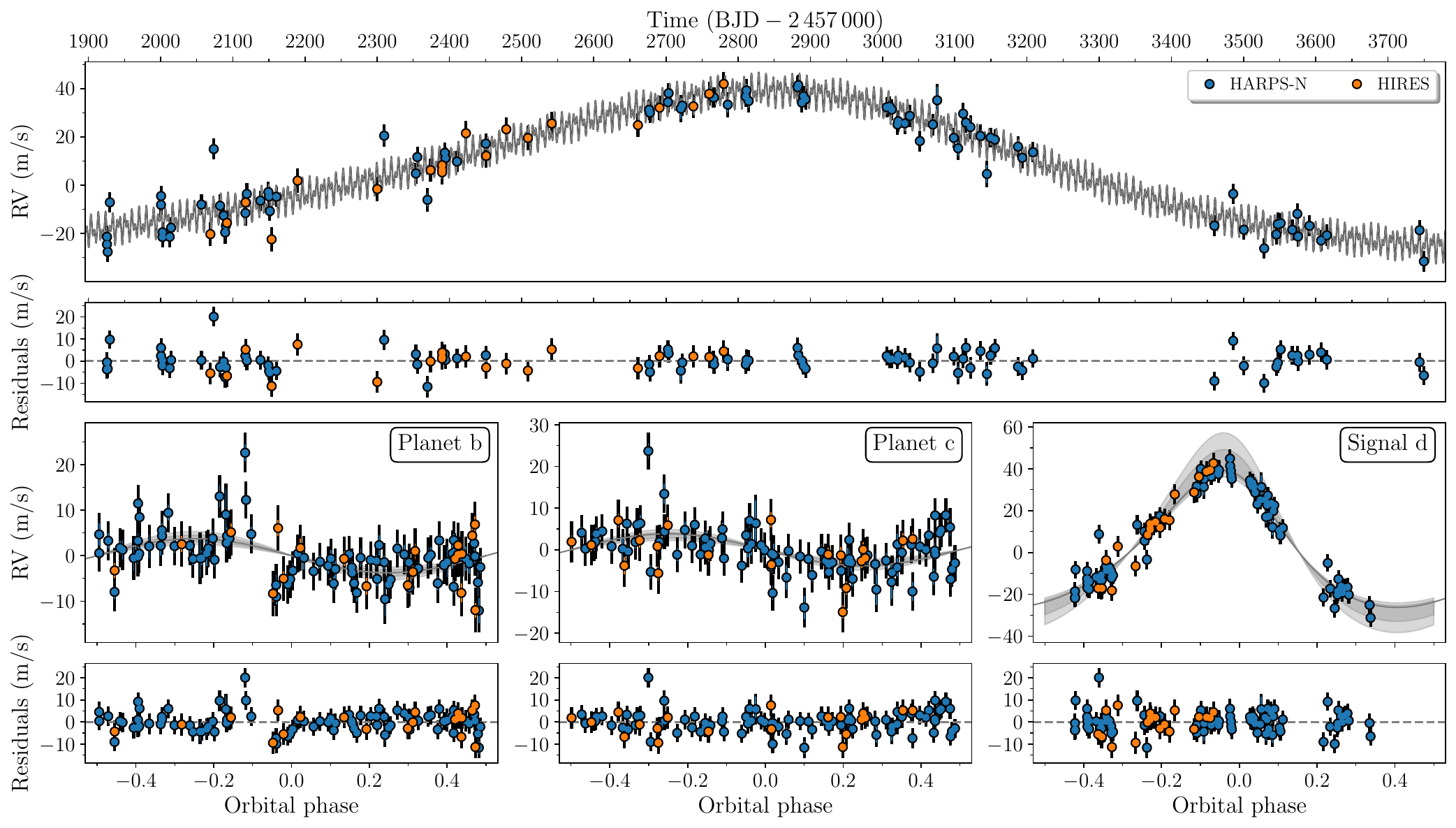}
    \caption{RVs of \targeta. Top: The HARPS-N (blue) and HIRES (orange) RV time series with the best-fitting three Keplerian model in grey with residuals shown in the panel below. Lower: The phasefolded RV curves for planet~b (left) and c (middle) as well as signal~d (right). The best-fitting models are shown as the grey lines with the shaded area denoting the $1$ and $2\,\sigma$ intervals in the $K$-amplitude. Residuals are given below.}
    \label{fig:rvs}
\end{figure*}

Modelling stellar activity through the use of multidimensional GP regression can significantly improve the orbital 
parameters \citep[e.g.][]{Rajpaul2015,2022MNRAS.509..866B}. 
This can be done by simultaneously modelling both the RVs and the activity indicator(s) displaying the signal with 
a quasi-periodic (QP) kernel \citep[see, e.g. Eq.~15 in][]{Aigrain2023} with hyperparameters describing the 
characteristic period ($\lambda_1$), the harmonic complexity ($\Gamma$), and the decoherence timescale ($\lambda_2$).
However, as pointed out by \citet{2022MNRAS.509..866B}, QP kernels should only be used in cases where 
the periodicity is smaller than the decoherence timescale (i.e, $\lambda_1<\lambda_2$), which in the present case constitutes a problem. 
This is because the periodicity of the signal seems to be longer than the baseline of the observations and is difficult to constrain. 
Obviously, it is therefore even more difficult to constrain the decoherence timescale in this case. 
Indeed, using the software {\tt{pyaneti}}\footnote{\url{https://github.com/oscaribv/pyaneti.}} \citep{2019MNRAS.482.1017B, 2022MNRAS.509..866B}, 
where we included the RVs and the dLW time series, we confirmed that it is  not possible to 
constrain $\lambda_1$ or $\lambda_2$ without applying  rather restrictive priors which 
goes against what we were trying to achieve by using multidimensional GPs,   i.e.  letting the data inform us. 
We were further able to confirm what is also evident from Fig.~\ref{fig:rvs}, namely that the harmonic 
complexity of this signal is very low, and it can therefore be exactly described by a sinusoid as discussed in \citet{Serrano2022}.

Given the low harmonic complexity, we decided to model the RVs  
as three Keplerians: one for each transiting planet and   one for the long-period signal, regardless 
of the origin of the signal  (planet or  stellar activity). 
We therefore proceeded without the inclusion of GPs and used our custom wrapper (i.e. the \texttt{batman} light curve and $N$ Keplerian model as outlined above) instead of \texttt{pyaneti}. 
In order to investigate if an even simpler model for the long-term trend is preferred, we   also tested a model 
including two Keplerians and  a quadratic trend. 
We find that the Bayesian information criterion (BIC) favours a three Keplerian model over a quadratic trend ($\Delta$BIC$=-47$).

The posterior distributions for the parameters were sampled through Markov chain Monte Carlo (MCMC) 
sampling through the {\tt{emcee}} code \citep{2013PASP..125..306F}. 
The MCMC was initialised with 100~walkers 
with a 150,000~steps each.   
The walkers were initialised with half of them starting in a tight Gaussian ball
around the best-fitting solution where the standard deviation is of the order of the uncertainty of the posterior for each parameter. The other half were spread uniformly across the allowed range. 
Convergence was assessed by calculating the rank normalised $\hat{R}$ 
diagnostic\footnote{\url{https://python.arviz.org/en/latest/api/generated/arviz.rhat.html}} and visually 
inspecting the chains in a \texttt{corner} plot \citep{2016JOSS....1...24F}.
When fitting for the eccentricity, we stepped in $\sqrt{e}\cos \omega$ and $\sqrt{e}\sin \omega$,  and   $\cos i$ instead of $i$. 
Furthermore, we used   a quadratic limb-darkening law, where we stepped in the sum ($q_{+}=q_1+q_2$) of the 
parameters to which we applied a Gaussian prior with a width of $\sigma=0.1$, while keeping the difference ($q_{-}=q_1-q_2$) fixed. At each step in $q_{+}$ we thus calculated $q_1=(q_++q_-)/2$ and $q_2=(q_+-q_-)/2$. 
The starting points for $q_1$ and $q_2$ were estimated by querying the table by \citet{Claret2018}.

We tested if any of the three signals displayed an appreciable amount of eccentricity
and the potential influence on the inferred parameters, especially the orbital period of planet d.
We began by making a model only including photometry  to constrain $P$ and $T_0$ for 
the two transiting planets. Those values were used as priors in the subsequent models only including RVs  
where we first did a run fixing the eccentricity to zero for all three Keplerians. 
In this case, the period for planet d came out to \mbox{$P=1798^{+62}_{-81}$~d} with a $K$-amplitude of \mbox{$28.8^{+0.9}_{-1.0}$~m~s$^{-1}$} 
meaning that with our 5~yr baseline 
we  barely covered the whole cycle of signal~d.
We then compared this to a model in which we allowed \mbox{$\sqrt{e}\cos \omega$} and \mbox{$\sqrt{e}\sin \omega$} to vary for all three. 
From this we found that the eccentricities for planets b and c  from this run were consistent with zero 
\mbox{($e_{\rm b}=0.038^{+0.020}_{-0.038}$} and \mbox{$e_{\rm c}=0.14^{+0.05}_{-0.13}$)}.
Moreover, we found a strong correlation between the orbital period and the eccentricity for planet~d 
where a tail extending beyond $e=0.5$, $P=5000$~d, and $K=40$~m~s$^{-1}$ is present in the posterior. 

This implies that with the RV data currently in hand, we cannot constrain the eccentricity of signal~d very well.
However, although these eccentric, long-period solutions could be consistent with the data, 
they suggest that the 5~yr (1824~d) of RVs we have collected so far,  were all acquired when planet~d was close to periastron. 
During this 5~yr time span,  we would  in this case  observe a large increase and decrease of 60~m~s$^{-1}$ in the RVs,    
while in the remaining   13~yr part of the orbit, the change in RVs   would only be about 20~m~s$^{-1}$. 
Such a scenario would mean   we have   have observed the system at a special time.
This effect is even more pronounced for longer  periods and  more eccentric solutions. 
Although it is not impossible that we   observed this system at epochs 
where it undergoes the most dramatic changes, it is unlikely.
Furthermore, the eccentric solutions
is to a large extent  driven by
the two most recent observations  as the posteriors before obtaining these last two 
RVs did not have these highly eccentric tails for planet~d. Particularly the most recent datum in 
Fig.~\ref{fig:rvs} is seen to be the lowest observed RV yet.
Finally, the quality of the derived orbit naturally depends on the number of cycles that has been 
covered \citep{Zakamska2011}, which presently in a best-case scenario for planet d means that 
we have only covered one cycle. Taken together we believe these arguments suggest that the 
eccentricity of planet d should be more modest. Rather than setting $e_{\rm d} = 0$, we applied a 
prior to $e_{\rm d}$ using the Beta-distribution from \citet[][]{2013MNRAS.434L..51K} with $\alpha=0.867$ and $\beta=3.03$. Applying this Beta-prior gives a period of \Pd~d ($\approx$ \Pdyear~years) for signal d.

  \begin{table}
 \centering
 \caption{Summary models of planets~b and c with free (model~\emph{i}) and fixed eccentricities (model~\emph{ii}).}
  \label{Table: eccentricity models planet b and c}
\begin{tabular}{lccc}
 \hline\hline
     \noalign{\smallskip}
Model    & $e$ &     $K_\mathrm{p}$ &     $M_\mathrm{p}$         \\
  & & (m~s$^{-1}$) & (\mearth)     \\
    \noalign{\smallskip}
     \hline
\noalign{\smallskip} 
\multicolumn{4}{c}{Model \emph{i}} \\
\noalign{\smallskip}
planet~b &   \ebfree  & $3.9^{+0.7}_{-0.8}$ & $9.7\pm1.8$   \\ \noalign{\smallskip} \noalign{\smallskip}
planet~c &   \ecfree & $3.8\pm0.7$ & $11.5\pm2.0$   \\
\noalign{\smallskip}\noalign{\smallskip} 
\multicolumn{4}{c}{Model \emph{ii}\,$\tablefootmark{a}$} \\ \noalign{\smallskip}
planet~b &  0 &  \kb& \mpb \\ 
\noalign{\smallskip}
planet~c &  0 &  \kc & \mpc \\
      \noalign{\smallskip} \noalign{\smallskip}
\hline 
\end{tabular}
\tablefoot{
\tablefoottext{a}{Adopted as our final model.} 
}
\end{table}

\begin{table*}
\centering
\caption{Best-fit transit and RV model of the \targeta~system, as described in Sect.~\ref{Section: transit and RV modelling}. }
\resizebox{2\columnwidth}{!}{%
\begin{tabular}{lcrrrr}
\hline
\hline
\noalign{\smallskip}
&&   \multicolumn{2}{c}{Planet b} & \multicolumn{2}{c}{Planet c} \\

\noalign{\smallskip}
Parameter & Units & Priors$\tablefootmark{a}$ & Final value & Priors$\tablefootmark{a}$ & Final value \\
\noalign{\smallskip}
\hline
\noalign{\smallskip}               
\multicolumn{3}{l}{\emph{Fitted parameters}}\\
\noalign{\smallskip}                
                ~~~$T_0$ &Transit epoch (\bjdtdb - 2\,457\,000)\dotfill &  $\mathcal{U}[1683.621,1683.630]$   & \Tzerob  &  $\mathcal{U}[1689.837,1689.984]$   &   \Tzeroc    \\
\noalign{\smallskip}                
                ~~~$P_\mathrm{orb}$ &Orbital period (d)\dotfill &  $\mathcal{U}[5.12,5.16]$  &\Pb &  $\mathcal{U}[9.32,9.50]$  &\Pc   \\
               
\noalign{\smallskip}                                
                ~~~$\cos i$  & Cosine of inclination \dotfill &  $\mathcal{U}$[0, 1]   & \cosib &  $\mathcal{U}$[0, 1] & \cosic   \\
\noalign{\smallskip}                
                ~~~$a/R_\star\,\tablefootmark{b}$  & Scaled semi-major axis\dotfill &  $\mathcal{U}$[10, 20]   &\arb   &  $\mathcal{U}$[15, 25]  & \arc  \\
 \noalign{\smallskip}                                      
                ~~~$R_{\mathrm{p}}/R_\star$ & Scaled planet radius\dotfill & $\mathcal{U}$[0.0, 0.2]    &\rrb& $\mathcal{U}$[0.0, 0.2]    & \rrc  \\
\noalign{\smallskip}
                ~~~$K $ & Doppler semi-amplitude   (\ms)\dotfill &   $\mathcal{U}$[0, 150]     &\kb &   $\mathcal{U}$[0, 150]       & \kc   \\
\noalign{\smallskip}
                ~~~$e$  &Eccentricity  \dotfill & $\mathcal{F}$   & $0$   & $\mathcal{F}$   & $0$  \\
\noalign{\smallskip}
                ~~~$\omega$    &Argument of periastron (deg)  \dotfill & $\mathcal{F}$  & $90$ & $\mathcal{F}$  & $90$    \\
 \noalign{\smallskip}
\multicolumn{3}{l}{\emph{Derived Parameters}}\\               
\noalign{\smallskip}                                             
                  ~~~$M_\mathrm{p}$ & Planet mass (\mearth)\dotfill   & \dots     & \mpb  &\dots & \mpc  \\
\noalign{\smallskip}                
                   ~~~$R_\mathrm{p}$ & Planet radius (\rearth)\dotfill &  \dots    & \rpb &\dots & \rpc  \\
\noalign{\smallskip}
                ~~~$i\, \,\tablefootmark{c}$  &Inclination (deg)\dotfill &   \dots   & \ib & \dots &  \ic  \\                
\noalign{\smallskip}                                
                ~~~$b$  &Impact parameter\dotfill &  \dots   & \bb &  \dots & \bc  \\
 \noalign{\smallskip}                                              
                  ~~~$a$ &Semi-major axis (au)\dotfill &  \dots   & \ab &  \dots & \ac   \\                   
 \noalign{\smallskip}                
           ~~~$F$ &Instellation  (\Fearth)\dotfill & \dots     &  \insolationb & \dots  & \insolationc  \\
  
   \noalign{\smallskip}                
           ~~~$\rho_\mathrm{p}$ & Planet density  (g~cm$^{-3}$)\dotfill &  \dots   & \denpb & \dots & \denpc \\

 \noalign{\smallskip}                
           ~~~$g_\mathrm{p}$ & Planet surface gravity  (cm~s$^{-2}$) \dotfill & \dots    &  \grapb & \dots & \grapc \\
\noalign{\smallskip}                
           ~~~$T_\mathrm{eq}\,\tablefootmark{d}$ &Equilibrium temperature (K)\dotfill & \dots    & \Teqb & \dots &  \Teqc   \\
           
 \noalign{\smallskip}                
           ~~~$\Lambda\,\tablefootmark{e} $ &  Jeans escape  (cm$^{2}$~g~erg$^{-1}$~s$^{-2}$) \dotfill & \dots    &  \jspb &  \dots & \jspc  \\ 
           
 \noalign{\smallskip}                
           ~~~TSM\tablefootmark{f} & Transmission spectroscopy metric \dotfill & \dots    &  \tsmb &  \dots & \tsmc  \\            
               
\noalign{\smallskip}                          
                ~~~$T_{14}$ &Total transit duration (h) \dotfill &  \dots   & \ttotb &  \dots & \ttotc    \\
 
 \noalign{\smallskip}                          
                ~~~$T_{23}$ &Full   transit  duration  (h)  \dotfill & \dots    & \tfulb &  \dots & \tfulc   \\               
\noalign{\smallskip} \hline \hline \noalign{\smallskip} 
&&   \multicolumn{2}{c}{Signal d} & & \\

\multicolumn{3}{l}{\emph{Fitted parameters}}\\      
\noalign{\smallskip} 
                ~~~$T_0$ &Transit epoch (\bjdtdb - 2\,457\,000)\dotfill &   $\mathcal{U}[3000,3500]$ & \Tzerod  &\ldots & \ldots   \\
\noalign{\smallskip}                
                ~~~$\ln P_\mathrm{orb}$ & Logarithm of orbital period (d)\dotfill &    $\mathcal{U}[0.8,10]$  &\lnPd &\ldots & \ldots  \\
\noalign{\smallskip}
                                       
                \noalign{\smallskip}
                ~~~$K $ & Doppler semi-amplitude   (\ms)\dotfill &     $\mathcal{U}$[0, 150]       & \kd  &\ldots & \ldots \\
\noalign{\smallskip}                                                
                ~~~$e$   &Eccentricity  \dotfill & $\mathcal{B}(0.867,3.03)$   & \ed &\ldots & \ldots  \\
\noalign{\smallskip}
                ~~~$\omega$    &Argument of periastron (deg)  \dotfill &   $\mathcal{U}(-180,180)$ & \wwd &\ldots & \ldots \\

\noalign{\smallskip}
\multicolumn{3}{l}{\emph{Derived Parameters (assuming planetary origin of signal~d)}}\\                      
\noalign{\smallskip}                
                ~~~$P_\mathrm{orb}$ &Orbital period (yr)\dotfill &    \dots  & \Pdyear &\ldots & \ldots  \\
\noalign{\smallskip}                                             
                  ~~~$M_\mathrm{p} \sin i$\,\tablefootmark{g} & Lower limit on planet mass (\mjup)\dotfill   &  \ldots  & \mpdjup &\ldots & \ldots  \\
\noalign{\smallskip}                                             
                  ~~~$a \sin i$ & Lower limit on semi-major axis (au)\dotfill   &  \ldots  & \ad &\ldots & \ldots  \\    
 \noalign{\smallskip} \hline \hline \noalign{\smallskip} 
\multicolumn{3}{l}{\emph{Additional Parameters}} &\ldots & \ldots \\      
\noalign{\smallskip}                              
                ~~~$\gamma_1$  &Systemic velocity HARPS-N (\ms)\dotfill &  $\mathcal{U}[-30000,-29000]$    &  \HARPSN      &\ldots & \ldots \\
               
\noalign{\smallskip}                              
                ~~~$\sigma_{1}$ &RV jitter HARPS-N (\ms)\dotfill &$\mathcal{U}[0, 50]$ &\jHARPSN     &\ldots & \ldots  \\        
\noalign{\smallskip}                              
                ~~~$\gamma_2$  &Systemic velocity HIRES  (\ms)\dotfill &  $\mathcal{U}[-1000,1000]$    &  \HIRES  &\ldots & \ldots    \\
               
\noalign{\smallskip}                              
                ~~~$\sigma_{2}$ &RV jitter HIRES (\ms)\dotfill &$\mathcal{U}[0, 50]$ &\jHIRES      &\ldots & \ldots  \\  
\noalign{\smallskip}  

                ~~~$q_1+q_2$ &Limb-darkening coeff. sum \dotfill &  $\mathcal{N}$[0.62, 0.10]   & \qsum  &\ldots & \ldots   \\
                
\noalign{\smallskip}                              
                ~~~$q_1-q_2$  &Limb-darkening coeff. difference  \dotfill &  $\mathcal{F}$   & 0.16   &\ldots & \ldots  \\                
 \noalign{\smallskip} \hline \noalign{\smallskip} 
\multicolumn{3}{l}{\emph{Adopted stellar parameters}}\\
\noalign{\smallskip}       
                ~~~$M_\star$   &Stellar mass (\Msun)  \dotfill &$\mathcal{F}$     & \smassariadne        &\ldots & \ldots      \\
\noalign{\smallskip}                
                ~~~$R_\star$ &Stellar radius (\Rsun)  \dotfill &  $\mathcal{F}$   &\sradiusariadne &\ldots & \ldots \\
\noalign{\smallskip}
                ~~~$T_{\rm eff}$  &Effective temperature (K) \dotfill & $\mathcal{F}$    & \steffariadne    &\ldots & \ldots \\

\noalign{\smallskip}                
\hline
\end{tabular}
}
\label{Table: Orbital and planetary parameters}
\tablefoot{The given values are the median and the uncertainty is the highest posterior density at a confidence level of 0.68.
\tablefoottext{a}{$\mathcal{U}$[a,b] refers to uniform priors in the range  \emph{a} --  \emph{b}, $\mathcal{F}$[a] to a fixed value $a$,  $\mathcal{N}$[a,b] 
to Gaussian priors with mean \emph{a} and standard deviation \emph{b}, and $\mathcal{B} (\alpha,\beta)$ refers to a Beta prior.} 
\tablefoottext{b}{A constraint on $a/R_\star$ is applied through the stellar density.}
\tablefoottext{c}{Orbit inclination   relative to the   plane of the sky.} 
\tablefoottext{d}{Dayside equilibrium temperature  without heat redistribution and  zero albedo.} 
\tablefoottext{e}{Jeans escape parameter defined as \mbox{$\Lambda =  G M_\mathrm{p} m_\mathrm{H} / (k_\mathrm{B} T_\mathrm{eq} R_\mathrm{p})$}   \citep{2017A&A...598A..90F}.}
\tablefoottext{f}{The transmission spectroscopy metric (TSM) is a proxy for the S/N of the James Webb Space Telescope (JWST) transmission spectroscopy  and recommended to be $>90$ \citep{2018PASP..130k4401K}.}
\tablefoottext{g}{Lower limit on the mass of planet~d assuming that   the long-period signal~d has a planetary origin.}
}
\end{table*}

In the following when we fitted the photometry and RVs jointly,
we therefore proceeded with the setup where we applied a Beta prior to $e_{\rm d}$. 
In order to investigate the eccentricities for planets~b and c, we ran two MCMC models summarised  in \mbox{Table~\ref{Table: eccentricity models planet b and c}}: \emph{(i)}  $\sqrt{e}\cos \omega$ and 
$\sqrt{e}\sin \omega$ were allowed to vary for both planets~b and c, and \emph{(ii)} with fixed eccentricities, 
$e_{\rm b}=e_{\rm c}=0$.  The   eccentricities in model \emph{(i)} came out to  \mbox{$e_b=$ \ebfree}~and \mbox{$e_c=$ \ecfree},  which again means that the orbit for planet b is completely consistent with 
being circular while a slightly eccentric model for planet c seems to be favoured. 
The resulting $K$-amplitudes and masses along with eccentricities for the two models are listed in Table~\ref{Table: eccentricity models planet b and c}.
The results are  fully consistent within the uncertainties. From a dynamical point of view 
(Sect.~\ref{sec-dynamical_analysis}) stable solutions are generally found for small eccentricities. 
Together with the fact that we find the eccentricity for planet b to be consistent with zero and only a very 
moderate eccentricity for planet c, we adopt  model  \emph{(ii)} in which the eccentricities for planets~b and c were fixed to zero as our final run
as we believe this to be the more conservative and simpler approach. 
Finally, we investigated the consequences of applying a different prior on $e_{\rm d}$, namely the ``Rayleigh+Exponential'' suggested by \citet{2025MNRAS.tmp..487S}, which is based on an updated sample of exoplanet eccentricities. The results of model \emph{(i)} and \emph{(ii)} with a ``Rayleigh+Exponential'' prior for planet~d were fully consistent with each other, and also with a Beta prior on planet~d.

 The final best-fitting transiting models for planets~b and c are shown in Fig.~\ref{fig:lcs} and the best-fitting Keplerians are shown in Fig.~\ref{fig:rvs}. 
We tabulate the posterior values in Table~\ref{Table: Orbital and planetary parameters} and show corner plots of the posterior distributions in Figs.~\ref{Figure: cornerplot planet b} - \ref{Figure: cornerplot signal d}.

The results stemming from the procedure outlined above was obtained using our own software. As an independent check, we modelled the photometry and RVs with {\tt{pyaneti}} \citep{2019MNRAS.482.1017B, 2022MNRAS.509..866B}, and a similar setup (i.e. a three-Keplerian model with a Beta-prior applied to $e_d$)  resulted in fully consistent parameters.

\subsection{Dynamical analysis}\label{sec-dynamical_analysis}
The influence that outer giant planets are expected to have on the orbital dynamics, especially the coplanarity, of inner transiting planets has been extensively studied \citep{2014ApJ...789..111B, 2017MNRAS.468..549B, 2017AJ....153...42L, Mustill2017, 2020MNRAS.498.5166P, 2021MNRAS.502.3746R, 2025ApJ...979..202L}. 
In the case of the \targeta~system, 
given the proximity of the  orbits of the innermost fairly massive planet to the 2:1 MMR region, 
the system may  be dynamically active, primarily  driven by the two mini-Neptunes. 
Any influence of the outer planet can likely be attributed to very long-term secular effects.
We have therefore focused on the short-term dynamics. 

The parameter space can be reduced to essentially three variables: the eccentricities forming a 
plane $(e_{\rm b}, e_{\rm c})$ and the difference of the periastron arguments 
$\Delta\varpi = \omega_{\rm c}-\omega_{\rm b}$, as the third dimension. 
In close in-systems, which are likely to have undergone inward migration, we can expect the 
most likely ``slices'' $\Delta\varpi=180^\circ$ or, depending on migration conditions, also 
$\Delta\varpi=0^\circ$ \citep[e.g.][]{Papaloizou2005,Hadden2020,Xu2019,Deck2022,Cresswell2022,Mills2023}. 
We also considered intermediate cases $\Delta\varpi=90^\circ$ and $270^\circ$, consistent 
with the best-fit values found in experiments with eccentricities as free parameters (Sect.~\ref{Section: transit and RV modelling}). 
Using this parametrisation, we can capture the essential features of the dynamics of the system 
by fixing relatively well-determined masses, inclinations, orbital periods, and mean anomalies 
at the epoch based on photometry data and further constrained by the joint RVs model. 
The eccentricity and pericentre arguments are allowed to vary as they have the largest 
uncertainties, reflecting the nature of the RV data.

\begin{figure}[!h]
\centering
\includegraphics[width=0.49\linewidth]{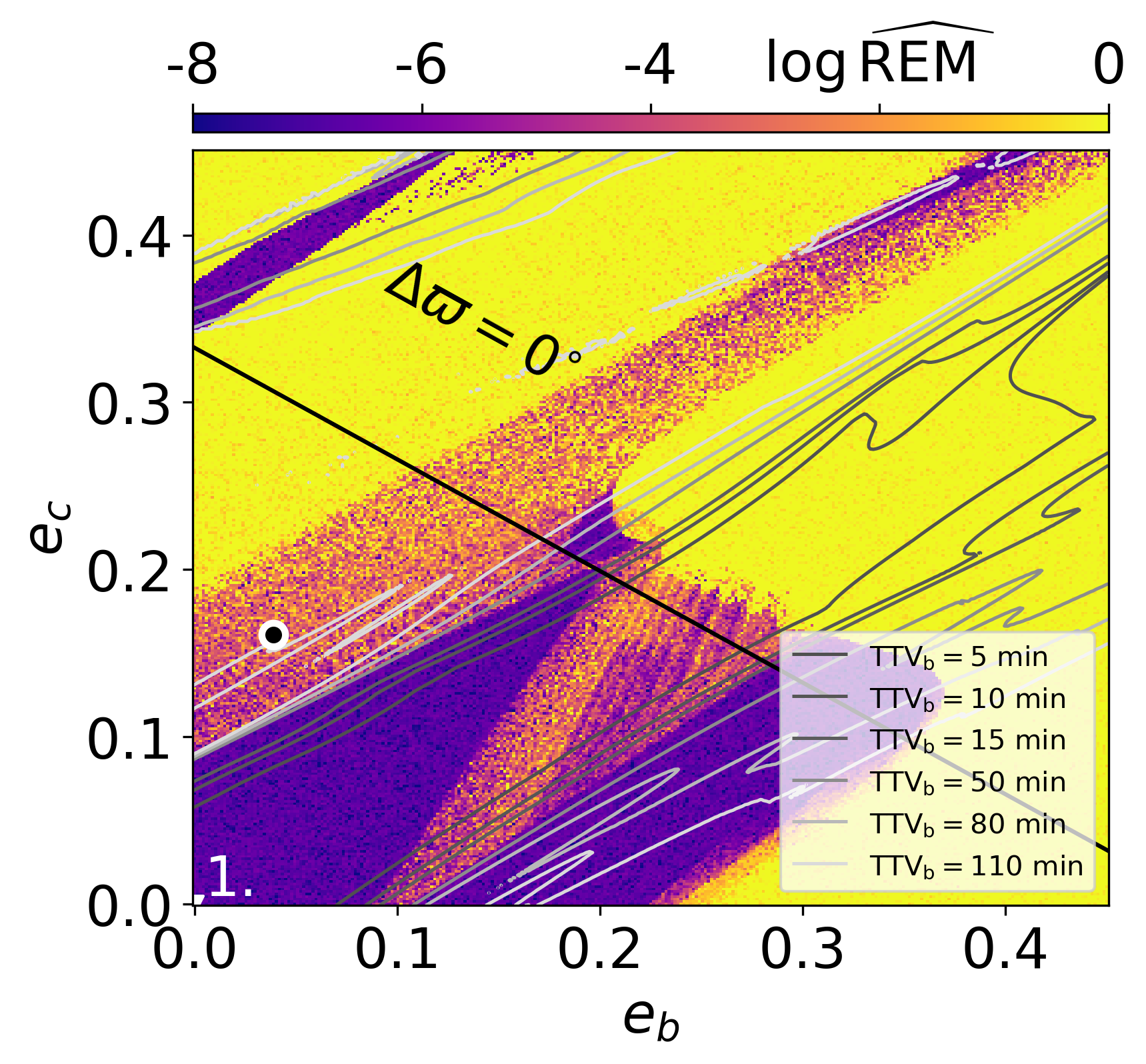}
\includegraphics[width=0.49\linewidth]{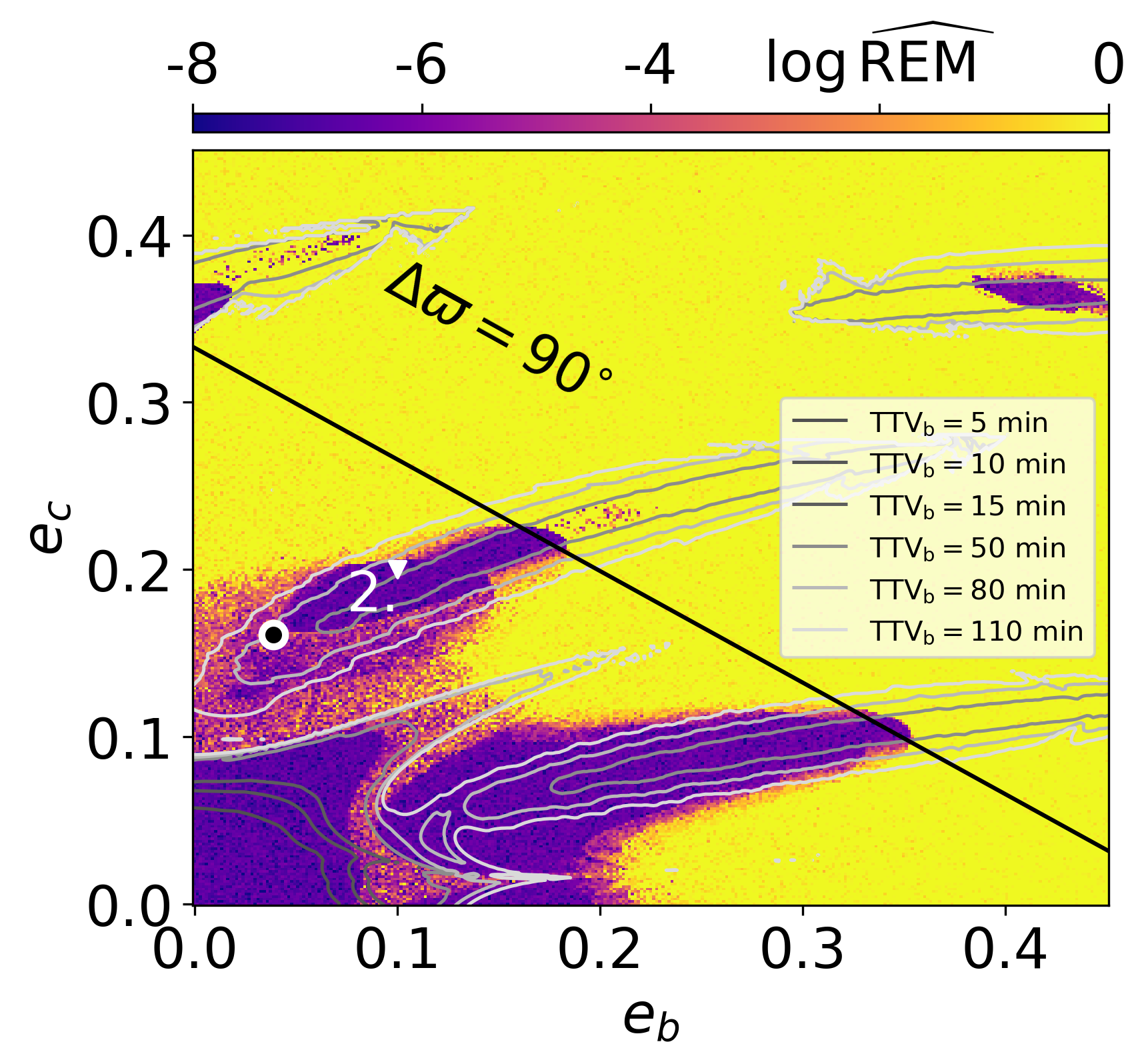}
\includegraphics[width=0.49\linewidth]{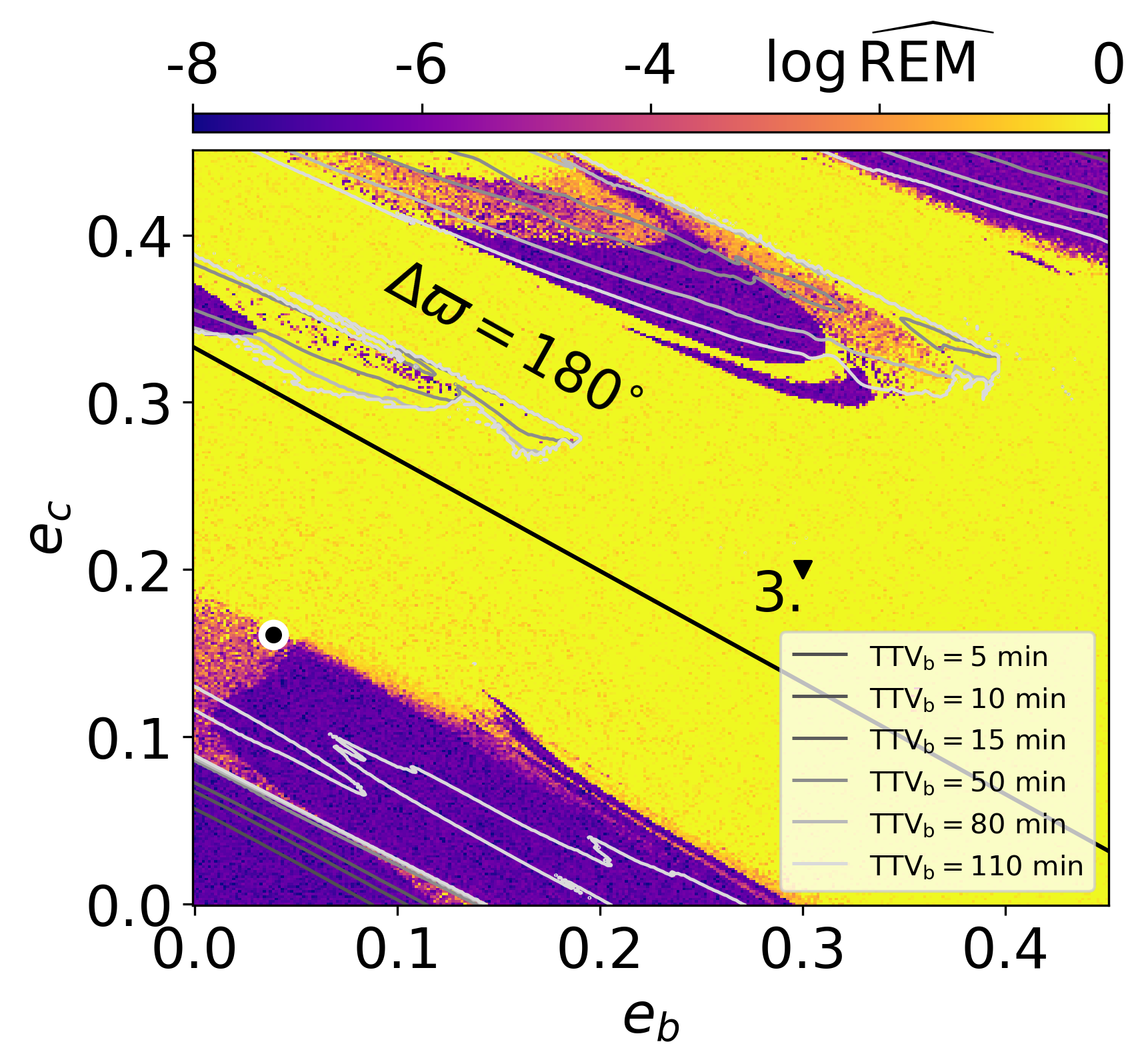}
\includegraphics[width=0.49\linewidth]{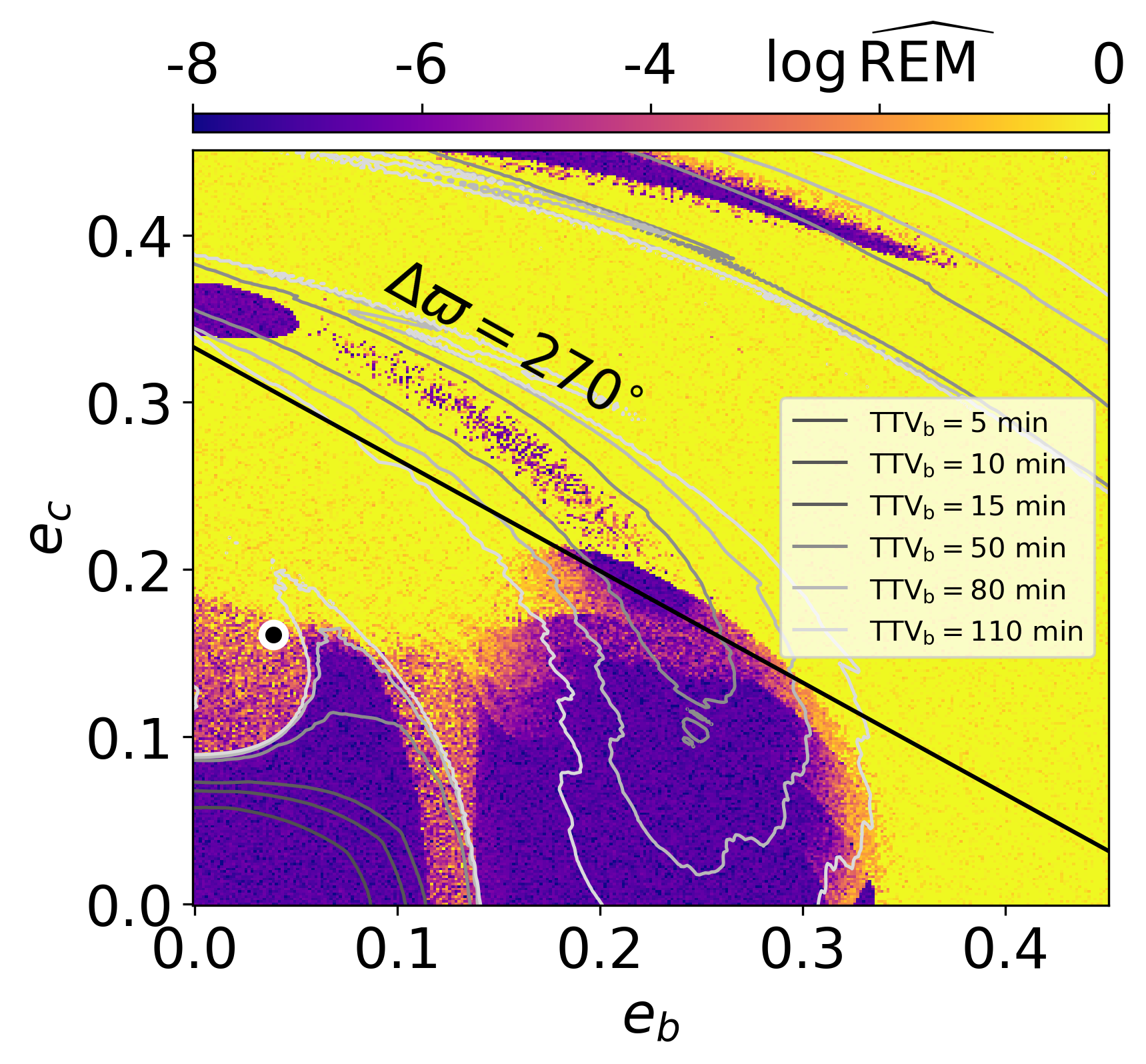}
    \caption{Dynamical maps of the eccentricities of the two inner planets ($e_b$, $e_c$) and fixed values of the argument of pericentre of \targetb ($\Delta \varpi =\omega_c-\omega_b$). The black filled circle with a white rim shows the location of the solution obtained by modelling with the eccentricities set as free parameters. Small values of the fast indicator $\log \mathrm{\widehat{REM}}$ characterises regular (long-term stable) solutions, which are marked with black and dark blue colour. Chaotic solutions are marked with brighter colours, up to yellow. The black line represents the so-called collision curve of orbits, defined by the condition: $a_b(1 + e_b) = a_c(1 - e_c)$.  The resolution of each plot is 301 $\times$ 301 points. Triangles  marked $1-3$ correspond to the solutions shown in Fig.~\ref{figure-TOI-2427-ebec}.
    Also, the labelled grey contours refer to the TTV amplitudes shown in Fig.~\ref{figure-TOI-2427-TTV} (details in Section \ref{sec-dynamical_analysis}).
    \label{figure-TOI-2427-dynamical_map}}
\end{figure}

To assess the dynamical stability of the solutions, we used the reversibility error method 
\citep[REM;][]{Panichi_Gozdziewski_Turchetti-2017MNRAS.468..469P}, which has been 
shown to be a close analog of the maximum Lyapunov exponent (MLE) and the well known 
indicator MEGNO \citep[e.g.][]{Gozdziewski2001,Cincotta2003}. In the analysis of multiple 
systems, the REM indicator relies on numerical integration schemes that are time-reversible, 
in particular symplectic algorithms. This method is based on the calculation of the difference 
between the initial state vector and the final state vector obtained by integrating the Newtonian 
equations of motion for a given interval and then returning to the initial epoch by integrating 
the system back for the same time. The difference depends on the dynamical nature of the 
system. $\mathrm{\widehat{REM}}=1$ or $\log\mathrm{\widehat{REM}}=0$ means that the 
difference reaches the size of the orbit. For regular (stable) orbits, $\mathrm{\widehat{REM}}$ 
grows with polynomial rate with time, and for chaotic (unstable) solutions it grows exponentially. 
The fast indicator is crucial for illustrating the phase space of the system, since it detects 
instability much faster than direct numerical integration could do.

We tested the dynamical stability of the solutions constructed around the nominal value of the best-fitted parameters from   
model \emph{(i)}  in Sect.~\ref{Section: transit and RV modelling} in the $(e_{\rm b}, e_{\rm c})$ plane for fixed, 
representative values of $\Delta\varpi$. 
We integrated the orbits with the \verb|whfast| integrator, the 17th-order 
corrector, and a fixed time step of 0.04~d as implemented in the \verb|REBOUND| 
package \citep{Rein_Liu-2012A&A...537A.128R,Rein_Spiegel-2015MNRAS.446.1424R} 
for 50,000 orbital periods of the 
second planet. Such a timescale is sufficient to detect short-term MMR-driven dynamics 
\citep{Panichi_Gozdziewski_Turchetti-2017MNRAS.468..469P}.

\begin{figure}[!h]
\centering
\includegraphics[width=\linewidth]{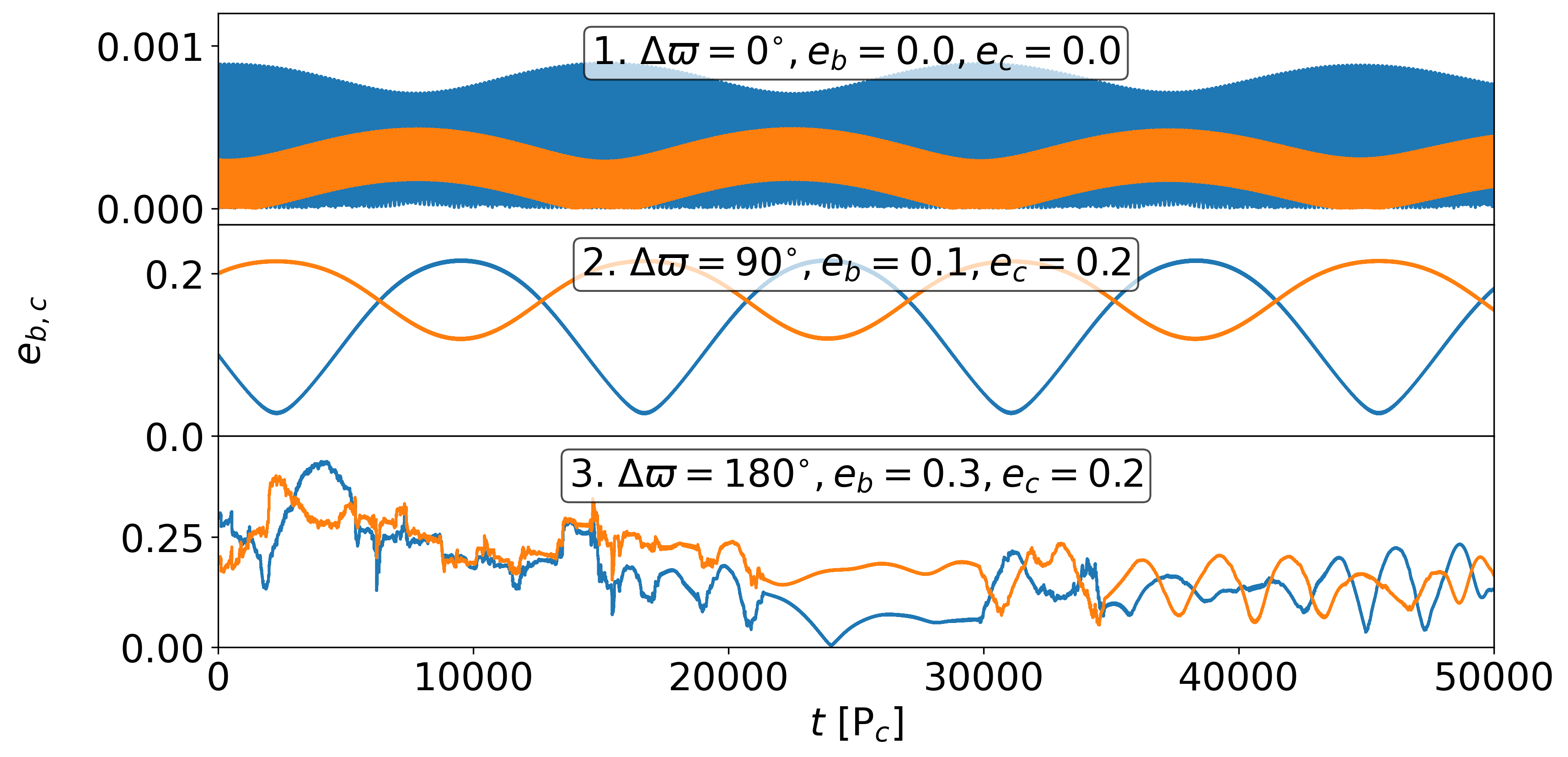}
    \caption{
    Temporal evolution of eccentricities for the representative solution selected in the dynamical maps shown in Fig.~\ref{figure-TOI-2427-dynamical_map}. The eccentricity of \targetb~and \targetc~are marked in blue and orange, respectively.
    \label{figure-TOI-2427-ebec}}
\end{figure}

\begin{figure}[!h]
\centering
\includegraphics[width=0.49\linewidth]{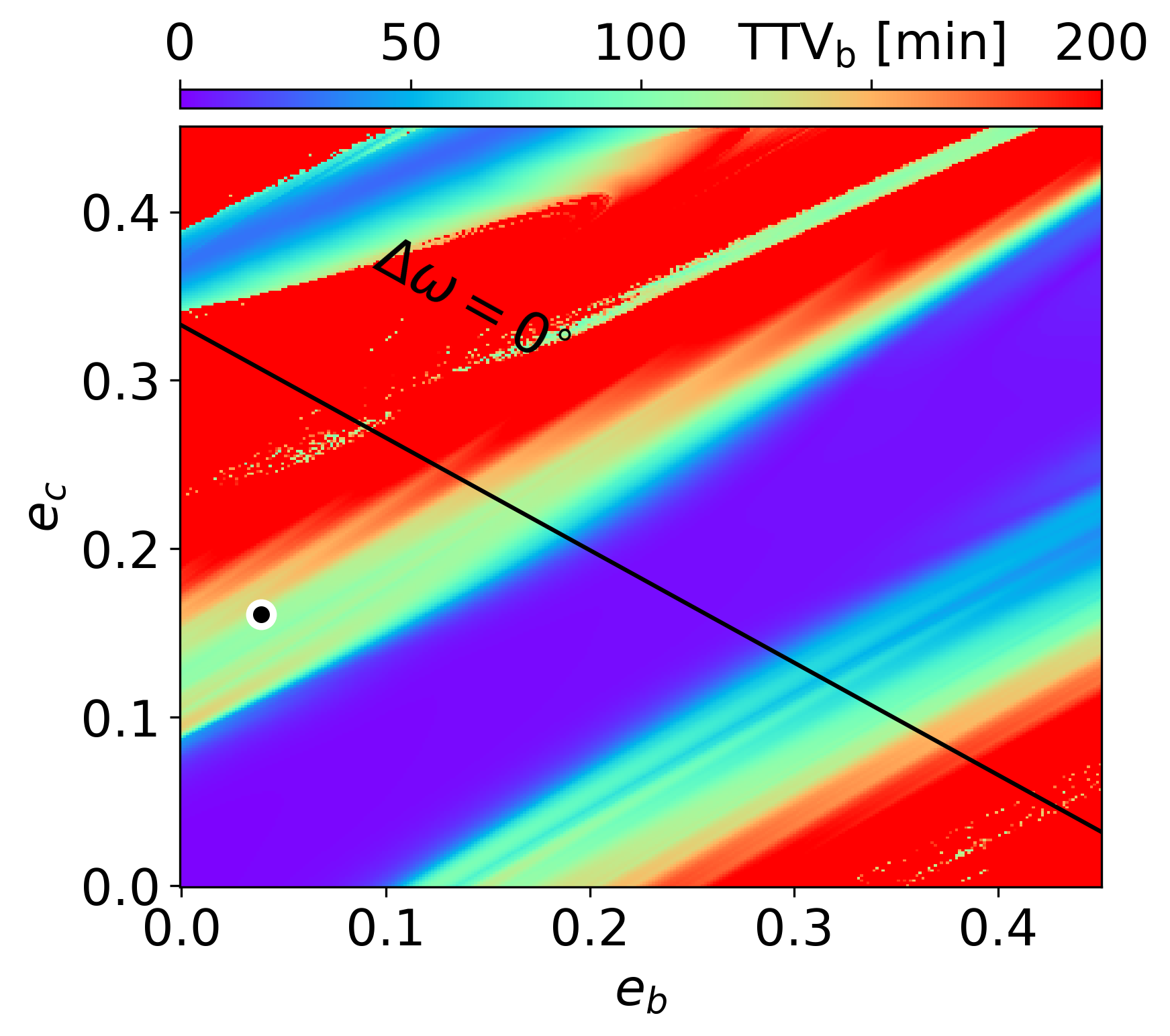}
\includegraphics[width=0.49\linewidth]{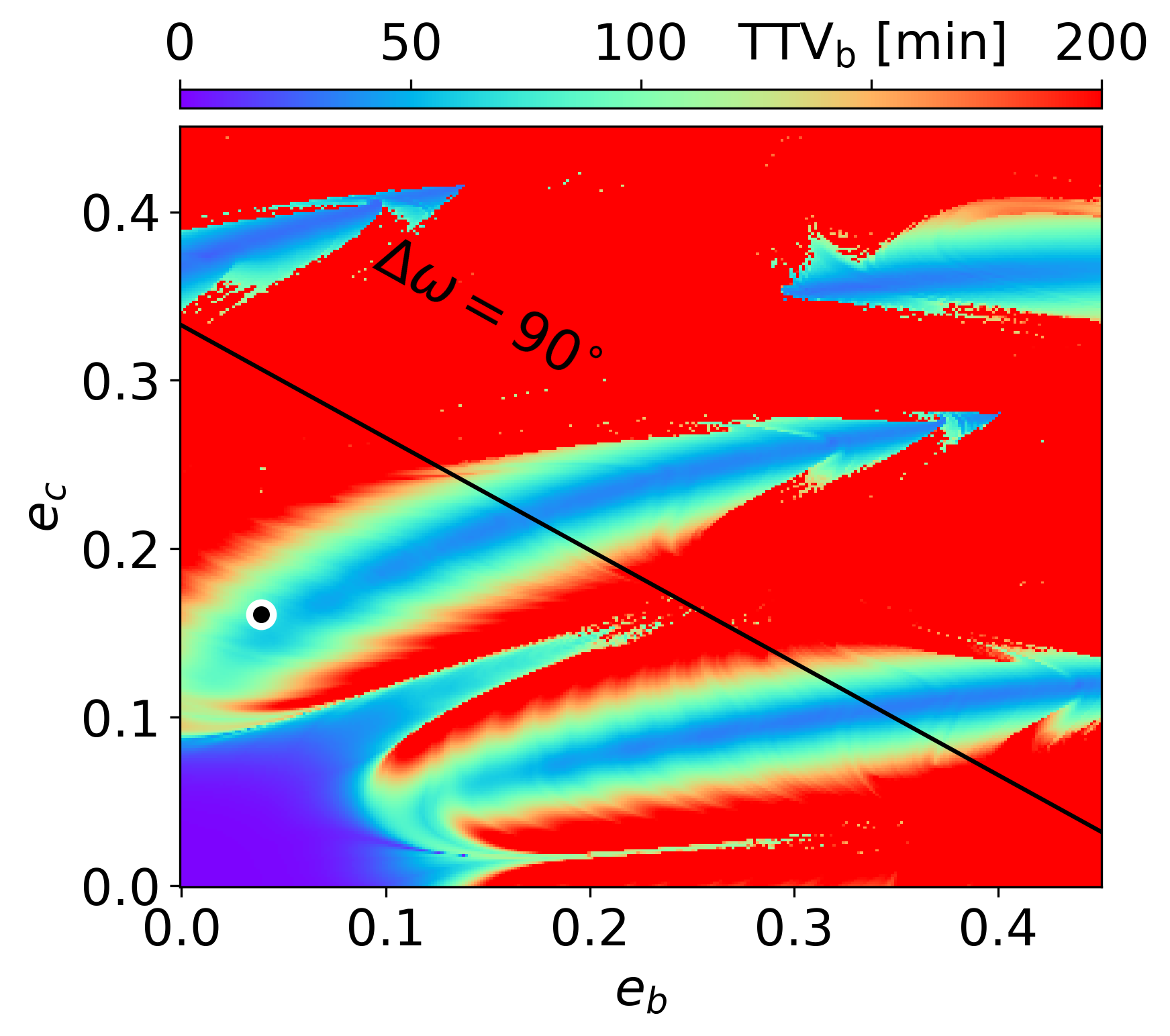}
\includegraphics[width=0.49\linewidth]{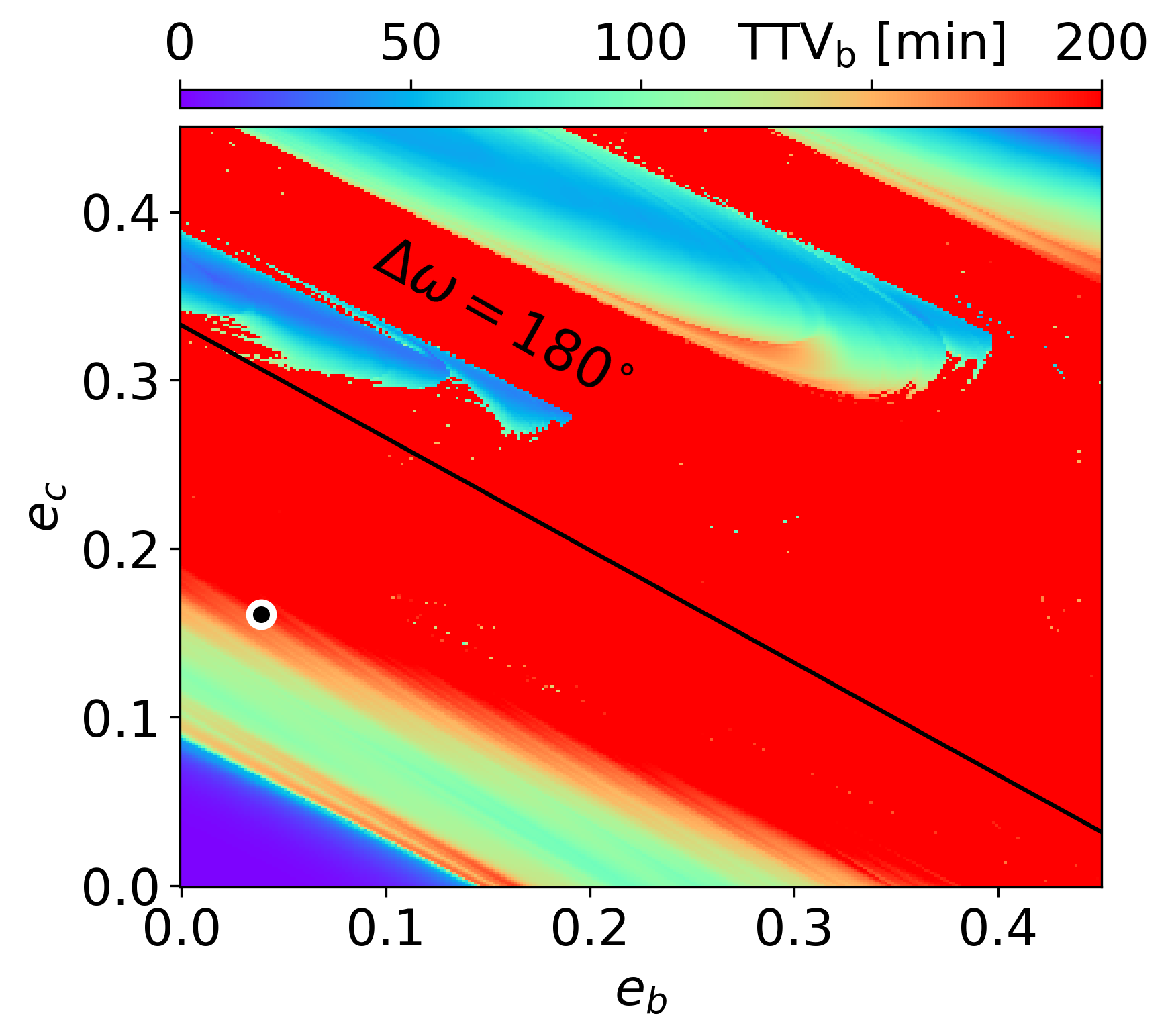}
\includegraphics[width=0.49\linewidth]{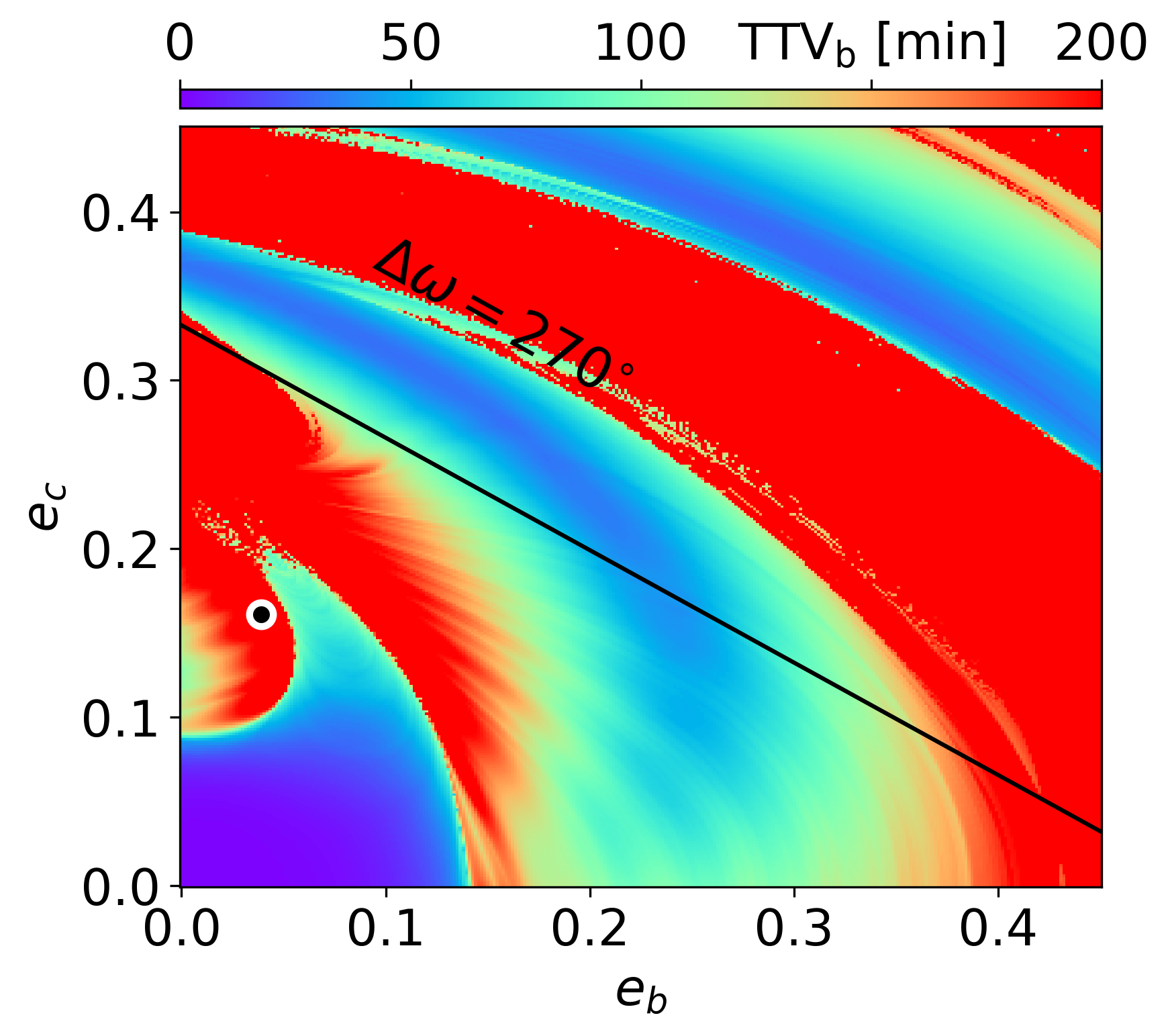}
    \caption{TTV amplitudes in the ($e_b$, $e_c$)-plane and fixed initial values of the argument of pericentre \targetb~($\Delta \varpi =\omega_c-\omega_b$), computed for the RV data time span. The black filled circle with a white rim shows the location of the solution obtained with a model where the eccentricities are set as free parameters.
    \label{figure-TOI-2427-TTV}}
\end{figure}

The results are shown in Fig.~\ref{figure-TOI-2427-dynamical_map}. Subsequent panels  are labeled with  $\Delta\varpi$, which is fixed in the given simulation. A black  curve marks the collision curve of the orbits.  The orbits can mutually cross already for moderate eccentricities and long-term stability of the  system is possible only in some regions of the $(e_{\rm b}, e_{\rm c})$-plane. In all  cases, the nominal system with $e_{\rm c} \simeq 0.16$  is close to the very edge of collision  zone and becomes unstable.  Furthermore, it is clear that the size of this zone and its shape depend strongly on  the initial relative orientation ($\Delta\varpi$) of the orbits.

\begin{figure}[!h]
\centering
\includegraphics[width=\linewidth]{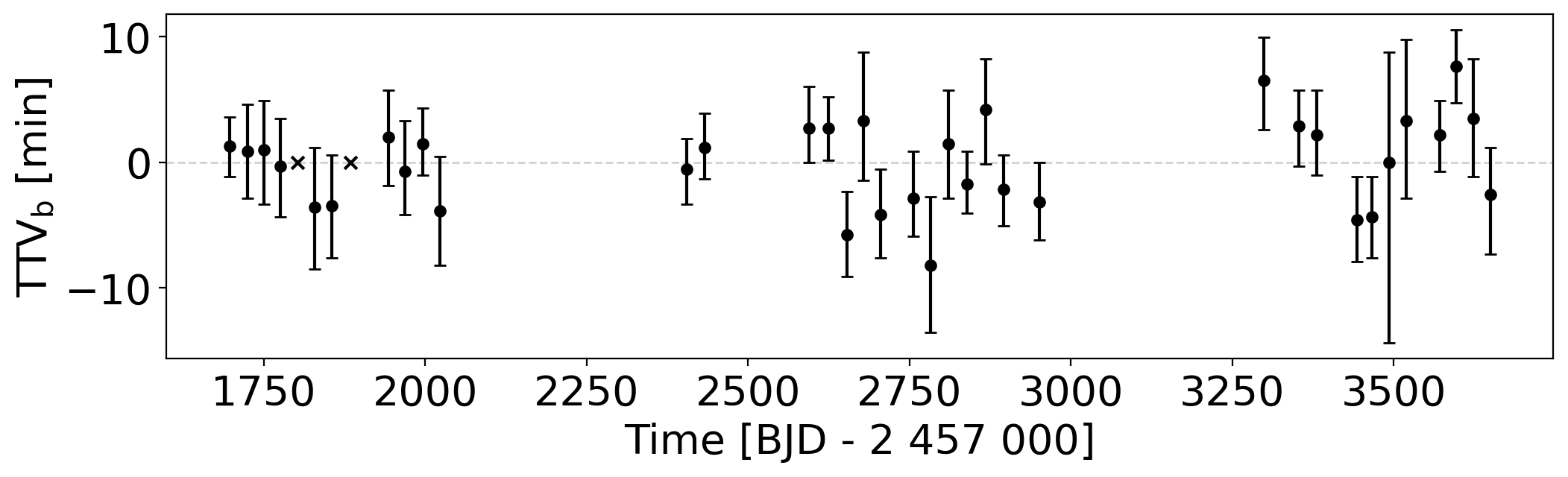}
    \caption{
    TTV measurements based on the TESS photometric data (and spanning all but two sectors) skipped due to excessive noise. 
    \label{figure-TOI-ttv}}
\end{figure}

We can also show that the REM signature of stability is closely reflected by the  geometric evolution of the orbits. This is illustrated with some examples in  Fig.~\ref{figure-TOI-2427-ebec}. The top panel is for a low eccentricity solution,  which is in the stability zone. The middle panel is for moderate eccentricities 
$e_{\rm b} = 0.1$, $e_{\rm c} = 0.2$ and $\Delta\varpi=90^\circ$ (near the origin),  the system is still strictly stable. The last panel shows a chaotic solution where  the eccentricity reaches the collision zone, such a system is unlikely to survive  long-term evolution.

The system does not seem to be involved in a strong, low-order resonance, as we have  verified with additional  direct integrations. Then the likely origin by convergent migration and the  final proximity of the planets to the star would also be consistent with low  eccentricities due to tidal damping. 

To verify this hypothesis more deeply on dynamical grounds, we simulated the range of  TTVs  for the inner pair (Fig.~\ref{figure-TOI-2427-TTV}), setting the same ($e_{\rm b}, e_{\rm c}$) plane and other parameters as in the dynamical maps in Fig.~\ref{figure-TOI-2427-dynamical_map}. Obviously, for eccentric systems, the TTV range could be as large as $\simeq 3$ hours and easily detectable in the available set of light curves.  However, this is not the case. Using {\tt{PyORBIT}} \citep{2016A&A...588A.118M, 2018AJ....155..107M} we have measured the TTVs for all light curves in all available TESS sectors (Fig.~\ref{figure-TOI-ttv}). In top of the strong noise, the measured TTV amplitude remains at the level of a few minutes with a large spread, making the detection of TTVs useful in orbital fitting practically impossible, despite (possibly) being a prior for dynamical RV modelling. 

Curiously, the simulated TTVs closely follow the structures in the REM map. In one case (top left panel, for $\Delta\varpi=0^{\circ}$) there is a diagonal band of small TTVs across all the $e_{\rm b} \simeq e_{\rm c}$.  However, in this region, the systems are stable only for small and moderate eccentricities, below the collision curve, hence only a corresponding triangular area in the TTV band is dynamically permitted. We illustrate this further with the TTV contours over-plotted on the REM maps (Fig.~\ref{figure-TOI-2427-dynamical_map}) to guide the eye. The correspondence of the two dynamical characteristics is striking  and for any tested $\Delta\varpi$, they overlap only in the region of small eccentricities. 

Moreover, in such a case, the Keplerian RV  model applied here remains valid, since the differences in the RV signals between  Keplerian and N-body models reach 1~m\,s$^{-1}$ after five~years of observations. Overall, our experiments support low-eccentricity orbits in the lower bound of $e_{\rm c}$, confirming the results of the analysis described in Sect.~\ref{Section: transit and RV modelling}.

Given the complex and active system described above, a detailed dynamical analysis of its state is beyond  the scope of this present paper. New RV observations would be crucial to constrain the inner eccentricities  and orbital elements of the outermost Jovian companion (especially its period) and allow us to infer hypothetical  low-mass planets in the apparently empty 
zone between the inner subsystem and this giant planet.

     \begin{figure*}[!ht]
     \centering
     \includegraphics[scale=0.5]{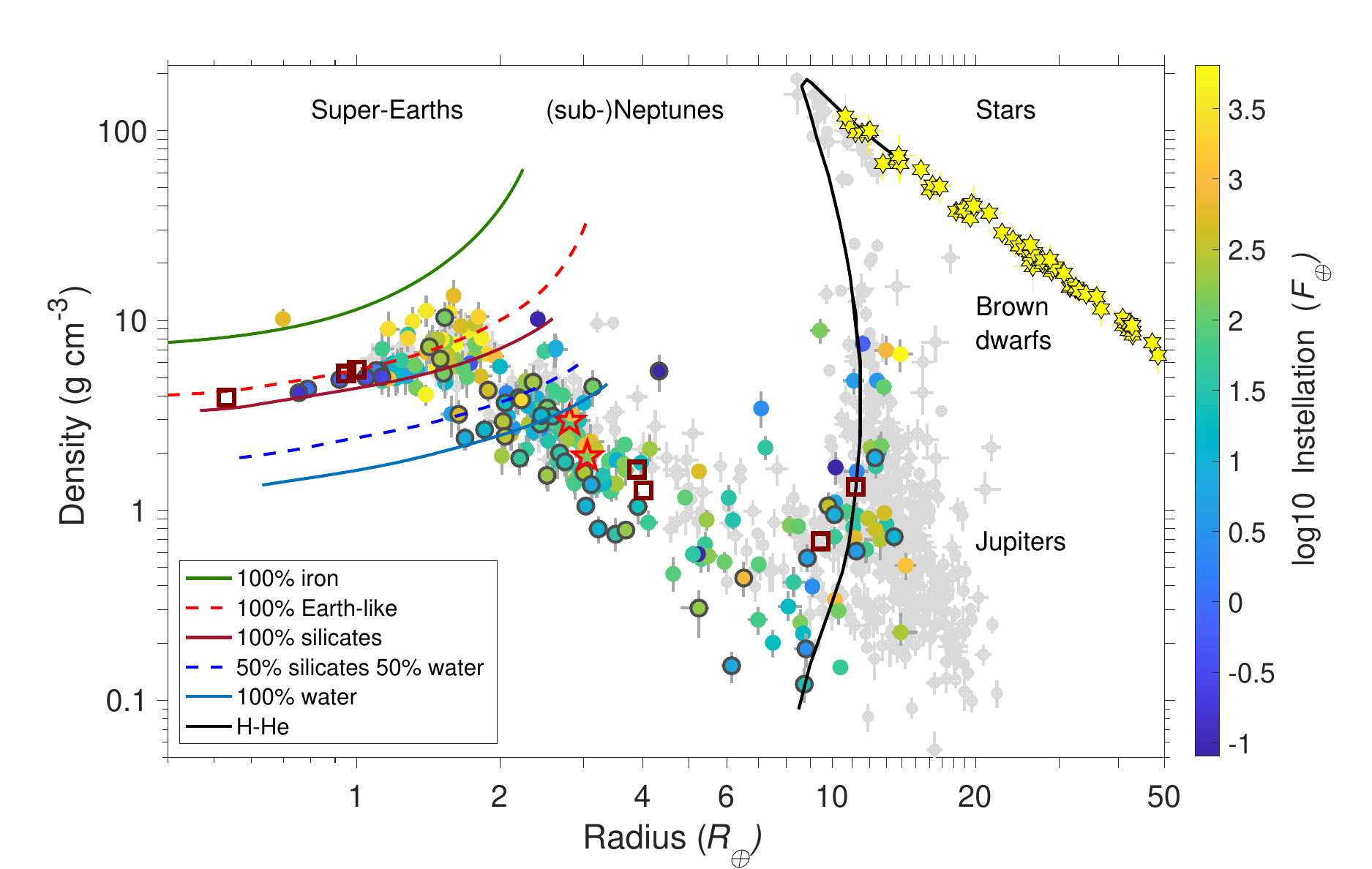}  
   \caption{Density vs radius diagram of the 230 planets in 159 multiplanet systems with $\geq 2$ planets from the NASA exoplanet archive with uncertainties better than 21~\% and 7~\% in mass and radius, respectively,  color-coded with instellation. The planets plotted in grey are single planet systems.  
   The masses are derived from RV measurements (181~planets) or from TTVs (49~planets outlined in dark grey),  and the radii from transit photometry.  Solar System planets are marked with brown squares. Low-mass stars are marked with yellow star symbols. \targeta~b and c are marked with red star symbols and fall within the diagonal strip of (sub-)Neptunes which displays decreasing density with increasing radius. The interior models for low mass-planets are from \citet{2019PNAS..116.9723Z}, and the solid   black line   shows the interior model of H/He dominated giant objects  with  \mbox{$Z = 0.02$}, 
\mbox{age = 5~Gyr}, and  no irradiation \citep{2003A&A...402..701B, 2008A&A...482..315B}.}
      \label{Fig: rho vs. radius} 
 \end{figure*}

\section{Discussion} \label{Section: Discussion}

\subsection{Interior structure models for planets b and c}
Figure~\ref{Fig: rho vs. radius} 
plots all planets from the NASA exoplanet archive with uncertainties better than 21~\% and 7~\% in mass and radius, respectively, as of 12~April 2025. The uncertainties are chosen to have the same impact on the bulk density.  
Also shown are brown dwarfs   \citep[Table~A.1 in][and references therein]{2025arXiv250219940B}, and eclipsing low-mass stars 
\citep[][and references therein]{2003A&A...398..239R, 2005A&A...431.1105B, 2005A&A...438.1123P, 2006A&A...447.1035P, 2009A&A...505..205D, 2013A&A...553A..30T, 2014MNRAS.437.2831Z, 2014A&A...572A.109D, 2016MNRAS.462..554C, 2017ApJ...849...11G, 2017A&A...604L...6V, 2017ApJ...847L..18S, 2018AJ....156...27C, 2019AJ....158...38C, 2021A&A...652A.127G,  2025arXiv250109795V, 2025arXiv250219940B}. 
\targetb~ and c nicely  fall within the diagonal strip of (sub-)Neptunes which displays decreasing density with increasing radius. 
The derived masses in this plot are mainly from RVs, but also from TTVs outlined in dark grey. Sub-Neptunes characterised by
TTVs   are puffier with lower densities compared to the bulk of the RV-characterised population of sub-Neptunes 
 clearly seen in Fig.~\ref{Fig: rho vs. radius}. This suggests that resonant planets have retained their lower initial density 
as compared to non-resonant sub-Neptunes  \citep{2024A&A...687L...1L}.  
The radius gap that separates rocky super-Earths with  volatile sub-Neptunes \citep{2017AJ....154..109F, 2018AJ....156..264F} is seen   between the 
water-rich and silicate interior composition models.   

To determine the planetary structure and composition of \targetb~and c, we performed an interior retrieval using the mass-radius table from \cite{Agui2021} as our initial approach. This table was calculated by assuming a three-layered structure: a Fe-rich core, a silicate mantle, and an irradiated water envelope. This interior model assumes hydrostatic equilibrium, conservation of mass, Gauss' theorem for the computation of gravity, and convection as the main heat transport mechanism. The density calculations use the Vinet equations of state (EOS) for iron and rock \citep[details in][]{Brugger16,Brugger17} and \cite{Mazevet19} for water under conditions exceeding the critical point. An atmospheric model provides the boundary conditions for pressure and temperature, adopting a wet/dry adiabatic profile near the surface and an isothermal mesosphere at low pressures. The atmosphere is heated by the stellar irradiation from the top isotropically, and internal heat sources coming from the interior are neglected.

\begin{table*}[]
\centering
\caption{Best fit compositional parameters, planetary masses, and equilibrium temperatures derived from our MCMC interior structure analysis. }
\label{tab:interior_retrieval_summary}
\begin{tabular}{cccccccc}
\hline \hline
              &                             & \multicolumn{2}{c}{Planet b} & \multicolumn{2}{c}{Planet c} \\
Parameter     & Description                 & Priors     &  Posteriors  & Priors     &  Posteriors   \\ \hline
CMF/(CMF+MMF) & Fe-to-refractory mass ratio &  $\mathcal{U}$[0, 1]           &    0.41$^{+0.32}_{-0.29}$       &   $\mathcal{U}$[0, 1]         &    0.41$^{+0.32}_{-0.29}$                \\
$M_{\rm p} $        & Planet mass [$M_{\oplus}$]                &      $\mathcal{N}$[9.4, 1.8]          &  9.7$\pm$1.7           &  $\mathcal{N}$[10.6, 2.1]          &    10.6$\pm$ 2.1                \\
WMF           & Water mass fraction         &   $\mathcal{U}$[0, 1]             &   0.82$^{+0.12}_{-0.15}$         &  $\mathcal{U}$[0, 1]          &  0.60$^{+0.15}_{-0.16}$                   \\
$T_{\rm eq}$          & Equilibrium temperature [K]     &   $\mathcal{N}$[975, 11]           &    975 $\pm$ 10   &     $\mathcal{N}$[794, 9]          &      794 $\pm$ 9        \\ \hline \hline
\end{tabular}
\tablefoot{The posterior columns represent the mean and uncertainties from the posterior distributions.} 
\end{table*}

\begin{figure*}[!ht]
\centering
\includegraphics[scale=0.5]{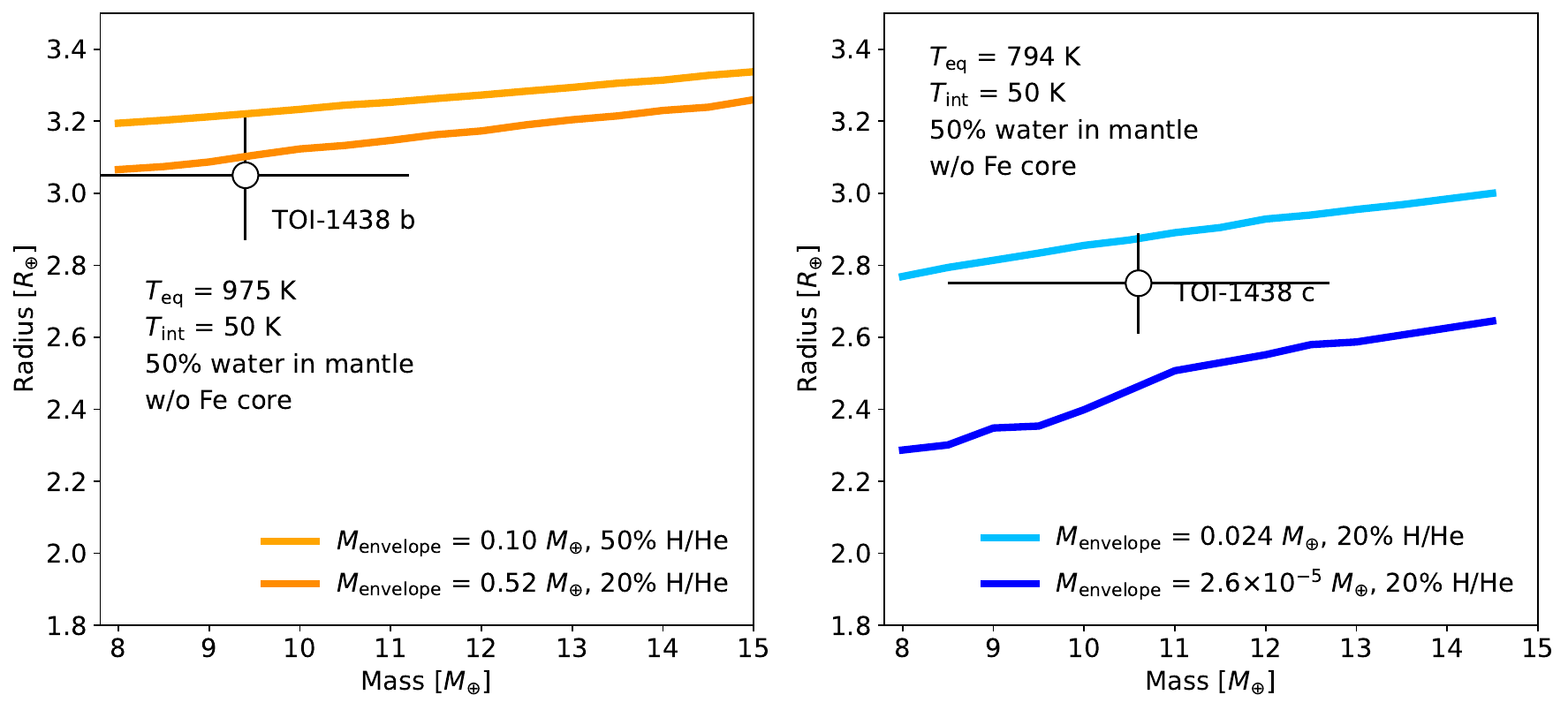} 
\caption{Mass-radius relationships for \targetb~(left panel) and c (right panel). The GASTLI interior structure models are flexible enough to incorporate an inner deep layer of equal parts rock and water (50~\% each), along with a H/He envelope containing variable metal content. We adopt an intrinsic temperature \mbox{$T_\mathrm{int}$ = 50}~K (see text) across all models. The observed masses and radii of \mbox{\targetb}~and c support multiple possible interior configurations, from planets with Fe-rich cores and isolated irradiated hydrospheres to those with mixed rock-water interiors covered by envelopes of varying H/He and metal proportions.}
      \label{fig:gastli_models} 
\end{figure*} 

The forward model incorporates four input parameters defined with their priors 
in Table \ref{tab:interior_retrieval_summary}. 
The Fe-to-refractory mass ratio, our first parameter, represents the 
core mass fraction (CMF) divided by the combined mass 
of the core and mantle (the sum of CMF and mantle mass fraction, MMF). 
The mantle and core together form the planet's 
refractory component (rocks and Fe), while volatiles exist separately in the envelope. 
For reference, the Earth presents a 
\mbox{CMF = 0.32}, and MMF + CMF $\approx 1$, yielding a Fe-to-refractory 
mass ratio of 0.32. Since water comprises the 
entire envelope, its mass fraction equals the water mass fraction (WMF).

We sampled the posterior using {\tt{emcee}} and computed the 
likelihood function  using the squared residuals of the 
observed mass and radius, according to Eq.~6 and 14 in 
\cite{Dorn15} and \cite{Acuna21}, respectively.  
For the MCMC retrieval, we used 32 walkers and $N = 10^{5}$ steps. 
The retrieval's autocorrelation 
time was $\tau = $ 70 and and 66 for planets b and c, respectively. The MCMC convergence criterion 
was \mbox{$\tau \ll N/50$}. In our case $N/50 = 2000$, showing that the chains are long enough to 
ensure convergence with the estimated autocorrelation times. We also ran retrievals assuming 
the mass values from different eccentricity models. The difference in WMF between these is less 
than 5 wt~\%, demonstrating that our conclusions on the interior composition of planets~b and c 
are robust against differences in eccentricities.

The posterior distribution functions of our MCMC retrievals are shown in Fig.~\ref{fig:interior_cornerplots}.  
Table \ref{tab:interior_retrieval_summary} shows the resulting mean and uncertainties of the 
posterior distribution functions. 
We observe that the retrieved Fe-to-refractory mass ratio spans CMF/(CMF + MMF) = [0.13, 0.73]. 
This range is wider than that reported in the super-Earth and hot rocky 
exoplanet population \citep{Schulze2021, Liu_Ni, brinkman24}, 
which is CMF/(CMF + MMF) $\sim$ [0.20,0.46] \citep{Plotnykov20}. 
Using only mass and radius measurements, 
we cannot effectively constrain the Fe-to-refractory ratio for sub-Neptunes.  
This is due to a change in CMF having a very small effect on radius compared 
to the volatile mass fraction, H/He and/or water, in sub-Neptunes and exoplanets 
with significant envelopes \citep[see][for a detailed discussion on this degeneracy]{Otegi20,acuna_thesis2022}.

We obtained well-defined 1$\sigma$ estimates  for  WMFs of  [0.67, 0.94] 
and \mbox{[0.44, 0.75]} for planets~b and c, respectively. We report a mean 
WMF for planet~b of $\approx20$ wt~\% higher than for planet~c, although their WMFs are 
consistent within $1.2~\sigma$. Given the current uncertainties, these planets might have 
comparable water mass fractions, or planet~b may be marginally more water-rich. 
Distinguishing between these scenarios would require improving the planetary radii 
precision from 6~\% to 2~\%, which would reduce WMF uncertainties from 18~wt~\% to 9~wt~\% 
which is a twofold improvement in precision. While JWST offers sufficient precision for such 
radius measurements, the TSMs of both planets (Table~\ref{Table: Orbital and planetary parameters}) 
are below the recommended cut-off for sub-Neptunes \citep[\mbox{TSM $>90$}, ][]{Kempton18}. 
Nonetheless, the atmospheric characterisation of ~\targetb~(\mbox{TSM = 64}), 
particularly the detection of H$_{2}$O and CH$_{4}$, may still be possible with a few transit 
observations, as suggested for planets with
similar TSMs \citep{Chaturvedi22}.

We used mass-radius curves by an interior structure model that makes two assumptions: 
volatile species have water-like densities and occupy an isolated envelope layer distinct 
from the mantle and core. However, actual planetary structures may be more complex. 
Recent JWST transmission spectra of sub-Neptunes reveal envelopes containing mixtures 
of H/He and high-molecular-weight species (H$_{2}$O, CO$_{2}$, CO, CH$_{4}$, NH$_{3}$) 
at envelope metal mass fractions $\sim50$~\% \citep{benneke_toi270d,Holmberg,PG_water_prog}. 
In addition,  rock mantles may be soluble and miscible with water and other volatiles \citep{Kite_fugacity,Vazan22,schlichting,Luo_Dorn_wet_rock}.

To demonstrate the degenerate nature of \targetb~and c's interiors as sub-Neptunes, 
we employed the open-source Python package GASTLI \citep[GAS gianT modeL for Interiors,][]{Acuna21,gastli} to generate forward interior structure models in Fig.~\ref{fig:gastli_models}. 
GASTLI models planetary structure using two distinct layers: a deep inner layer 
containing equal parts rocks and water by mass, 
and an outer envelope composed of H/He and water serving as a proxy for metals. 
GASTLI uses state-of-the-art EOS for rock, water and H/He. 
We couple it to its default atmospheric grid to calculate the envelope boundary conditions. 
This grid contains pressure-temperature profiles for warm ($T_\mathrm{eq}<1000$~K) 
volatile-rich exoplanets generated 
with the self-consistent radiative-convective model petitCODE \citep[details in][]{gastli}.

Given  the wide range of possible  ages within $1\,\sigma$  for   
\targeta~(Sect.~\ref{Subsection: Stellar modelling}) and that 
sub-Neptunes cool faster than their gas giant counterparts \citep{ChenRogers16}, 
we adopted an intrinsic temperature of $T_\mathrm{int} = 50$~K. Fig. \ref{fig:gastli_models} 
shows that both planets could have a mixed interior of volatiles and rock, 
overlaid by a metal-rich, H/He envelope. In particular, planet b could have an envelope as 
 massive as 0.5~$M_{\oplus}$, representing 5~\% of the planet's total mass. 
Its composition ranges from an equal mix of H/He and metals to a high mean molecular 
weight atmosphere composed purely of metals. 
In contrast, planet c's envelope is significantly less massive, $< 1$~\% of its total mass, 
with a minimum envelope metal content of 80~\% by mass. 
External heat sources, such as tidal heating from non-zero eccentricity \citep{Agundez14}, 
would produce effects similar to increasing the H/He fraction in the envelope. 
Consequently, if internal heat sources are present, envelopes containing less than 20~\% 
H/He might be necessary to explain the observed masses and radii of both planets.

\subsection{Signal d} \label{Subsection: signal d}
\begin{figure}
    \centering
    \includegraphics[width=\linewidth]{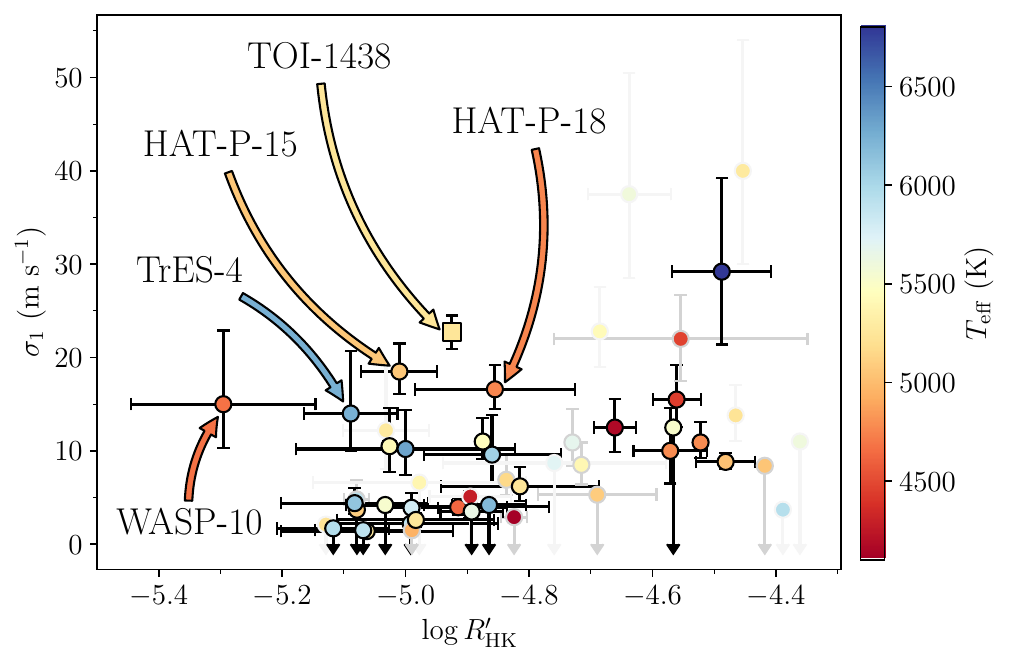}
    \caption{RV jitter ($\sigma_1$) as derived in \citet{Bonomo2017} as a function of $\log R^\prime_{\rm HK}$ activity index calculated in \citet{Claudi2024} for exoplanet host stars with $M_\star<1.1$~\Msun. Light-grey error bars denote systems with a baseline shorter than 1~yr, grey error bars are systems with a baseline  $1-2$~yr, and black error bars are systems with   baselines $>2$~yr. The error bars with arrows are upper limits on the RV jitter. The markers are colour coded with the stellar $T_{\rm eff}$. \targeta~is marked with a square. The marked systems HAT-P-15, HAT-P-19, TrES-4, and WASP-10 are discussed in Sect.~\ref{Subsection: system architecture}.
        Adapted from Fig.~3 in  \citet{Hekker2006}.}
    \label{fig:jitter}
\end{figure}

If signal~d is shown to be due to stellar activity, 
a striking feature of this signal would be the large amplitude 
of \kd~m~s$^{-1}$ (or potentially higher, as explained in Sect.~\ref{Section: transit and RV modelling}), 
especially considering that the star is not very active
($\log R^\prime_{\rm HK}=-4.925\pm0.013$). 
To investigate if an amplitude of this magnitude is typical for stars 
with this activity level, 
we cross-matched the exoplanet hosts $\log R^\prime_{\rm HK}$ values derived in 
\citet{Claudi2024} with the jitter ($\sigma_1$) values from \citet{Bonomo2017},
both based on observations with HARPS-N. 
In Fig.~\ref{fig:jitter}, we show stars appearing
in both aforementioned studies with $M_\star<1.1$~\Msun~and at least ten~RV 
observations. The light-grey error bars are systems with a baseline shorter than 1~yr, 
 the black ones have a baseline longer than 2~yr,  
and the grey ones have baselines in between.
To set  \targeta~into this context, we fitted the RVs assuming a two-planet model (planets~b and c) 
and  let the jitter term absorb the long-period signal.  
The result was $\sigma_1=22.7\pm1.8$~m~s$^{-1}$,   a rather large value among these stars as is evident from Fig.~\ref{fig:jitter}.

One aspect this plot obviously fails to convey is the complexity of the signal(s)
responsible for the jitter.  
Signal~d in \targeta~has a low harmonic complexity 
and can be nicely modelled as a simple Keplerian orbit. 
The figure also does not provide any information on the potential periodicity 
of the phenomenon.
We therefore took a closer look at four of the systems closest to \targeta~in $\log R^\prime_{\rm HK} - \sigma_1$ space, 
namely, \mbox{HAT-P-15} \citep[\mbox{$T_{\rm eff}=5568\pm90$~K}, $\log g=4.38\pm0.03$;][]{Kovacs2010}, 
\mbox{HAT-P-18} \citep[\mbox{$T_{\rm eff}=4750\pm100$~K}, $\log g=4.50\pm0.25$;][]{Hartman2011}, 
\mbox{WASP-10} \citep[$T_{\rm eff}=4675\pm100$~K, $\log g=4.40\pm0.20$;][]{Christian2009}, 
and \mbox{TrES-4} \citep[$T_{\rm eff}=6100\pm150$~K, $\log g=4.045\pm0.034$;][]{Mandushev2007}.
All four targets have been monitored 
with HARPS-N and/or HIRES and have baselines well over 1500~d \citep{2014ApJ...785..126K, Bonomo2017}; 
however, the sampling for \targeta~is much better 
(105~RVs compared to at most 48 for HAT-P-18).
We collected the RVs from these two studies, along with RVs for WASP-10
obtained with 
the SOPHIE and FIES spectrographs and looked at the residuals after 
subtracting the best-fitting models.
Figure~\ref{Figure: residuals} shows the residuals and their associated periodograms.

The morphology of the residuals from these four targets appear quite different
from that of  
\targeta,~as they do not seem to display any coherent signal on long timescales.
Indeed, from calculating the GLS, the most prominent peaks had periodicities of
165~d or less.
Moreover, the peaks in the periodograms do not have a ${\rm FAP}<0.1$~\%;  therefore, they are not considered to be significant.
We further note that for WASP-10, the jitter stemming from HIRES 
was significantly lower than that coming from SOPHIE and FIES. 
This is also quite apparent by eye and from the root-mean-square
(RMS$_{\rm HIRES}=4.26$~m~s$^{-1}$, 
RMS$_{\rm SOPHIE}=29.39$~m~s$^{-1}$,
RMS$_{\rm FIES}=47.73$~m~s$^{-1}$). 
The large value reported for the jitter 
for this system might therefore be explained (at least partly) by
instrumental noise.
Taken together, we believe that the source of the jitter in \targeta~is inherently different in both 
morphology and periodicity compared to the other four systems. Overall, this supports a planetary origin of signal~d.
However, to confidently confirm the existence of  planet~d, we need more data over a longer baseline to  resolve the long-period signals of stellar activity.

\subsection{System architecture} \label{Subsection: system architecture}
\begin{figure}
    \centering
    \includegraphics[width=\linewidth]{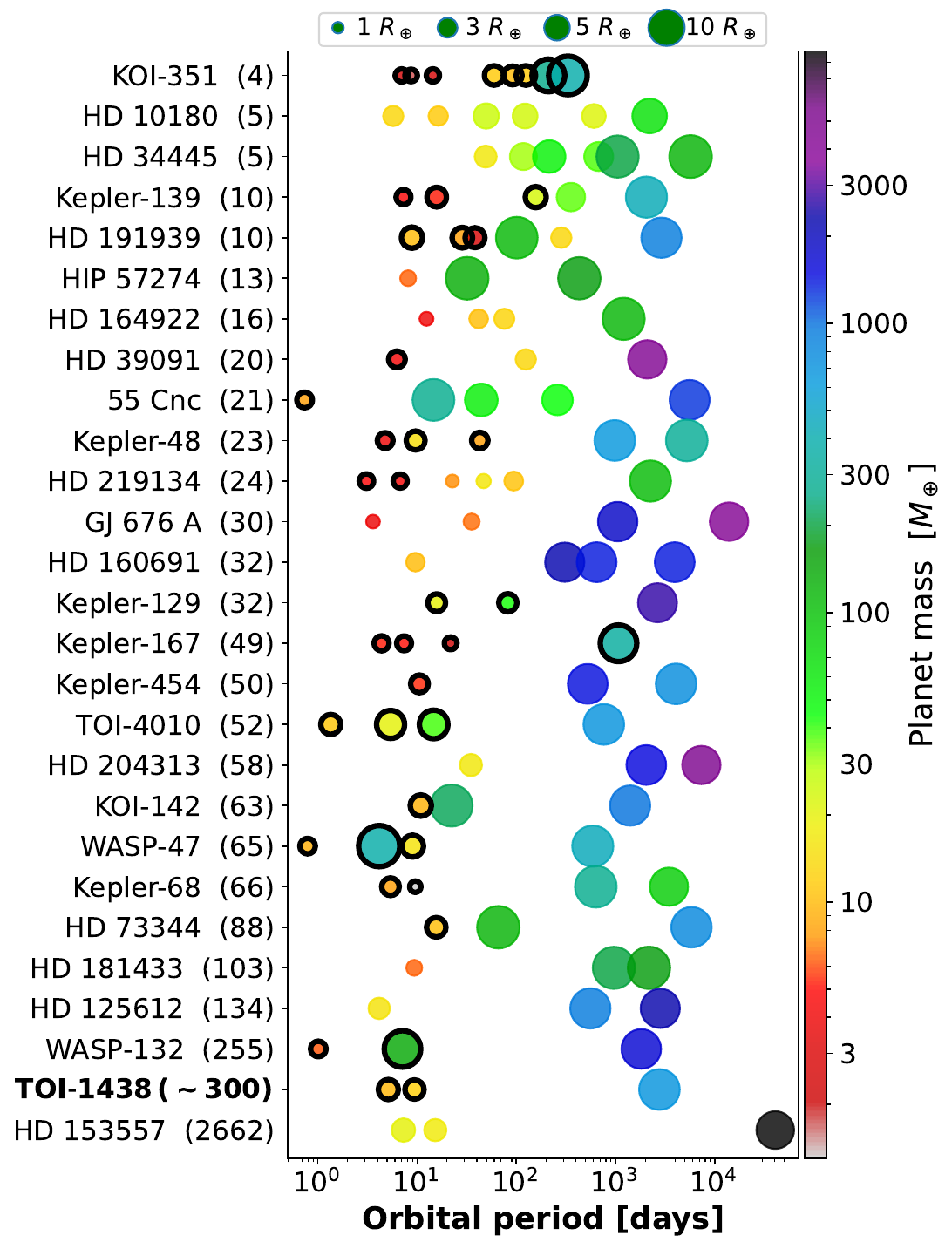} 
\caption{Architectures of all   27 detected systems with minimum three planets and at least one inner small planet and one outer giant, including the \targeta~system assuming a planetary nature of signal~d.
The largest orbital period ratio between two adjacent planets in each system increases from top to bottom and is 
written in parentheses after the name of the host star.  
The sizes and colors 
of the circles indicate the radii and masses of the planets, respectively. Planets with transit measurements are outlined in black.
All   planets have RV measurements except for HD~73344~b, \emph{Kepler}-167~b, c, and d, as well as the eight planets orbiting KOI-351.}
\label{fig:architecture}
\end{figure}

The architecture of the \targeta~system is   composed of an inner system with two sub-Neptunes 
and likely an outer system  harbouring a giant planet. 
If a three-planet system,   \targeta~belongs to a small group of detected systems with inner, 
small planets and outer long-period giant planets; for instance, \mbox{Kepler-48} 
\citep{steffen2013, marcy2014} and \mbox{TOI-4010} \citep{kunimoto2023}.
To compare the  architecture of \targeta~to all the detected  systems  so far,  
we selected systems from the 
NASA Exoplanet Archive  that contain a 
minimum of three planets, with at least one confirmed small \mbox{($R< 4$~\rearth)} or low-mass planet 
\mbox{($M< 20$~\mearth)} with an orbital \mbox{period < 10 days} and an outer confirmed  giant planet 
\mbox{($R> 8$~\rearth}~or $M> 90$~\mearth) with \mbox{$P$ > 300 days}. 
We applied the mass-radius relationships from \citet{muller} to convert the measured masses to radii 
and vice versa for the planets that did not have both radius and mass values.
There are only 26~systems that fulfill these criteria as of 12~April 2025, most of 
which  do not display a large gap 
between the outer giant and its inner adjacent planet similar to  \targeta, as
visualised in  Fig.~\ref{fig:architecture}.  
The largest  orbital period ratio  between two adjacent planets in each system is given 
in parentheses after each 
host star's name and increases from top to bottom in the figure.
It is clear that the \targeta~system has a unique architecture with two small, low-mass, and 
tightly packed planets orbiting close to the host star and a Jupiter-like planet residing 
in the outer region of the system  
separated from planet~c by an orbital period ratio of $\sim 300$. Only one system, HD\,153557 \citep{feng2022}, 
 has a larger orbital period ratio  between the inner and outer system; however,
 the outer object  is a 
brown dwarf, or possibly a low-mass star, with a minimum mass of 27~\mjup.

Previous studies have shown that inner small planets in observed multi-planetary systems 
with outer giants tend to have more irregular orbital spacings than the planets in systems without 
any long-period giants \citep{2023AJ....166...36H, 2024A&A...692A.122M}. 
This could suggest   that \targeta~may host at least one (or even several)  
undetected planets  adjacent to planet~c with a different orbital spacing  compared to the planet pair~b and c. 
For example,  if  a planet  were to hide in the \targeta~system with   an orbital period of approximately 
150~days, we could place an upper limit 
on its mass of $29\pm8$~\mearth, assuming an orbit that is circular and coplanar with the two transiting planets.  
Continued RV observations would provide even stronger constraints on, or discovery of, potentially hidden planets.

While the number of known systems with both small (super-Earths or sub-Neptunes) inner planets and 
outer gas giants is relatively small, owing to the relative difficulty of detecting the latter, several 
studies suggest that these two populations are correlated. \cite{Uehara2016} estimated that 20~\% 
of the systems with inner planets host outer giants, based on mono-transits in \emph{Kepler} light curves, 
while \cite{Bryan2019} estimated a rate of 39~\% based on published RV datasets for a heterogeneous 
sample of inner systems detected either in transit or in RV. However, with dedicated follow-up of 
38~\emph{Kepler}/K2 systems, \cite{Bonomo2023} found a lower rate of 9~\% for the conditional 
occurrence of outer giants given the presence of small inner planets,   consistent with the 
overall population of such giants. \cite{Bryan2024} found a strong metallicity dependency of the 
conditional occurrence rate, and \cite{Zhu2024} argued that this accounted for the \cite{Bonomo2023} 
finding as a manifestation of Simpson's paradox. Nevertheless, with a smaller, but more homogeneous 
sample than \cite{Bryan2024}, \cite{vanZandt2024} did not recover the metallicity effect, although 
\cite{2025arXiv250106342V} have recovered the inner small--outer giant correlation. Overall, larger 
but more inhomogeneous samples seem to indeed suggest a correlation between outer giants and
 inner small planets. This is, however, not always seen in smaller more controlled samples, although these 
 may lack the statistical power to identify this correlation significantly. In light of this, more such systems, 
 and their adequate characterisation, are clearly valuable.

\subsection{Formation  of systems with small inner planets and outer giants}
Both sub-Neptunes and Jovian planets are thought to mainly 
form via core accretion, where envelope accretion follows the accumulation of a solid core of 
$\sim10$~\mearth~ \citep{Pollack1996}. 
Variants are pebble accretion and planetesimal accretion, 
depending on the size of the dominant source of accreted solids. Modern versions of these models 
suggest that any correlation between close-in planets and outer gas giants arises because of the 
disc-wide dependence of planet formation arising from the global budget of solids in the disc. From planetesimal 
accretion models, \cite{Schlecker2021} found that co-existence of outer giant planets and smaller inner 
planets occurs preferentially for intermediate values of the quantity of solids: too little, and cores cannot 
grow large enough to begin gas accretion; too much, and multiple gas giants form, which typically 
destabilise the inner planets during instabilities \citep[see also][]{Mustill2017,Pu2021}. In models of 
pebble accretion, \citet{Bitsch2023} found frequent co-existence of inner and outer planets if the envelope 
contraction is not overly efficient (i.e. avoiding growth of the inner sub-Neptunes to full gas giants). However, again, 
where multiple outer planets form, the instabilities among them can often reduce the multiplicity of the inner system
 \citep{Bitsch2020}. In the case of \targeta, the large separation between the dynamically cold inner 
 sub-Neptunes and the outer giant may have helped preserve them from any instability (should one have
 occurred). The low (albeit non-zero) eccentricity of the outer planet might suggest a mild instability, with relatively 
 little effect on the inner system \citep{Pu2021}; this could indicate, for instance, the prior ejection of a relatively low-mass 
 outer giant planet \citep{Kokaia2020,Pu2021}.

\section{Conclusions}\label{Section: Conclusions}
In this paper, we present  the detection and   characterisation of  the  \mbox{\targeta}~system with two short-orbital period sub-Neptunes 
and a tentative  Jupiter-like planet with an orbital period of \Pdyear~yr. If the latter is indeed a planet,  it
has a minimum mass of  \mbox{\mpdjup~\mjup}. 
Overall, given the constraints on the orbital model parameters that make it possible 
to construct a global view of the dynamics, our stability analysis  combined with the 
measured small TTV amplitudes and their predicted values in the parameter space   
favours an orbital model with nearly circular orbits of the inner subsystem. 
Long-period magnetic stellar activity seems to be present at some level, as indicated by several activity indicators. Thus, we cannot confirm the existence of planet~d which remains a tentative discovery. 
More data are required to fully cover the entire orbital period of signal~d and to resolve the long-period stellar activity signals. 
If it is confirmed as a triple-planet system, the architecture of \targeta~would exhibit one of the largest gaps between  inner small planets and outer giant planets of all known systems so far. This particular aspect could point to further, undetected planets. 

In our interior structure analysis, we considered both a water world structure with an iron core and a silicate-rich mantle, as well as a range of sub-Neptune structures with varying H/He mass fractions and mixed rock-water cores. \targetb~ and c present volatile-rich envelopes, which are a mixture of H/He and water. This is due to the fact that pure water envelopes require very high envelope mass fractions to explain their low densities ($>$ 50~\%). Planets b and c could contain up to 2.5~\% and 0.2~\% of H/He in mass, respectively. The compositions of their deep cores could not be constrained from mass and radius data alone due to degeneracies, which may range from cores constituted by rock and Fe to silicate melts with miscible water and other volatiles.

 \begin{acknowledgements}
 
 This paper includes data collected by the TESS mission. Funding for the TESS mission is provided by the NASA Explorer Program. We acknowledge the use of public TESS data from pipelines at the TESS Science Office and at the TESS Science Processing Operations Center. 
 Funding for the TESS mission is provided by NASA's Science Mission Directorate. 
 Resources supporting this work were provided by the NASA High-End Computing (HEC) Program through the NASA Advanced Supercomputing (NAS) Division at Ames Research Center for the production of the SPOC data products. 
  We also acknowledge the use of TESS High-Level Science Products (HLSP) produced by the Quick-Look Pipeline (QLP) at the TESS Science Office at MIT, which are publicly available from the Mikulski Archive for Space Telescopes (MAST) at the Space Telescope Science Institute (STScI). STScI is operated by the Association of Universities for Research in Astronomy, Inc., under NASA contract NAS 5–26555. 
 This research has made use of the Exoplanet Follow-up Observation Program (ExoFOP; DOI: 10.26134/ExoFOP5) website, which is operated by the California Institute of Technology, under contract with the National Aeronautics and Space Administration under the Exoplanet Exploration Program. 
 Some of the observations in this paper made use of the High-Resolution Imaging instrument ‘Alopeke and were obtained under Gemini LLP Proposal Number: GN/S-2021A-LP-105. ‘Alopeke was funded by the NASA Exoplanet Exploration Program and built at the NASA Ames Research Center by Steve B. Howell, Nic Scott, Elliott P. Horch, and Emmett Quigley. Alopeke was mounted on the Gemini North telescope of the international Gemini Observatory, a program of NSF’s OIR Lab, which is managed by the Association of Universities for Research in Astronomy (AURA) under a cooperative agreement with the National Science Foundation. on behalf of the Gemini partnership: the National Science Foundation (United States), National Research Council (Canada), Agencia Nacional de Investigación y Desarrollo (Chile), Ministerio de Ciencia, Tecnología e Innovación (Argentina), Ministério da Ciência, Tecnologia, Inovações e Comunicações (Brazil), and Korea Astronomy and Space Science Institute (Republic of Korea).
This work has made use of SME package, which benefits from the continuing development work by J. Valenti and N. Piskunov 
and we gratefully acknowledge their continued support. 
We acknowledge  fruitful discussions with Andreas Quirrenbach on stellar activity and the effect on RVs. 
C.M.P., M.F.  and   A.J.M. acknowledge the support of the  Swedish National Space Agency (DNR 65/19,  177/19, and  2023-00146).  C.M.P., E.K. and A.M. acknowledge the support of GENIE at Chalmers.
Funding for the Stellar Astrophysics Centre is provided by The Danish National Research Foundation (Grant agreement no.: DNRF106).
C.N.W. acknowledge support from the TESS mission via subaward s3449 from MIT. 
T.M. acknowledges support from the Spanish Ministry of Science and Innovation with the 
the proyecto plan nacional \textit{PLAtoSOnG} (grant no. PID2023-146453NB-100). 
R.A.G., D.B.P., and L.B. acknowledge the support from the GOLF and PLATO Centre National D'{\'{E}}tudes Spatiales grants. 
S.M. and H.J.D. acknowledges support by the Spanish Ministry of Science and Innovation with the grants number PID2019-107061GB-C66 and PID2023-149439NB-C41, and through AEI under the Severo Ochoa Centres of Excellence Programme 2020--2023 (CEX2019-000920-S). 
K.G., D.J. and G.N. gratefully acknowledges the Centre of Informatics Tricity Academic Supercomputer and networK (CI TASK, Gda\'nsk, Poland) for computing resources (grant no. PT01187). G.N. thanks for the research funding from the Ministry of Science and Higher Education programme the "Excellence Initiative - Research University" conducted at the Centre of Excellence in Astrophysics and Astrochemistry of the Nicolaus Copernicus University in Toru\'n, Poland. 
G.M. acknowledges financial support from the Severo Ochoa grant
CEX2021-001131-S and from the Ram\' on y Cajal grant RYC2022-037854-I
funded by MCIN/AEI/1144 10.13039/501100011033 and FSE+. 
F.M. acknowledges the financial support from the Agencia Estatal de Investigaci\'{o}n del Ministerio de Ciencia, Innovaci\'{o}n y Universidades (MCIU/AEI) through grant PID2023-152906NA-I00.
E.P. acknowledge financial support from the Agencia Estatal de Investigaci\'on of the Ministerio de Ciencia e Innovaci\'on MCIN/AEI/10.13039/501100011033 and the ERDF “A way of making Europe” through project PID2021-125627OB-C32, and from the Centre of Excellence  “Severo Ochoa”  award to the Instituto de Astrofisica de Canarias.
R.L. acknowledges support for this work by NASA through the NASA Hubble Fellowship grant HST-HF2-51559.001-A awarded by the Space Telescope Science Institute, which is operated by the Association of Universities for Research in Astronomy, Inc., for NASA, under contract NAS5-26555. 
K.W.F.L. acknowledge support by European Union’s Horizon Europe Framework Programme under the Marie Sk\`lodowska-Curie Actions grant agreement No. 101086149 (EXOWORLD). 
 J.K. acknowledge support from the Swiss NCCR PlanetS and the Swiss National Science Foundation. This work has been carried out within the framework of the NCCR PlanetS supported by the Swiss National Science Foundation under grants 51NF40182901 and 51NF40205606. J.K. acknowledges support from the Swedish National Space Agency (SNSA; DNR 2020-00104) and of the Swiss National Science Foundation under grant number TMSGI2\_211697. 
VVE. has been supported by UK's Science \& Technology Facilities Council through the STFC grants ST/W001136/1
and ST/S000216/1. 
  \end{acknowledgements}

\bibliographystyle{aa}
\bibliography{references}

 \begin{appendix}

\section{Radial velocity data}

\begin{table*}
\begin{center}
\caption{Radial velocities and spectral activity indicators measured from TNG/HARPS-N spectra with the DRS.
\label{Table: HARPSN RVs drs}}
\begin{tabular}{cccrccccccc}
\hline
\hline
\noalign{\smallskip}
\multicolumn{1}{c}{BJD$_\mathrm{TBD}$} &
\multicolumn{1}{c}{RV} &
\multicolumn{1}{c}{$\sigma_\mathrm{RV}$} &
\multicolumn{1}{c}{BIS} &
\multicolumn{1}{c}{$\sigma_\mathrm{BIS}$} &
\multicolumn{1}{c}{FWHM} &
\multicolumn{1}{c}{Contrast} &
\multicolumn{1}{c}{$\log{R^\prime_{HK}}$} &
\multicolumn{1}{c}{$\sigma_{\log{R^\prime_{HK}}}$} &
\multicolumn{1}{c}{SNR} &
\multicolumn{1}{c}{$\mathrm{T_{exp}}$}\\ 
\multicolumn{1}{c}{-2457000} &
\multicolumn{1}{c}{($\mathrm{m\,s^{-1}}$)} &
\multicolumn{1}{c}{($\mathrm{m\,s^{-1}}$)} &
\multicolumn{1}{c}{($\mathrm{m\,s^{-1}}$)} &
\multicolumn{1}{c}{($\mathrm{m\,s^{-1}}$)} &
\multicolumn{1}{c}{($\mathrm{km\,s^{-1}}$)} &
\multicolumn{1}{c}{(\%)} &
\multicolumn{1}{c}{---} &
\multicolumn{1}{c}{---} &
\multicolumn{1}{c}{(@550nm)} &
\multicolumn{1}{c}{(s)}\\ 
\noalign{\smallskip}
\hline
\noalign{\smallskip}
     1925.74746 &      -29497.725 &           1.589 &          13.015 &           2.248 &           6.322 &          41.977 &         -4.8790 &          0.0096 &            48.5 &          1800.0\\
     1925.76981 &      -29494.617 &           2.134 &           8.604 &           3.018 &           6.325 &          41.570 &         -4.8767 &          0.0143 &            38.0 &          1800.0\\
     1926.74886 &      -29500.857 &           1.445 &           8.112 &           2.044 &           6.325 &          42.023 &         -4.8789 &          0.0083 &            52.7 &          1800.0\\
     1929.72901 &      -29480.355 &           1.203 &          13.034 &           1.701 &           6.331 &          41.979 &         -4.8585 &          0.0062 &            61.4 &          1800.0\\
     2000.57290 &      -29481.379 &           1.608 &          10.923 &           2.274 &           6.317 &          41.931 &         -4.8616 &          0.0090 &            48.1 &          1800.0\\
     2000.65755 &      -29477.725 &           1.631 &          12.939 &           2.307 &           6.308 &          41.981 &         -4.8632 &          0.0090 &            47.5 &          1500.0\\
     2002.57645 &      -29494.714 &           1.224 &          10.899 &           1.731 &           6.327 &          41.952 &         -4.8769 &          0.0064 &            60.8 &          1800.0\\
     2002.66203 &      -29492.744 &           1.366 &           9.476 &           1.932 &           6.329 &          41.982 &         -4.8759 &          0.0076 &            55.1 &          1500.0\\
     2012.58204 &      -29494.634 &           1.713 &           6.329 &           2.422 &           6.303 &          42.049 &         -4.8992 &          0.0113 &            45.4 &          1800.0\\
     2014.58299 &      -29490.865 &           1.385 &           4.459 &           1.959 &           6.304 &          42.106 &         -4.9019 &          0.0082 &            54.8 &          1800.0\\
     2056.55645 &      -29481.247 &           1.595 &           3.506 &           2.256 &           6.304 &          42.006 &         -4.8839 &          0.0093 &            48.3 &          1700.0\\
     2073.62803 &      -29458.288 &           1.846 &           5.402 &           2.611 &           6.361 &          41.757 &         -4.8491 &          0.0109 &            43.8 &          1700.0\\
     2082.43490 &      -29481.769 &           2.806 &          17.562 &           3.968 &           6.314 &          41.841 &         -4.8589 &          0.0192 &            29.7 &          1700.0\\
     2087.42720 &      -29485.814 &           2.174 &           9.230 &           3.075 &           6.297 &          41.957 &         -4.8776 &          0.0146 &            37.7 &          1800.0\\
     2089.45674 &      -29492.806 &           2.337 &           7.992 &           3.305 &           6.303 &          41.839 &         -4.8704 &          0.0163 &            35.6 &          1800.0\\
     2090.48671 &      -29490.027 &           1.754 &           7.070 &           2.480 &           6.306 &          41.858 &         -4.8997 &          0.0108 &            45.0 &          1800.0\\
     2117.48442 &      -29484.790 &           3.287 &          21.457 &           4.649 &           6.312 &          41.728 &         -4.8784 &          0.0249 &            28.0 &          1700.0\\
     2119.45691 &      -29476.857 &           1.802 &          13.978 &           2.548 &           6.322 &          41.857 &         -4.8761 &          0.0122 &            42.4 &          1700.0\\
     2138.34160 &      -29479.644 &           1.466 &          18.542 &           2.073 &           6.375 &          41.996 &         -4.8497 &          0.0077 &            53.5 &          1800.0\\
     2149.32319 &      -29476.090 &           1.414 &          14.044 &           2.000 &           6.312 &          41.994 &         -4.9083 &          0.0089 &            53.6 &          1800.0\\
     2150.31933 &      -29477.876 &           1.298 &           4.593 &           1.835 &           6.303 &          42.046 &         -4.9070 &          0.0073 &            58.8 &          1800.0\\
     2151.31789 &      -29483.965 &           1.301 &           8.004 &           1.840 &           6.288 &          42.053 &         -4.9102 &          0.0074 &            58.8 &          1800.0\\
     2160.31282 &      -29478.019 &           1.385 &          14.277 &           1.958 &           6.310 &          41.998 &         -4.8869 &          0.0082 &            55.4 &          1800.0\\
     2309.71597 &      -29452.750 &           2.408 &           5.609 &           3.405 &           6.302 &          41.895 &         -4.8939 &          0.0169 &            35.0 &          1800.0\\
     2353.62295 &      -29468.402 &           1.670 &          11.403 &           2.362 &           6.298 &          42.073 &         -4.9091 &          0.0111 &            47.5 &          2400.0\\
     2355.57948 &      -29461.566 &           1.571 &          20.962 &           2.222 &           6.301 &          42.014 &         -4.9149 &          0.0107 &            50.5 &          3300.0\\
     2369.67671 &      -29479.314 &           2.723 &           7.205 &           3.851 &           6.283 &          42.130 &         -4.9448 &          0.0251 &            31.5 &          1800.0\\
     2393.66078 &      -29459.814 &           1.079 &          21.883 &           1.526 &           6.308 &          42.028 &         -4.9123 &          0.0059 &            68.3 &          1800.0\\
     2394.57379 &      -29462.001 &           1.298 &          23.662 &           1.836 &           6.302 &          42.022 &         -4.9082 &          0.0073 &            59.0 &          1800.0\\
     2410.55303 &      -29463.457 &           1.841 &           4.650 &           2.604 &           6.297 &          42.075 &         -4.9114 &          0.0130 &            43.6 &          1800.0\\
     2450.48219 &      -29456.002 &           1.100 &          -0.128 &           1.556 &           6.285 &          42.079 &         -4.9163 &          0.0057 &            68.7 &          1800.0\\
     2676.71340 &      -29442.017 &           1.618 &          11.327 &           2.288 &           6.300 &          42.451 &         -4.9317 &          0.0139 &            52.3 &          1800.0\\
     2677.72392 &      -29443.183 &           1.506 &           7.817 &           2.129 &           6.302 &          42.468 &         -4.9271 &          0.0125 &            54.9 &          1800.0\\
     2702.67857 &      -29438.826 &           1.319 &          12.015 &           1.865 &           6.322 &          42.318 &         -4.9068 &          0.0089 &            63.1 &          1800.0\\
     2703.69387 &      -29435.126 &           1.453 &          16.453 &           2.055 &           6.315 &          42.309 &         -4.9160 &          0.0107 &            57.9 &          1800.0\\
     2720.63146 &      -29441.535 &           4.126 &          -2.316 &           5.835 &           6.288 &          42.301 &         -4.9615 &          0.0477 &            23.9 &          1800.0\\
     2722.64219 &      -29440.295 &           1.322 &           6.234 &           1.870 &           6.296 &          42.510 &         -4.9656 &          0.0100 &            60.9 &          1800.0\\
     2765.55642 &      -29436.920 &           1.154 &           6.621 &           1.632 &           6.314 &          42.415 &         -4.9547 &          0.0079 &            68.2 &          1800.0\\
     2766.55074 &      -29436.916 &           1.433 &           5.804 &           2.027 &           6.313 &          42.413 &         -4.9439 &          0.0110 &            55.5 &          1800.0\\
     2785.57158 &      -29439.898 &           3.300 &          -0.012 &           4.667 &           6.265 &          42.179 &         -4.9384 &          0.0350 &            29.2 &          2100.0\\
     2810.52211 &      -29436.483 &           1.435 &           2.821 &           2.030 &           6.295 &          42.526 &         -4.9811 &          0.0117 &            56.8 &          1800.0\\
     2811.51754 &      -29434.007 &           2.698 &           4.481 &           3.815 &           6.294 &          42.423 &         -4.9758 &          0.0280 &            33.8 &          1800.0\\
     2814.51752 &      -29438.375 &           2.588 &          -3.295 &           3.660 &           6.299 &          42.407 &         -4.9655 &          0.0269 &            35.0 &          1800.0\\
     2882.32639 &      -29432.663 &           2.253 &          24.821 &           3.186 &           6.137 &          44.443 &         -4.9587 &          0.0230 &            35.4 &          1800.0\\
     2883.32580 &      -29431.768 &           1.301 &           5.275 &           1.840 &           6.303 &          42.443 &         -4.9415 &          0.0093 &            62.0 &          1800.0\\
     2888.32352 &      -29439.078 &           0.921 &           8.556 &           1.303 &           6.282 &          42.541 &         -4.9814 &          0.0057 &            83.5 &          1800.0\\
     2890.34981 &      -29436.360 &           1.090 &           6.537 &           1.541 &           6.282 &          42.514 &         -4.9975 &          0.0080 &            72.2 &          1800.0\\
     2893.38029 &      -29437.654 &           1.566 &           4.158 &           2.215 &           6.276 &          42.515 &         -5.0140 &          0.0152 &            53.3 &          1800.0\\
\noalign{\smallskip}
\hline \noalign{\smallskip}
\multicolumn{11}{c}{Continued on next page} \\
\noalign{\smallskip}
\hline
\end{tabular}
\end{center}
\end{table*}

\begin{table*}
\begin{center}
  \caption*{Table~A.1 continued from previous page.}
\begin{tabular}{cccrccccccc}
\hline
\hline
\noalign{\smallskip}
\multicolumn{1}{c}{BJD$_\mathrm{TBD}$} &
\multicolumn{1}{c}{RV} &
\multicolumn{1}{c}{$\sigma_\mathrm{RV}$} &
\multicolumn{1}{c}{BIS} &
\multicolumn{1}{c}{$\sigma_\mathrm{BIS}$} &
\multicolumn{1}{c}{FWHM} &
\multicolumn{1}{c}{Contrast} &
\multicolumn{1}{c}{$\log{R^\prime_{HK}}$} &
\multicolumn{1}{c}{$\sigma_{\log{R^\prime_{HK}}}$} &
\multicolumn{1}{c}{SNR} &
\multicolumn{1}{c}{$\mathrm{T_{exp}}$}\\ 
\multicolumn{1}{c}{-2457000} &
\multicolumn{1}{c}{($\mathrm{m\,s^{-1}}$)} &
\multicolumn{1}{c}{($\mathrm{m\,s^{-1}}$)} &
\multicolumn{1}{c}{($\mathrm{m\,s^{-1}}$)} &
\multicolumn{1}{c}{($\mathrm{m\,s^{-1}}$)} &
\multicolumn{1}{c}{($\mathrm{km\,s^{-1}}$)} &
\multicolumn{1}{c}{(\%)} &
\multicolumn{1}{c}{---} &
\multicolumn{1}{c}{---} &
\multicolumn{1}{c}{(@550nm)} &
\multicolumn{1}{c}{(s)}\\ 
\noalign{\smallskip}
\hline
\noalign{\smallskip}
     3005.74562 &      -29440.950 &           1.465 &           3.540 &           2.072 &           6.294 &          42.507 &         -4.9723 &          0.0119 &            57.6 &          1800.0\\
     3009.74370 &      -29440.668 &           1.229 &          12.586 &           1.739 &           6.284 &          42.573 &         -4.9848 &          0.0096 &            66.5 &          1800.0\\
     3012.77108 &      -29441.929 &           1.697 &           5.040 &           2.401 &           6.277 &          42.422 &         -5.0017 &          0.0163 &            50.5 &          1800.0\\
     3020.73597 &      -29448.097 &           1.290 &           0.765 &           1.824 &           6.275 &          42.611 &         -4.9908 &          0.0108 &            62.8 &          1800.0\\
     3021.74745 &      -29446.739 &           2.129 &           6.644 &           3.010 &           6.277 &          42.535 &         -4.9823 &          0.0233 &            41.0 &          1800.0\\
     3030.73930 &      -29447.647 &           1.934 &          -2.799 &           2.736 &           6.286 &          42.515 &         -4.9649 &          0.0187 &            45.3 &          1800.0\\
     3037.70755 &      -29444.546 &           2.205 &           5.187 &           3.118 &           6.306 &          42.371 &         -4.9576 &          0.0215 &            40.5 &          1800.0\\
     3051.63949 &      -29454.950 &           1.945 &           8.662 &           2.750 &           6.291 &          42.484 &         -5.0070 &          0.0220 &            42.7 &          1800.0\\
     3069.60902 &      -29448.123 &           2.179 &           7.183 &           3.081 &           6.270 &          42.255 &         -4.9886 &          0.0224 &            40.6 &          1800.0\\
     3075.60948 &      -29438.077 &           5.609 &          -9.552 &           7.933 &           6.296 &          42.255 &         -4.9206 &          0.0697 &            20.1 &          1800.0\\
     3098.58714 &      -29453.647 &           1.204 &          14.537 &           1.703 &           6.296 &          42.474 &         -4.9804 &          0.0091 &            66.8 &          1800.0\\
     3104.63774 &      -29457.966 &           3.017 &         -14.175 &           4.266 &           6.277 &          42.389 &         -5.0288 &          0.0399 &            32.1 &          1800.0\\
     3111.60322 &      -29443.598 &           1.867 &          12.298 &           2.641 &           6.290 &          42.453 &         -4.9458 &          0.0161 &            46.0 &          1800.0\\
     3115.58560 &      -29447.312 &           1.201 &          10.362 &           1.699 &           6.276 &          42.597 &         -4.9764 &          0.0082 &            68.3 &          1800.0\\
     3121.61404 &      -29449.166 &           2.769 &          -9.202 &           3.916 &           6.288 &          42.531 &         -4.9575 &          0.0316 &            33.6 &          1800.0\\
     3135.61160 &      -29452.855 &           2.477 &           8.004 &           3.503 &           6.291 &          42.498 &         -4.9714 &          0.0270 &            36.7 &          1800.0\\
     3144.52786 &      -29468.595 &           3.696 &          -2.169 &           5.228 &           6.296 &          42.375 &         -4.9246 &          0.0409 &            27.1 &          1800.0\\
     3149.55484 &      -29453.540 &           1.316 &           5.928 &           1.862 &           6.288 &          42.559 &         -4.9833 &          0.0110 &            61.6 &          1800.0\\
     3155.52354 &      -29454.356 &           1.279 &          12.910 &           1.809 &           6.306 &          42.417 &         -4.9331 &          0.0093 &            63.2 &          1800.0\\
     3187.51982 &      -29457.266 &           1.565 &          15.232 &           2.213 &           6.304 &          42.303 &         -4.9362 &          0.0129 &            54.2 &          1800.0\\
     3193.50137 &      -29461.872 &           1.628 &           0.632 &           2.302 &           6.281 &          42.606 &         -4.9657 &          0.0152 &            51.4 &          1800.0\\
     3208.39713 &      -29459.586 &           1.287 &           6.144 &           1.820 &           6.283 &          42.552 &         -4.9682 &          0.0102 &            61.8 &          1800.0\\
     3459.61353 &      -29490.030 &           1.553 &           8.829 &           2.196 &           6.284 &          42.604 &         -4.9720 &          0.0129 &            52.8 &          1800.0\\
     3485.62276 &      -29476.836 &           1.222 &          17.363 &           1.729 &           6.328 &          42.296 &         -4.9324 &          0.0085 &            64.7 &          1800.0\\
     3500.42314 &      -29491.683 &           1.349 &           9.616 &           1.908 &           6.289 &          42.517 &         -4.9655 &          0.0098 &            60.3 &          1800.0\\
     3528.56655 &      -29499.446 &           1.706 &          -0.710 &           2.412 &           6.292 &          42.568 &         -4.9752 &          0.0149 &            49.1 &          1800.0\\
     3545.46888 &      -29493.647 &           1.238 &           6.004 &           1.751 &           6.330 &          42.378 &         -4.9288 &          0.0082 &            64.3 &          1800.0\\
     3547.44564 &      -29489.409 &           2.627 &           2.530 &           3.715 &           6.335 &          42.273 &         -4.8917 &          0.0242 &            34.7 &          2400.0\\
     3551.47102 &      -29488.998 &           1.353 &          17.949 &           1.913 &           6.341 &          42.325 &         -4.9010 &          0.0087 &            60.5 &          1800.0\\
     3567.46300 &      -29491.699 &           1.723 &           0.508 &           2.437 &           6.311 &          42.387 &         -4.9204 &          0.0141 &            49.0 &          1800.0\\
     3574.49216 &      -29485.107 &           1.659 &           9.930 &           2.346 &           6.322 &          42.331 &         -4.9085 &          0.0124 &            51.5 &          1800.0\\
     3575.50971 &      -29494.507 &           1.669 &           6.374 &           2.361 &           6.328 &          42.325 &         -4.9214 &          0.0134 &            51.4 &          1800.0\\
     3591.35651 &      -29490.064 &           1.996 &          23.268 &           2.823 &           6.334 &          42.265 &         -4.9013 &          0.0160 &            43.7 &          1800.0\\
     3607.35216 &      -29496.138 &           1.980 &          13.788 &           2.800 &           6.319 &          42.299 &         -4.9002 &          0.0166 &            44.0 &          1800.0\\
     3615.32162 &      -29494.017 &           2.043 &          13.806 &           2.890 &           6.318 &          42.322 &         -4.8962 &          0.0182 &            41.9 &          1800.0\\
     3743.77648 &      -29491.888 &           1.671 &          15.914 &           2.364 &           6.327 &          42.016 &         -4.9023 &          0.0129 &            50.9 &          1800.0\\
     3749.72589 &      -29504.810 &           1.541 &          17.444 &           2.180 &           6.326 &          42.285 &         -4.9178 &          0.0121 &            54.2 &          1800.0\\
\noalign{\smallskip}
\hline
\end{tabular}
\end{center}
\end{table*}

\begin{table*}
\begin{tiny}
\begin{center}
\caption{Radial velocities and spectral activity indicators measured from TNG/HARPS-N spectra measured with SERVAL.
\label{Table: HARPSN RVs serval}}
\begin{tabular}{crccccccccccccc}
\hline
\hline
\noalign{\smallskip}
\multicolumn{1}{c}{BJD$_\mathrm{TBD}$} &
\multicolumn{1}{c}{RV} &
\multicolumn{1}{c}{$\sigma_\mathrm{RV}$} &
\multicolumn{1}{c}{CRX} &
\multicolumn{1}{c}{$\sigma_\mathrm{CRX}$} &
\multicolumn{1}{c}{dlW} &
\multicolumn{1}{c}{$\sigma_\mathrm{dlW}$} &
\multicolumn{1}{c}{$\mathrm{H_{\alpha}}$} &
\multicolumn{1}{c}{$\mathrm{\sigma_{H_{\alpha}}}$} &
\multicolumn{1}{c}{$\mathrm{NaD_{1}}$} &
\multicolumn{1}{c}{$\mathrm{\sigma_{NaD_{1}}}$} &
\multicolumn{1}{c}{$\mathrm{NaD_{2}}$} &
\multicolumn{1}{c}{$\mathrm{\sigma_{NaD_{2}}}$} &
\multicolumn{1}{c}{SNR} &
\multicolumn{1}{c}{$\mathrm{T_{exp}}$}\\ 
\multicolumn{1}{c}{-2457000} &
\multicolumn{1}{c}{($\mathrm{m\,s^{-1}}$)} &
\multicolumn{1}{c}{($\mathrm{m\,s^{-1}}$)} &
\multicolumn{1}{c}{($\mathrm{m\,s^{-1}\,Np^{-1}}$)} &
\multicolumn{1}{c}{($\mathrm{m\,s^{-1}\,Np^{-1}}$)} &
\multicolumn{1}{c}{($\mathrm{m^2\,s^{-2}}$)} &
\multicolumn{1}{c}{($\mathrm{m^2\,s^{-2}}$)} &
\multicolumn{1}{c}{---} &
\multicolumn{1}{c}{---} &
\multicolumn{1}{c}{---} &
\multicolumn{1}{c}{---} &
\multicolumn{1}{c}{---} &
\multicolumn{1}{c}{---} &
\multicolumn{1}{c}{(@600 nm)} &
\multicolumn{1}{c}{(s)}\\ 
\noalign{\smallskip}
\hline
\noalign{\smallskip}
     1925.74746 &         -58.275 &           1.321 &           6.212 &          10.496 &          15.618 &           2.371 &          0.4968 &          0.0017 &          0.2272 &          0.0017 &          0.2969 &          0.0020 &            54.2 &          1800.0\\
     1925.76981 &         -58.422 &           1.812 &           8.577 &          14.375 &          62.829 &           3.390 &          0.4934 &          0.0021 &          0.2319 &          0.0022 &          0.2971 &          0.0027 &            42.4 &          1800.0\\
     1926.74886 &         -61.283 &           1.513 &          22.328 &          11.769 &          13.405 &           1.987 &          0.4968 &          0.0016 &          0.2250 &          0.0015 &          0.2945 &          0.0018 &            58.6 &          1800.0\\
     1929.72901 &         -40.940 &           1.453 &          12.858 &          11.901 &          15.160 &           2.278 &          0.4947 &          0.0015 &          0.2224 &          0.0013 &          0.2934 &          0.0016 &            65.0 &          1800.0\\
     2000.57290 &         -40.901 &           1.376 &           2.619 &          10.164 &          18.338 &           2.410 &          0.4916 &          0.0014 &          0.2253 &          0.0015 &          0.2913 &          0.0019 &            59.0 &          1800.0\\
     2000.65755 &         -39.948 &           1.529 &         -11.548 &          10.989 &          11.009 &           2.873 &          0.4907 &          0.0013 &          0.2276 &          0.0015 &          0.2917 &          0.0018 &            59.5 &          1500.0\\
     2002.57645 &         -54.200 &           1.228 &          -6.362 &           9.810 &          23.383 &           1.988 &          0.4953 &          0.0014 &          0.2263 &          0.0013 &          0.2943 &          0.0016 &            66.0 &          1800.0\\
     2002.66203 &         -51.402 &           1.217 &         -15.061 &           9.676 &          20.393 &           2.062 &          0.4968 &          0.0016 &          0.2263 &          0.0015 &          0.2990 &          0.0018 &            59.2 &          1500.0\\
     2012.58204 &         -54.526 &           1.526 &          10.390 &          11.945 &           5.879 &           2.589 &          0.4915 &          0.0018 &          0.2275 &          0.0018 &          0.2965 &          0.0022 &            51.0 &          1800.0\\
     2014.58299 &         -49.916 &           1.450 &         -12.812 &          11.065 &           5.918 &           2.639 &          0.4902 &          0.0014 &          0.2277 &          0.0014 &          0.2958 &          0.0017 &            63.2 &          1800.0\\
     2056.55645 &         -42.220 &           1.523 &          -4.091 &          11.436 &          12.660 &           3.082 &          0.4996 &          0.0014 &          0.2297 &          0.0016 &          0.2957 &          0.0019 &            58.4 &          1700.0\\
     2073.62803 &         -18.352 &           1.548 &          -7.259 &          11.486 &          39.235 &           3.197 &          0.4997 &          0.0015 &          0.2314 &          0.0017 &          0.2983 &          0.0021 &            54.6 &          1700.0\\
     2082.43490 &         -41.922 &           2.379 &           3.632 &          17.045 &          22.558 &           4.127 &          0.4962 &          0.0018 &          0.2321 &          0.0025 &          0.2957 &          0.0030 &            39.6 &          1700.0\\
     2087.42720 &         -46.273 &           1.745 &          20.737 &          13.131 &          11.655 &           3.241 &          0.4964 &          0.0019 &          0.2347 &          0.0021 &          0.2973 &          0.0026 &            44.8 &          1800.0\\
     2089.45674 &         -53.195 &           1.769 &         -17.926 &          14.088 &          23.930 &           3.426 &          0.5040 &          0.0025 &          0.2387 &          0.0025 &          0.2961 &          0.0030 &            39.3 &          1800.0\\
     2090.48671 &         -49.324 &           1.431 &          -9.934 &          10.832 &          19.296 &           2.841 &          0.4986 &          0.0016 &          0.2377 &          0.0018 &          0.3012 &          0.0021 &            53.8 &          1800.0\\
     2117.48442 &         -43.299 &           2.619 &          12.153 &          19.942 &          38.291 &           4.046 &          0.5002 &          0.0026 &          0.2373 &          0.0031 &          0.3048 &          0.0036 &            34.2 &          1700.0\\
     2119.45691 &         -37.407 &           1.484 &          13.059 &          13.036 &          28.503 &           2.691 &          0.5098 &          0.0029 &          0.2339 &          0.0023 &          0.3021 &          0.0028 &            40.9 &          1700.0\\
     2138.34160 &         -39.819 &           1.333 &          -8.592 &          10.456 &          19.834 &           2.815 &          0.4987 &          0.0015 &          0.2293 &          0.0015 &          0.3035 &          0.0018 &            60.6 &          1800.0\\
     2149.32319 &         -35.374 &           1.133 &          -8.010 &           9.471 &          15.718 &           2.373 &          0.4999 &          0.0019 &          0.2305 &          0.0016 &          0.3002 &          0.0020 &            55.5 &          1800.0\\
     2150.31933 &         -39.134 &           1.337 &          -1.465 &          10.596 &          10.203 &           1.991 &          0.4967 &          0.0014 &          0.2260 &          0.0013 &          0.2950 &          0.0016 &            66.3 &          1800.0\\
     2151.31789 &         -44.622 &           1.134 &          -3.436 &           8.841 &          44.381 &           2.824 &          0.5018 &          0.0013 &          0.2239 &          0.0013 &          0.2977 &          0.0016 &            68.7 &          1800.0\\
     2160.31282 &         -39.621 &           1.302 &          -2.518 &          10.850 &          48.176 &           2.543 &          0.4931 &          0.0017 &          0.2288 &          0.0015 &          0.2950 &          0.0018 &            58.5 &          1800.0\\
     2309.71597 &         -13.907 &           1.925 &         -12.318 &          14.015 &          21.823 &           3.573 &          0.4951 &          0.0017 &          0.2252 &          0.0021 &          0.3034 &          0.0026 &            44.9 &          1800.0\\
     2353.62295 &         -28.158 &           1.101 &          10.332 &           8.271 &          10.046 &           2.745 &          0.4919 &          0.0015 &          0.2295 &          0.0016 &          0.2950 &          0.0020 &            56.9 &          2400.0\\
     2355.57948 &         -23.799 &           1.284 &          -8.869 &          10.087 &           8.420 &           2.567 &          0.4926 &          0.0015 &          0.2229 &          0.0016 &          0.2999 &          0.0019 &            57.4 &          3300.0\\
     2369.67671 &         -41.436 &           2.272 &         -29.796 &          18.341 &           0.648 &           3.289 &          0.4954 &          0.0029 &          0.2251 &          0.0028 &          0.2923 &          0.0034 &            34.0 &          1800.0\\
     2393.66078 &         -21.165 &           1.049 &          -4.267 &           8.678 &          16.496 &           1.969 &          0.4876 &          0.0014 &          0.2296 &          0.0012 &          0.2945 &          0.0015 &            71.3 &          1800.0\\
     2394.57379 &         -23.361 &           1.041 &          -5.895 &           7.857 &          18.006 &           2.197 &          0.4895 &          0.0012 &          0.2196 &          0.0013 &          0.2913 &          0.0015 &            69.7 &          1800.0\\
     2410.55303 &         -24.617 &           1.553 &          15.435 &          12.058 &           3.439 &           2.506 &          0.4950 &          0.0018 &          0.2319 &          0.0019 &          0.3040 &          0.0023 &            49.4 &          1800.0\\
     2450.48219 &         -14.659 &           0.980 &          -1.783 &           7.469 &           5.014 &           1.974 &          0.4925 &          0.0011 &          0.2312 &          0.0011 &          0.3008 &          0.0013 &            81.7 &          1800.0\\
     2676.71340 &          -3.926 &           1.443 &          -2.263 &          12.322 &         -24.412 &           2.129 &          0.4839 &          0.0018 &          0.2265 &          0.0016 &          0.3010 &          0.0020 &            55.0 &          1800.0\\
     2677.72392 &          -4.019 &           1.154 &           7.789 &          10.021 &         -25.385 &           1.838 &          0.4842 &          0.0019 &          0.2252 &          0.0016 &          0.3010 &          0.0020 &            55.1 &          1800.0\\
     2702.67857 &           0.988 &           1.258 &          -6.798 &          10.183 &         -10.708 &           1.878 &          0.4757 &          0.0012 &          0.2216 &          0.0012 &          0.3009 &          0.0015 &            71.8 &          1800.0\\
     2703.69387 &           3.387 &           1.122 &          -7.957 &           9.127 &          -8.679 &           2.127 &          0.4740 &          0.0014 &          0.2228 &          0.0014 &          0.3009 &          0.0017 &            65.1 &          1800.0\\
     2720.63146 &           1.830 &           3.074 &          -5.060 &          26.616 &         -14.054 &           4.249 &          0.4914 &          0.0044 &          0.2234 &          0.0041 &          0.2967 &          0.0050 &            24.8 &          1800.0\\
     2722.64219 &          -0.941 &           1.193 &          -8.573 &           9.671 &         -32.924 &           1.527 &          0.4864 &          0.0014 &          0.2260 &          0.0013 &          0.2939 &          0.0016 &            67.6 &          1800.0\\
     2765.55642 &           2.744 &           0.872 &          -5.543 &           7.270 &         -23.788 &           1.353 &          0.4887 &          0.0013 &          0.2227 &          0.0012 &          0.2920 &          0.0014 &            73.1 &          1800.0\\
     2766.55074 &           3.164 &           1.072 &           4.021 &           9.134 &         -18.619 &           1.716 &          0.4940 &          0.0017 &          0.2217 &          0.0015 &          0.2970 &          0.0019 &            57.6 &          1800.0\\
     2785.57158 &          -2.700 &           2.533 &          21.913 &          20.610 &          -5.799 &           4.507 &          0.5021 &          0.0029 &          0.2396 &          0.0031 &          0.3002 &          0.0037 &            32.9 &          2100.0\\
     2810.52211 &           3.257 &           1.152 &          -6.779 &           9.339 &         -35.837 &           1.926 &          0.4884 &          0.0014 &          0.2279 &          0.0014 &          0.2928 &          0.0017 &            64.1 &          1800.0\\
     2811.51754 &           6.822 &           2.007 &           2.552 &          16.042 &         -30.098 &           3.129 &          0.4981 &          0.0023 &          0.2335 &          0.0025 &          0.3018 &          0.0030 &            39.4 &          1800.0\\
     2814.51752 &           2.882 &           2.114 &          16.941 &          17.132 &         -21.153 &           2.582 &          0.4991 &          0.0024 &          0.2315 &          0.0025 &          0.2956 &          0.0030 &            39.5 &          1800.0\\
     2882.32639 &           6.311 &           1.567 &           6.952 &          15.070 &         -26.522 &           2.724 &          0.4998 &          0.0041 &          0.2310 &          0.0030 &          0.2957 &          0.0037 &            31.2 &          1800.0\\
     2883.32580 &           7.861 &           1.132 &          -8.783 &           9.328 &         -26.060 &           1.860 &          0.4957 &          0.0015 &          0.2289 &          0.0013 &          0.2989 &          0.0016 &            66.0 &          1800.0\\
     2888.32352 &           0.393 &           0.909 &           1.079 &           7.361 &         -38.113 &           1.290 &          0.4927 &          0.0011 &          0.2259 &          0.0009 &          0.2976 &          0.0012 &            89.6 &          1800.0\\
     2890.34981 &           4.393 &           1.045 &         -10.164 &           8.637 &         -35.652 &           1.662 &          0.4955 &          0.0014 &          0.2254 &          0.0011 &          0.2992 &          0.0014 &            74.9 &          1800.0\\
     2893.38029 &           2.334 &           1.330 &          -8.531 &          11.125 &         -32.818 &           2.387 &          0.4924 &          0.0018 &          0.2241 &          0.0016 &          0.3000 &          0.0020 &            56.2 &          1800.0\\
     3005.74562 &          -0.949 &           1.138 &          16.213 &           8.866 &         -30.972 &           2.107 &          0.4935 &          0.0014 &          0.2246 &          0.0013 &          0.2979 &          0.0017 &            64.9 &          1800.0\\
\noalign{\smallskip}
\hline \noalign{\smallskip}
\multicolumn{15}{c}{Continued on next page} \\
\noalign{\smallskip}
\hline
\end{tabular}
\end{center}
\end{tiny}
\end{table*}   
     
  \begin{table*}
\begin{tiny}
\begin{center}
\caption*{Table A.2 continued from previous page.}
\begin{tabular}{crcrrrrcccccccc}
\hline
\hline
\noalign{\smallskip}
\multicolumn{1}{c}{BJD$_\mathrm{TBD}$} &
\multicolumn{1}{c}{RV} &
\multicolumn{1}{c}{$\sigma_\mathrm{RV}$} &
\multicolumn{1}{c}{CRX} &
\multicolumn{1}{c}{$\sigma_\mathrm{CRX}$} &
\multicolumn{1}{c}{dlW} &
\multicolumn{1}{c}{$\sigma_\mathrm{dlW}$} &
\multicolumn{1}{c}{$\mathrm{H_{\alpha}}$} &
\multicolumn{1}{c}{$\mathrm{\sigma_{H_{\alpha}}}$} &
\multicolumn{1}{c}{$\mathrm{NaD_{1}}$} &
\multicolumn{1}{c}{$\mathrm{\sigma_{NaD_{1}}}$} &
\multicolumn{1}{c}{$\mathrm{NaD_{2}}$} &
\multicolumn{1}{c}{$\mathrm{\sigma_{NaD_{2}}}$} &
\multicolumn{1}{c}{SNR} &
\multicolumn{1}{c}{$\mathrm{T_{exp}}$}\\ 
\multicolumn{1}{c}{-2457000} &
\multicolumn{1}{c}{($\mathrm{m\,s^{-1}}$)} &
\multicolumn{1}{c}{($\mathrm{m\,s^{-1}}$)} &
\multicolumn{1}{c}{($\mathrm{m\,s^{-1}\,Np^{-1}}$)} &
\multicolumn{1}{c}{($\mathrm{m\,s^{-1}\,Np^{-1}}$)} &
\multicolumn{1}{c}{($\mathrm{m^2\,s^{-2}}$)} &
\multicolumn{1}{c}{($\mathrm{m^2\,s^{-2}}$)} &
\multicolumn{1}{c}{---} &
\multicolumn{1}{c}{---} &
\multicolumn{1}{c}{---} &
\multicolumn{1}{c}{---} &
\multicolumn{1}{c}{---} &
\multicolumn{1}{c}{---} &
\multicolumn{1}{c}{(@600 nm)} &
\multicolumn{1}{c}{(s)}\\ 
\noalign{\smallskip}
\hline
\noalign{\smallskip}   
     3009.74370 &          -1.866 &           1.165 &          22.996 &           9.295 &         -35.345 &           2.232 &          0.4968 &          0.0014 &          0.2235 &          0.0012 &          0.3017 &          0.0015 &            71.1 &          1800.0\\
     3012.77108 &          -4.465 &           1.631 &           9.145 &          13.145 &         -26.715 &           3.292 &          0.4953 &          0.0016 &          0.2231 &          0.0016 &          0.3045 &          0.0020 &            56.7 &          1800.0\\
     3020.73597 &          -7.707 &           1.197 &          -7.168 &          10.213 &         -40.757 &           2.134 &          0.4875 &          0.0016 &          0.2244 &          0.0013 &          0.2961 &          0.0017 &            65.2 &          1800.0\\
     3021.74745 &          -8.214 &           1.509 &          18.594 &          12.818 &         -35.262 &           2.924 &          0.4912 &          0.0025 &          0.2223 &          0.0021 &          0.3029 &          0.0027 &            42.4 &          1800.0\\
     3030.73930 &          -8.337 &           1.616 &          -6.522 &          13.323 &         -31.950 &           2.795 &          0.4940 &          0.0019 &          0.2219 &          0.0018 &          0.3072 &          0.0023 &            49.8 &          1800.0\\
     3037.70755 &          -5.684 &           1.884 &          -9.743 &          15.263 &         -16.892 &           2.994 &          0.4918 &          0.0020 &          0.2259 &          0.0020 &          0.3038 &          0.0026 &            45.4 &          1800.0\\
     3051.63949 &         -16.670 &           1.663 &         -27.142 &          14.922 &         -28.957 &           2.343 &          0.4899 &          0.0028 &          0.2229 &          0.0022 &          0.3009 &          0.0028 &            40.7 &          1800.0\\
     3069.60902 &          -8.267 &           1.655 &         -14.675 &          13.848 &          -7.250 &           3.204 &          0.4881 &          0.0023 &          0.2263 &          0.0021 &          0.3065 &          0.0027 &            43.5 &          1800.0\\
     3075.60948 &           2.803 &           3.739 &           1.835 &          31.252 &         -14.509 &           4.986 &          0.4923 &          0.0045 &          0.2248 &          0.0048 &          0.2996 &          0.0060 &            22.4 &          1800.0\\
     3098.58714 &         -14.844 &           0.880 &          12.737 &           7.512 &         -24.249 &           1.652 &          0.4869 &          0.0014 &          0.2161 &          0.0012 &          0.3060 &          0.0016 &            70.5 &          1800.0\\
     3104.63774 &         -16.445 &           1.953 &          25.181 &          16.390 &         -21.243 &           3.619 &          0.4851 &          0.0030 &          0.2288 &          0.0028 &          0.3076 &          0.0035 &            34.4 &          1800.0\\
     3111.60322 &          -4.882 &           1.497 &          25.364 &          11.569 &         -25.566 &           2.589 &          0.4818 &          0.0015 &          0.2263 &          0.0016 &          0.3022 &          0.0021 &            55.5 &          1800.0\\
     3115.58560 &          -8.926 &           1.074 &           4.013 &           8.760 &         -43.005 &           1.928 &          0.4871 &          0.0011 &          0.2241 &          0.0011 &          0.3008 &          0.0014 &            80.6 &          1800.0\\
     3121.61404 &          -8.448 &           2.017 &          16.857 &          16.810 &         -33.875 &           3.373 &          0.4860 &          0.0026 &          0.2272 &          0.0026 &          0.3057 &          0.0033 &            37.1 &          1800.0\\
     3135.61160 &         -12.987 &           2.049 &          15.754 &          17.081 &         -27.922 &           3.215 &          0.4838 &          0.0023 &          0.2306 &          0.0023 &          0.2968 &          0.0029 &            41.0 &          1800.0\\
     3144.52786 &         -26.415 &           2.471 &          15.237 &          20.654 &         -25.410 &           4.316 &          0.4841 &          0.0033 &          0.2318 &          0.0035 &          0.2986 &          0.0043 &            29.8 &          1800.0\\
     3149.55484 &         -14.042 &           1.255 &           2.053 &          11.093 &         -38.942 &           1.709 &          0.4796 &          0.0016 &          0.2258 &          0.0013 &          0.3074 &          0.0017 &            64.8 &          1800.0\\
     3155.52354 &         -14.325 &           1.054 &          -4.303 &           9.257 &         -21.395 &           1.574 &          0.4801 &          0.0015 &          0.2269 &          0.0013 &          0.3048 &          0.0016 &            67.2 &          1800.0\\
     3187.51982 &         -19.166 &           1.254 &           6.756 &          10.673 &          -4.466 &           2.629 &          0.4809 &          0.0015 &          0.2311 &          0.0015 &          0.2997 &          0.0019 &            60.9 &          1800.0\\
     3193.50137 &         -21.217 &           1.160 &          -0.004 &          10.140 &         -43.288 &           2.723 &          0.4751 &          0.0018 &          0.2333 &          0.0017 &          0.3056 &          0.0021 &            54.4 &          1800.0\\
     3208.39713 &         -19.206 &           0.949 &           3.642 &           8.346 &         -32.809 &           1.845 &          0.4715 &          0.0016 &          0.2277 &          0.0014 &          0.3034 &          0.0017 &            64.0 &          1800.0\\
     3459.61353 &         -52.078 &           1.310 &           4.341 &          10.607 &         -42.304 &           2.225 &          0.4957 &          0.0016 &          0.2204 &          0.0015 &          0.2990 &          0.0019 &            58.6 &          1800.0\\
     3485.62276 &         -38.872 &           1.077 &           5.922 &           9.300 &          -8.398 &           1.535 &          0.4921 &          0.0016 &          0.2206 &          0.0013 &          0.2995 &          0.0016 &            65.2 &          1800.0\\
     3500.42314 &         -50.967 &           1.095 &           4.076 &           8.523 &         -36.104 &           2.069 &          0.4894 &          0.0013 &          0.2294 &          0.0013 &          0.2955 &          0.0016 &            69.5 &          1800.0\\
     3528.56655 &         -60.873 &           1.067 &           4.704 &           8.706 &         -41.303 &           2.200 &          0.4977 &          0.0018 &          0.2313 &          0.0017 &          0.2980 &          0.0021 &            53.6 &          1800.0\\
     3545.46888 &         -53.273 &           1.274 &           3.849 &          10.543 &         -19.152 &           1.510 &          0.4973 &          0.0015 &          0.2345 &          0.0013 &          0.2997 &          0.0016 &            68.7 &          1800.0\\
     3547.44564 &         -46.680 &           1.837 &         -10.564 &          14.983 &         -10.822 &           3.583 &          0.4976 &          0.0025 &          0.2279 &          0.0025 &          0.3004 &          0.0031 &            38.6 &          2400.0\\
     3551.47102 &         -51.567 &           1.026 &         -14.010 &           8.010 &         -13.583 &           1.997 &          0.4957 &          0.0014 &          0.2278 &          0.0013 &          0.2990 &          0.0016 &            68.3 &          1800.0\\
     3567.46300 &         -51.510 &           1.587 &          28.544 &          12.813 &         -21.465 &           2.167 &          0.4894 &          0.0019 &          0.2302 &          0.0017 &          0.3011 &          0.0022 &            52.8 &          1800.0\\
     3574.49216 &         -44.526 &           1.452 &          -3.743 &          11.394 &         -10.662 &           2.466 &          0.4925 &          0.0015 &          0.2287 &          0.0015 &          0.3004 &          0.0019 &            60.4 &          1800.0\\
     3575.50971 &         -54.586 &           1.265 &          -6.372 &          10.222 &         -12.718 &           2.308 &          0.4925 &          0.0016 &          0.2290 &          0.0015 &          0.2959 &          0.0019 &            58.2 &          1800.0\\
     3591.35651 &         -51.910 &           1.727 &         -28.188 &          13.318 &          -5.723 &           2.846 &          0.4929 &          0.0018 &          0.2307 &          0.0019 &          0.3082 &          0.0023 &            50.3 &          1800.0\\
     3607.35216 &         -57.440 &           1.741 &          -4.145 &          14.314 &         -13.714 &           2.250 &          0.4913 &          0.0019 &          0.2263 &          0.0019 &          0.3017 &          0.0023 &            48.9 &          1800.0\\
     3615.32162 &         -58.845 &           1.363 &          15.653 &          12.109 &          -9.196 &           2.824 &          0.4862 &          0.0027 &          0.2291 &          0.0022 &          0.3046 &          0.0028 &            41.5 &          1800.0\\
     3743.77648 &         -52.815 &           1.528 &         -12.326 &          13.008 &          18.302 &           2.771 &          0.4796 &          0.0018 &          0.2258 &          0.0016 &          0.3012 &          0.0021 &            54.5 &          1800.0\\
     3749.72589 &         -67.494 &           1.276 &         -11.571 &          10.990 &         -12.047 &           2.403 &          0.4788 &          0.0018 &          0.2243 &          0.0016 &          0.2991 &          0.0020 &            56.6 &          1800.0\\
\noalign{\smallskip}
\hline
\end{tabular}
\end{center}
\end{tiny}
\end{table*}

 \clearpage
\begin{table*}
\begin{center}
  \caption{Radial velocities and spectral activity indicators extracted from  HIRES spectra. The spectral resolution is 50\,000.}   \label{Table: HIRES RVs}
  \begin{tabular}{rrrrr}
    \hline
    \hline
     \noalign{\smallskip}
    \multicolumn{1}{c}{BJD$_\mathrm{TBD}$ (d)} &
    \multicolumn{1}{c}{RV} &
    \multicolumn{1}{c}{$\sigma_\mathrm{RV}$} &
    \multicolumn{1}{c}{S-index} &
    \multicolumn{1}{c}{$\sigma_\mathrm{S-index}$}   \\ 
    \noalign{\smallskip}
    \multicolumn{1}{c}{-2457000} &
    \multicolumn{1}{c}{(\ms)} &
   \multicolumn{1}{c}{(\ms)} &
    \multicolumn{1}{c}{---} &
    \multicolumn{1}{c}{---}  \\ 
    \noalign{\smallskip}
    \hline
    \noalign{\smallskip}
2459068.996039	&	-31.9223870565839	&	1.58609867095947	&	0.2404	&	0.001	\\
2459091.903590	&	-27.254684787594	&	2.07991886138916	&	0.2347	&	0.001	\\
2459117.746010	&	-18.7973207807876	&	1.65706002712250	&	0.2517	&	0.001	\\
2459153.764358	&	-34.0642727947058	&	1.58761537075043	&	0.2302	&	0.001	\\
2459189.737038	&	-9.74054052172774	&	1.83101606369019	&	0.2670	&	0.001	\\
2459300.142053	&	-13.2534923563946	&	1.73066008090973	&	0.2536	&	0.001	\\
2459373.945057	&	-5.41569305693773	&	1.49286580085754	&	0.2423	&	0.001	\\
2459389.920888	&	-6.32071842831923	&	1.59697532653809	&	0.2646	&	0.001	\\
2459389.961768	&	-3.35709306000303	&	1.67562794685364	&	0.2619	&	0.001	\\
2459390.011571	&	-5.20555394934138	&	1.60369610786438	&	0.2607	&	0.001	\\
2459422.937937	&	9.87880840430022	&	1.96272063255310	&	0.2095	&	0.001	\\
2459450.973157	&	0.50471086946222	&	1.75168633460999	&	0.2575	&	0.001	\\
2459478.824326	&	11.5097598084982	&	1.77714145183563	&	0.2610	&	0.001	\\
2459508.776629	&	7.90569467907215	&	1.63071000576019	&	0.2525	&	0.001	\\
2459541.773792	&	13.9118396269181	&	1.77524840831757	&	0.2389	&	0.001	\\
2459661.152785	&	13.2454792755569	&	1.77994811534882	&	0.2337	&	0.001	\\
2459691.118694	&	20.4637476019482	&	1.95632743835449	&	0.2350	&	0.001	\\
2459737.952171	&	21.0261062658166	&	1.61103308200836	&	0.2245	&	0.001	\\
2459760.017662	&	26.2139979769045	&	1.62301576137543	&	0.2425	&	0.001	\\
2459780.008834	&	30.3119877119156	&	1.69775712490082	&	0.2288	&	0.001	\\
    \hline
  \end{tabular}
\end{center}
\end{table*}

\clearpage


\section{Figures}

\noindent\begin{minipage}{\textwidth}
     \centering
     \includegraphics[scale=0.55]{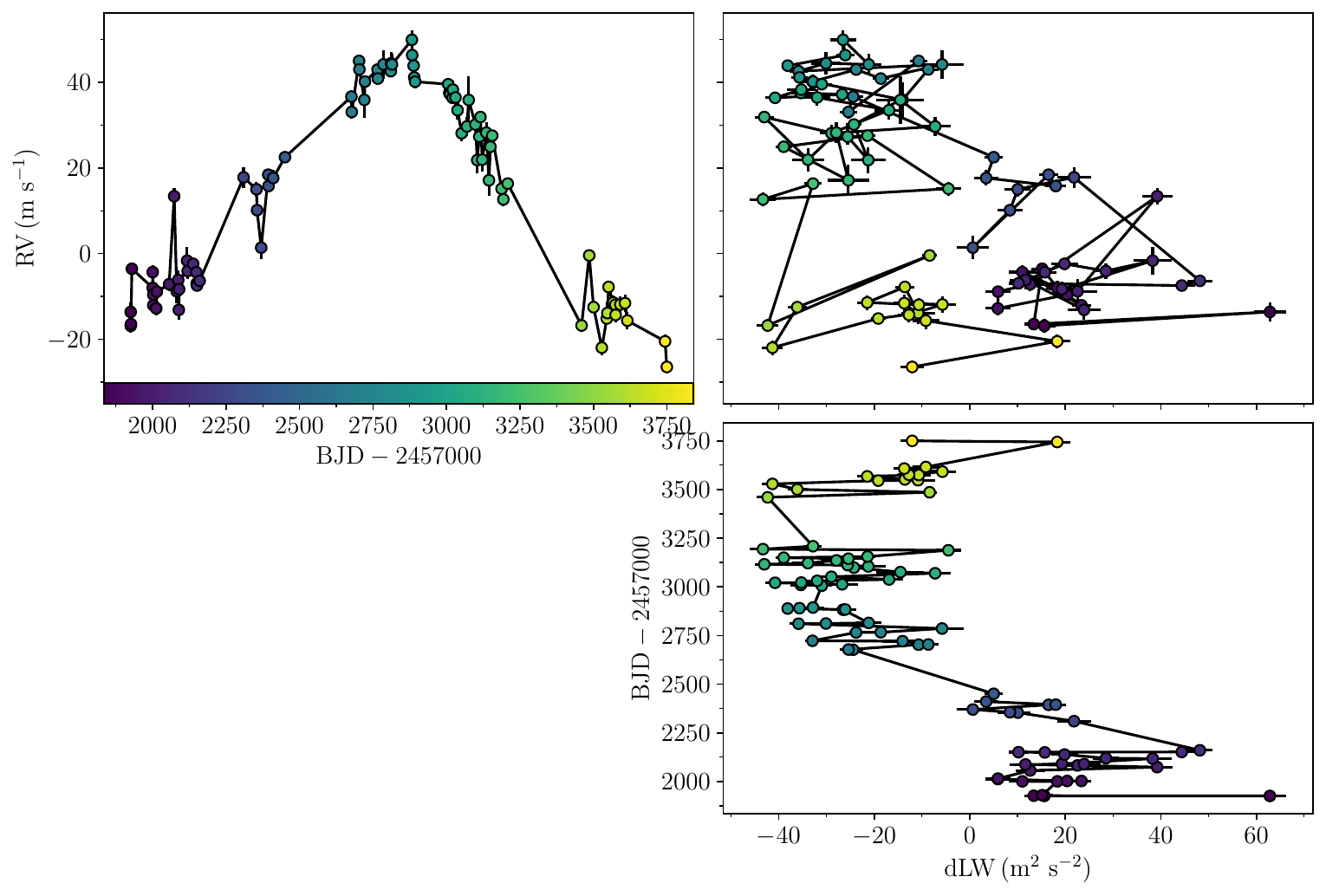} 
   \captionof{figure}{RVs vs time, differential line width (dLW) vs RVs, and dLW vs time. The models of planets~b and c have been subtracted from the RVs. }
      \label{Fig: correlation plot RV vs dLW}

\vspace{1cm}
 \centering
   \includegraphics[scale=0.55]{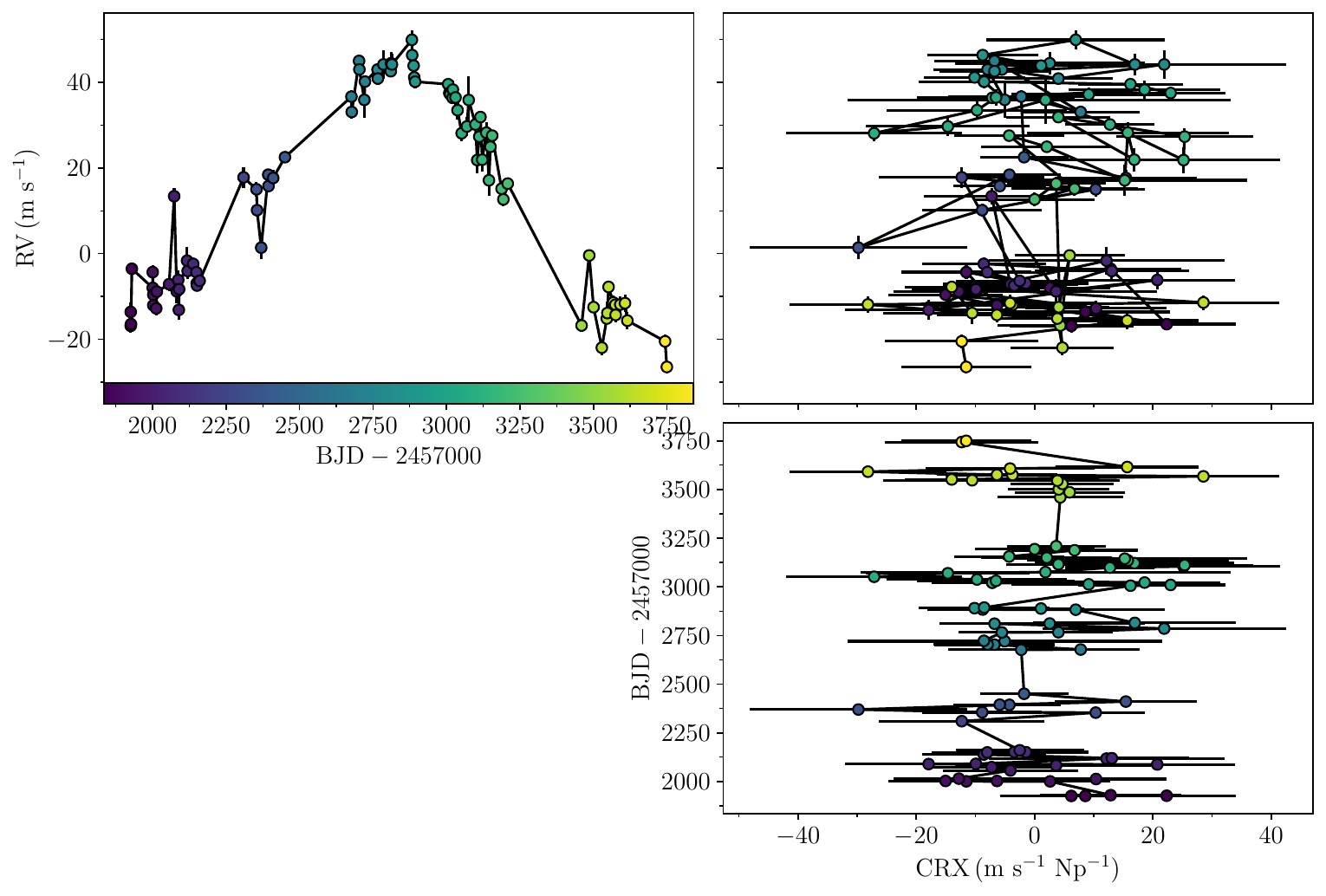}  
   \caption{RVs vs time, RV chromatic    index  (CRX) vs RVs, and CRX vs time. The models of planets~b and c have been subtracted from the RVs.} 
      \label{Figure: Fig: correlation plot RV vs CRX}
 \end{minipage}

  \begin{figure*}[!ht]
 \centering
  \includegraphics[scale=0.35]{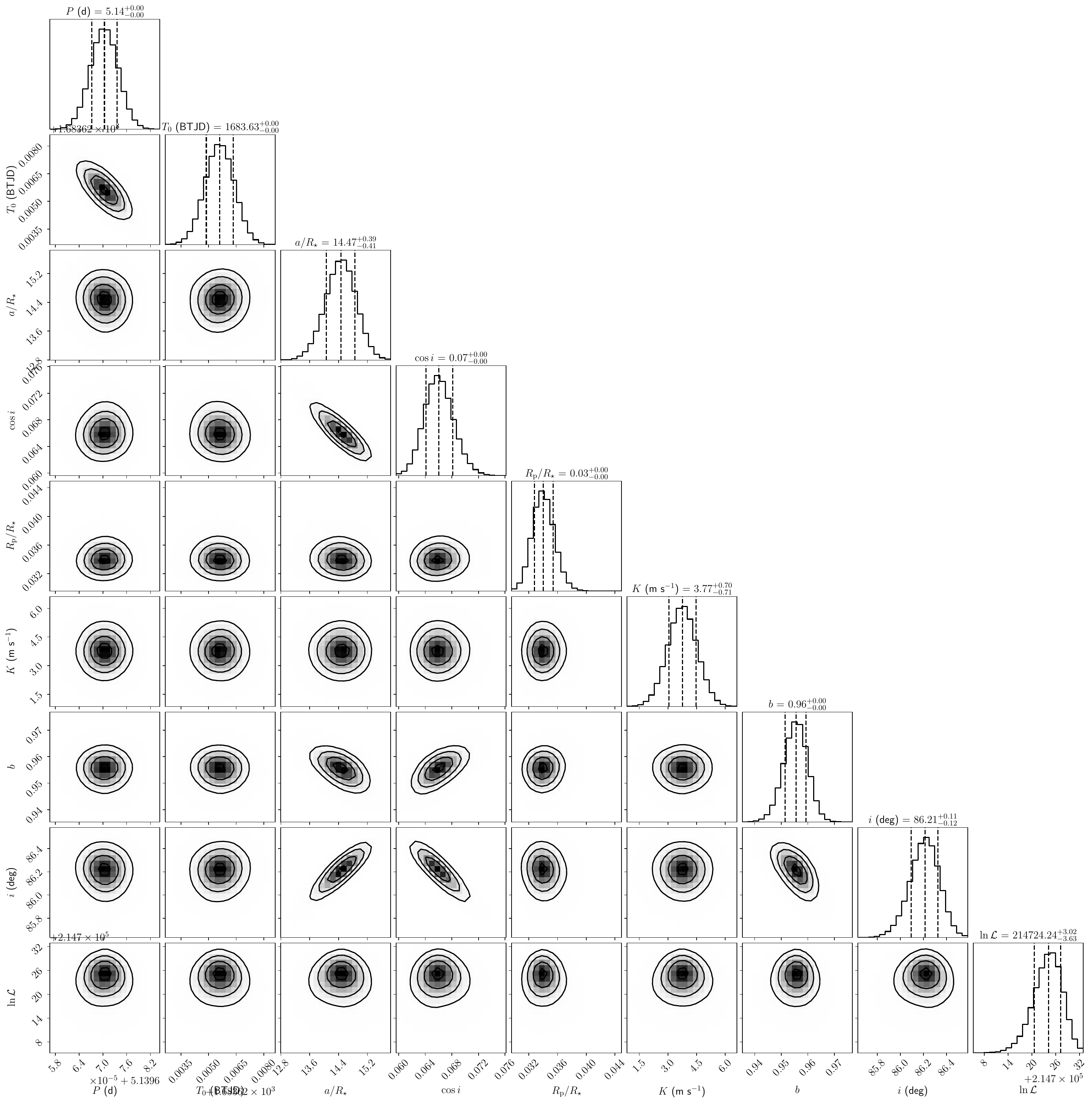} 
   \caption{Corner plot of the posterior distributions of the joint transit and RV model of planet~b.} 
      \label{Figure: cornerplot planet b}
 \end{figure*}

   \begin{figure*}[!ht]
 \centering
  \includegraphics[scale=0.35]{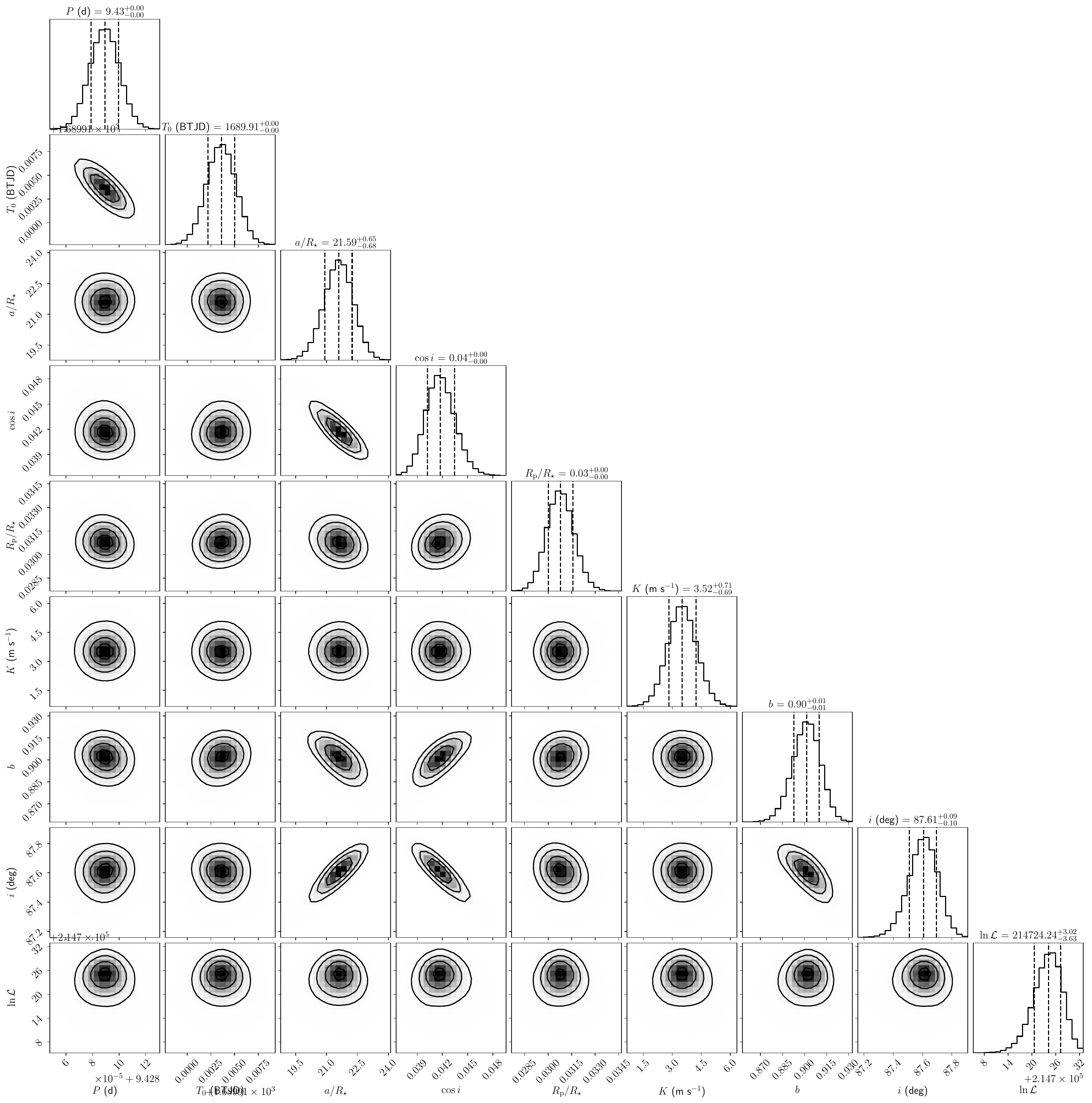} 
   \caption{Corner plot of the posterior distributions of the joint transit and RV model of planet~c.} 
      \label{Figure: cornerplot planet c}
 \end{figure*}

   \begin{figure*}[!ht]
 \centering
  \includegraphics[scale=0.45]{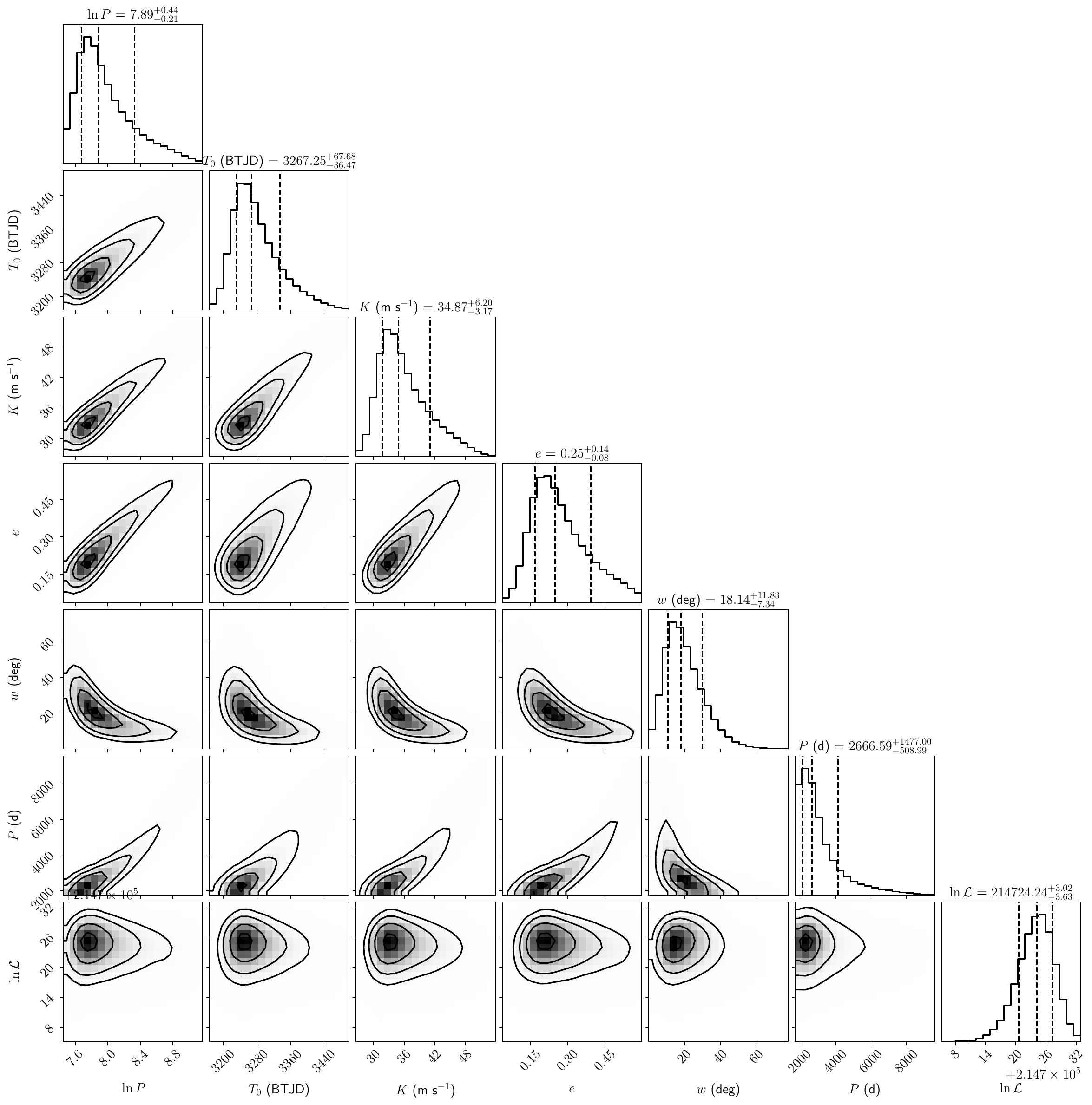}
   \caption{Corner plot of the posterior distributions of the   RV model of signal~d.} 
      \label{Figure: cornerplot signal d}
 \end{figure*}

 \begin{figure*}[!ht]
\centering
\includegraphics[scale=0.4]{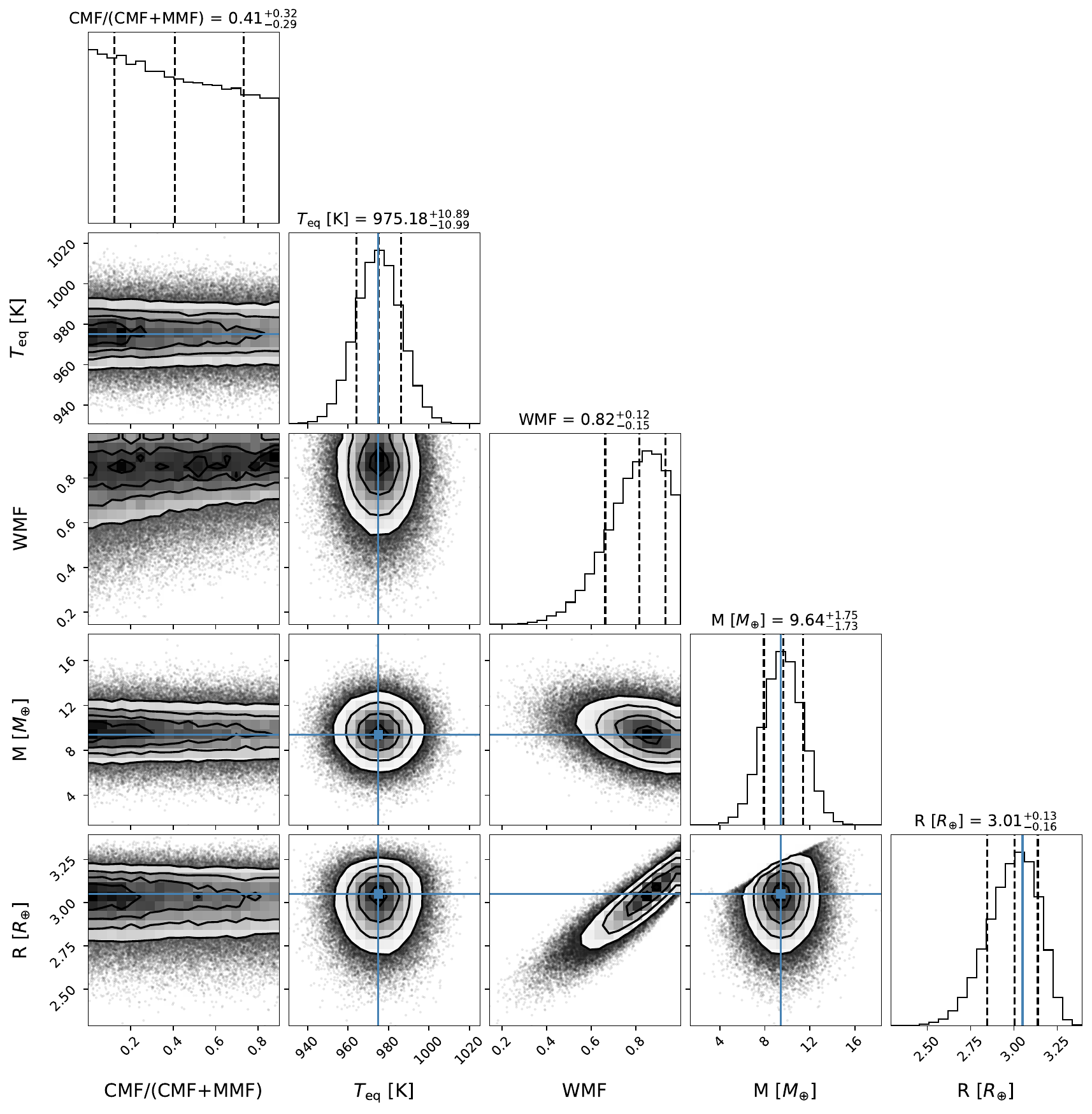} 
\includegraphics[scale=0.4]{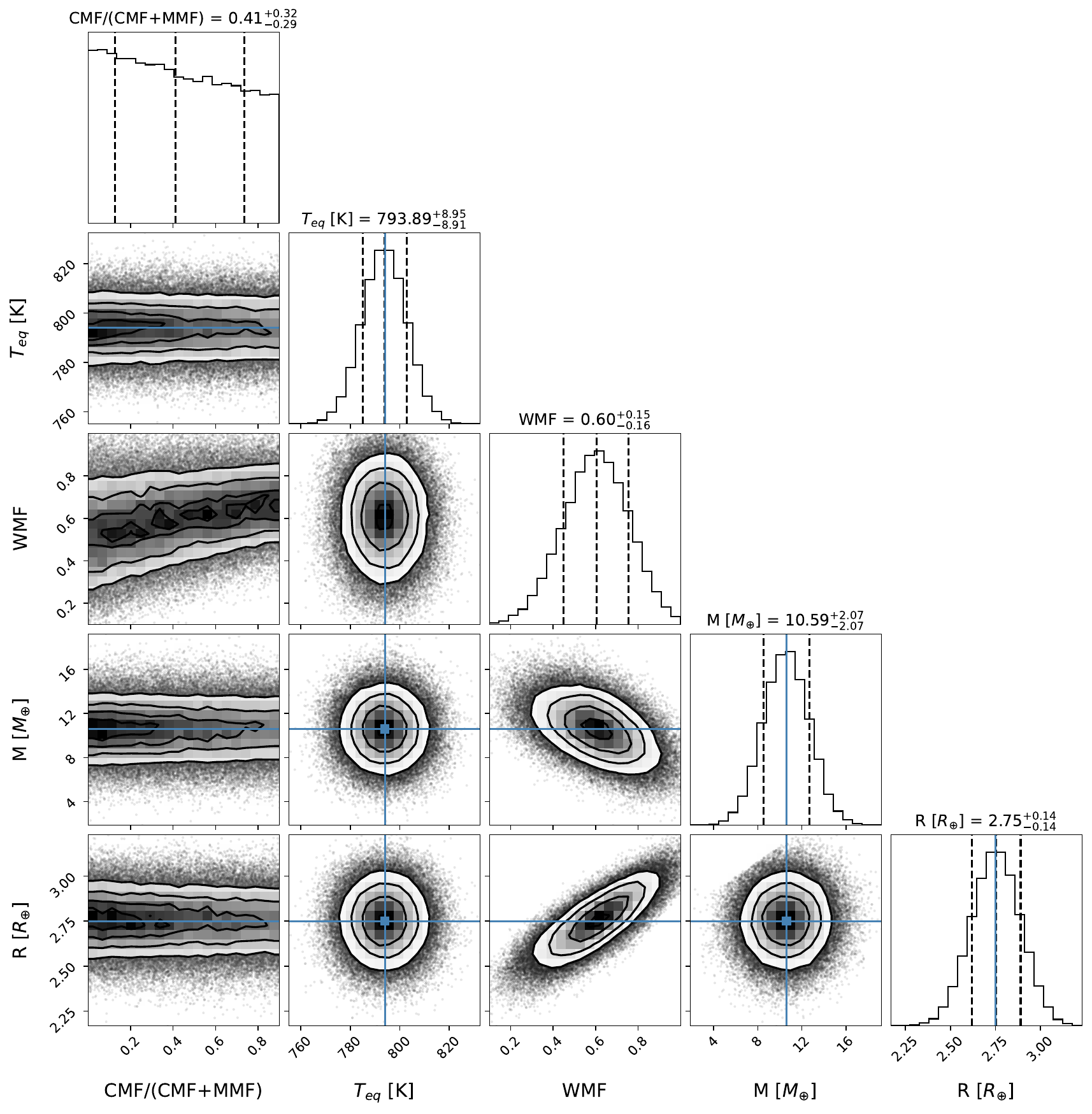} 
\caption{Corner plot of the posteriors of our MCMC retrieval of \targetb~ (top) and planet~c (bottom) interior structures and compositions.} 
      \label{fig:interior_cornerplots} 
\end{figure*} 

  \begin{figure*}[!ht]
 \centering
  \includegraphics[scale=0.55]{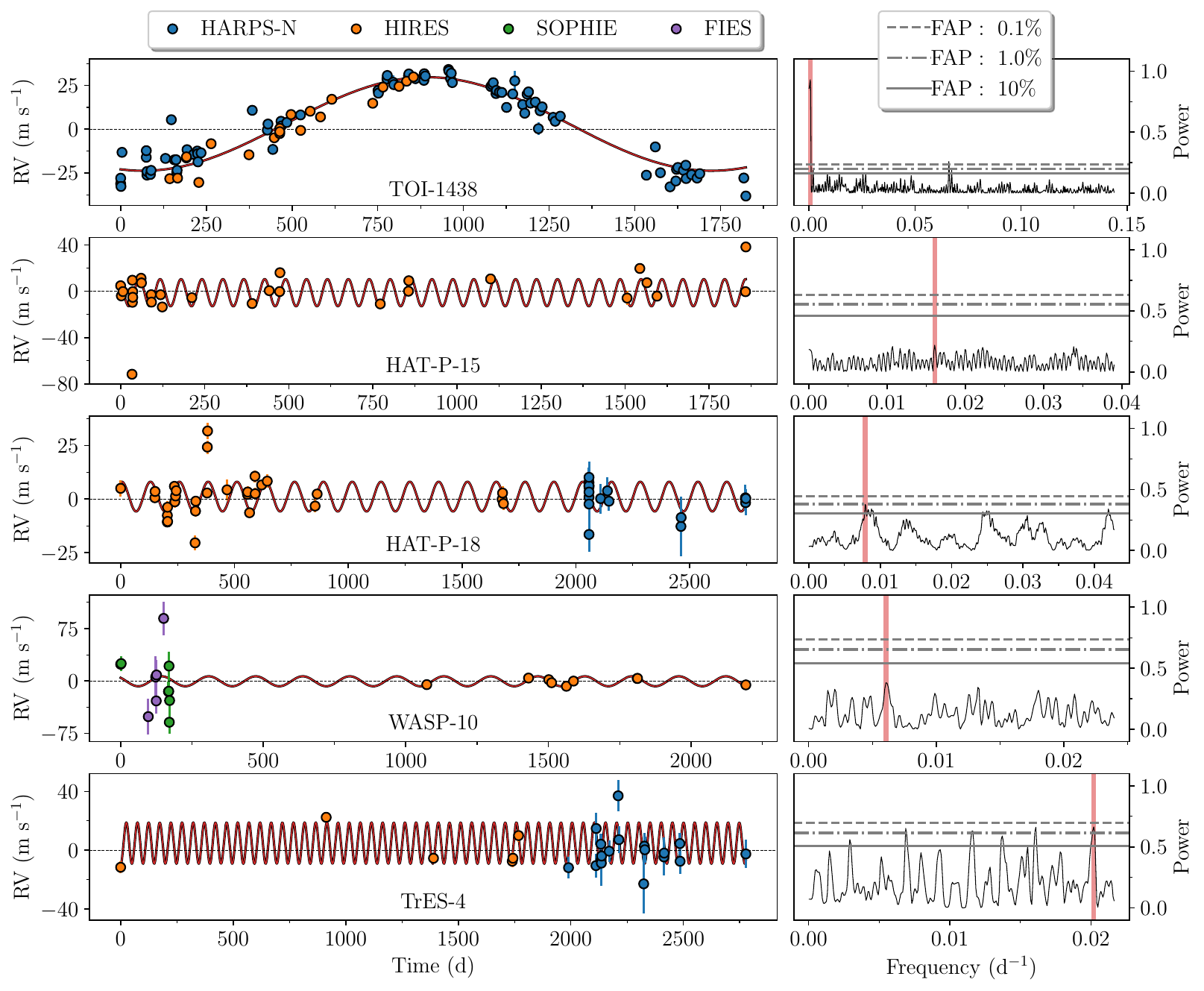}
   \caption{RV residuals for \targeta~and the four systems highlighted in Fig.~\ref{fig:jitter}  are shown in the plots to the left. The sinusoid shown in red in each of these is calculated from the most prominent peak highlighted in red in the associated GLS to the right. The false alarm probabilities are shown as horizontal lines, with values indicated in the legend.}
      \label{Figure: residuals}
 \end{figure*}
 
\end{appendix}

\end{document}